\documentclass[apj,twocolumn,twocolappendix,numberedappendix]{openjournal}
\usepackage{enumitem,multirow}
\usepackage{ctable}

\usepackage{amsmath}
\usepackage{amssymb}
\usepackage{graphicx} 
\usepackage{comment}
\usepackage{url}
\usepackage{hyperref}
\usepackage{color}
\usepackage{gensymb}
\usepackage{bm}
\usepackage{upgreek}
\usepackage{makecell}
\usepackage{placeins}
\hypersetup{
     colorlinks   = true,
     citecolor    = blue,
     linkcolor = blue
}
\usepackage[utf8]{inputenc}
\usepackage[T1]{fontenc}

\newcommand\msun{\ensuremath{\mathrm{M}_{\odot}}}
\newcommand\Gaia{\textit{Gaia}}

\shortauthors{LU ET AL.}
\shorttitle{DETECTABILITY OF SUBHALOS IN STREAMS}

\begin{document}

\title{Detectability of dark matter subhalo impacts in Milky Way stellar streams\vspace{-1.5cm}}
\author{Junyang Lu$^1$ }
\author{Tongyan Lin$^1$ }
\author{Mukul Sholapurkar$^{1,2}$ }
\author{Ana Bonaca$^3$}
\thanks{$^*$E-mail: jul143@ucsd.edu}
\affiliation{$^1$ Department of Physics, University of California, San Diego, CA 92093, USA \\
  $^2$ Institute of High Energy Physics, Austrian Academy of Sciences, Dominikanerbastei 16, 1010 Vienna, Austria\\
  $^3$  The Observatories of the Carnegie Institution for Science, 813 Santa Barbara Street, Pasadena, CA 91101, USA
}


\begin{abstract}\noindent
Stellar streams are a promising way to probe the gravitational effects of low-mass dark matter (DM) subhalos. In recent years, there has been a remarkable explosion in the number of stellar streams detected in the Milky Way, and hundreds more may be discovered with future surveys such as LSST. Studies of DM subhalo impacts on streams have so far focused on a few of the thinnest and brightest streams, and it is not known how much information can be gained from the others. In this work, we develop a method to quickly estimate the minimum detectable DM subhalo mass of a given stream, depending on its width, length, distance, and stellar density. We use an analytic model for the impacts and apply a test statistic to determine whether they are detectable. We consider several observational scenarios, based on current and future surveys including \Gaia, DESI, Via, and LSST. We find that at 95\% confidence level, a stream like GD-1 has a minimum detectable subhalo mass of $\sim 6\times 10^6~\msun$ in \Gaia\ data and $\sim 8\times 10^5~\msun$ with LSST 10 year sensitivity. 
Applying our results to confirmed Milky Way streams, we rank order them by their sensitivity to DM subhalos and identify promising ones for further study.

\end{abstract}

\maketitle

\tableofcontents

\section{Introduction}

The predictions of the standard cosmological model, $\Lambda$CDM, match well with measurements of the matter power spectrum $P(k)$ at large scales ($k \lesssim 10~\mathrm{h/Mpc}$)~\citep{chabanier2019}. At small scales, however, we currently lack reliable measurements. Various dark matter (DM) models with different properties can have drastically different behaviors at small scales ($k \gtrsim 10~\mathrm{h/Mpc}$)~\citep{2017ARA&A..55..343B,bechtol2023snowmass2021}. Probing small scales will test the paradigm of cold dark matter in a new regime, and could reveal hints for the underlying particle physics of dark matter.

In the late universe, density perturbations with $k \gtrsim 10~\mathrm{h/Mpc}$ have become nonlinear and clustered into bound Galactic-scale structures and substructures.
Measuring DM subhalo abundances in the range $10^5\,\msun-10^8\,\msun$ can effectively put constraints on small scales ($10^1-10^3~\mathrm{h/Mpc}$) of the matter power spectrum~\citep{2017ARA&A..55..343B}. 
Several observational probes have been proposed to determine the abundance of DM subhalos in this mass range, including ultra-faint dwarf galaxies (e.g.~\cite{2010MNRAS.404L..16M,Nadler_2024}), strong gravitational lensing (e.g.~\cite{2020MNRAS.491.6077G,keeley2024jwstlensedquasardark}) and stellar streams (e.g.~\cite{Ibata_2002,Johnston_2002,Banik:2019smi}).

The focus of this paper is on the potential for stellar streams to probe these low mass DM subhalos. Stellar streams are formed from the tidal
debris of dwarf galaxies and globular clusters in the Galactic halo~\citep{10.1093/mnras/275.2.429,1996ApJ...465..278J,1999MNRAS.307..495H,bonaca2024stellarstreamsgaiaera}. The resulting structures, especially those formed from globular clusters, are thin and dynamically cold, making them  sensitive to the gravitational effects of low-mass DM subhalos~\citep{Ibata_2002, Johnston_2002}. A subhalo passing near a stream, referred to as an impact, induces a perturbation, with measurements of streams potentially providing a record of past impacts.

The most commonly discussed signature of a subhalo impact is a density fluctuation or gap~\citep{Yoon2011ApJ...731...58Y,Carlberg2012}. The rates for impacts to cause gaps in streams have been studied both analytically and in simulations~\citep{Carlberg2012,Erkal_2016,Barry:2023ksd,menker2024,Adams:2024zhi}, 
and it is estimated that a gap from a subhalo as low as a few $\times 10^5 \msun$ could be detectable with LSST photometry~\citep{drlicawagner2019probing}.
Rather than identifying individual gaps, the cumulative history of subhalo impacts can also be probed with the power spectrum of density fluctuations~\citep{Bovy2017}. This technique similarly could be sensitive to subhalos down to $10^5 \msun$.
Some of the above density features have already been found in well-studied streams such as GD-1, and shown to be consistent with impacts from DM subhalos at the $\sim 10^7 \msun$ mass scale~\citep{Bonaca2019,Banik2021a,Banik:2019smi}. However, other effects also give rise to perturbations in streams, including epicyclic density variations~\citep{2012MNRAS.420.2700K,Ibata2020}, progenitor disruption~\citep{Webb2019}, and other satellites and baryonic structures in the Milky Way~\citep{2017MNRAS.470...60E,pearson2017gapslengthasymmetrystellar,2019MNRAS.484.2009B,deBoer2020,Bonaca2020}.

In addition to gaps, DM subhalos leave other signatures such as velocity perturbations and angular deflections. Accounting for these can add valuable information to help disentangle the potential sources of a stream perturbation. Previous works \citep{Erkal_2015_2,hilmi2024inferringdarkmattersubhalo} have developed frameworks to infer impact properties from observed density and kinematic data. Precise kinematics would aid in reconstruction of a perturber's orbit and mass, in principle allowing for its identification. Photometric surveys like \Gaia~\citep{GaiaEDR3} and LSST~\citep{lsstsciencebook2009} can provide proper motions for the stars. Meanwhile, spectroscopic surveys like S5~\citep{Li_2022_S5}, DESI~\citep{Cooper_2023}, Via\footnote{\url{https://via-project.org}}, 4MOST~\citep{de_Jong_2012}, and WEAVE~\citep{WEAVE} can provide confirmation of stream stars and line-of-sight velocities.

A limitation of prior studies is that they have all focused on DM subhalo detection in a few exemplar streams, in particular assuming a cold and long stream, of which the prototypical example is GD-1~\citep{Grillmair_2006}. This ideal stream is certainly well-suited to the problem. At the same time, there has been a rapid growth in the number of Milky Way stellar streams identified in the era of \Gaia~\citep{bonaca2024stellarstreamsgaiaera} with over 100 confirmed~\citep{2023MNRAS.520.5225M,Ibata_2023_STREAMFINDER}. There may be many more in \Gaia\ data waiting to be confirmed~\citep{Shih:2023jfv} and hundreds more discovered with LSST~\citep{pearson2024forecastingpopulationglobularcluster}. It is essential to understand how much information we can gain from all of the streams combined, and which of the observed streams are the most promising.

In this paper, we build a statistical framework for rapidly estimating the detectability of DM subhalo impacts in Milky Way streams, accounting for stream properties such as density, internal dispersion or width, and distance. Our focus is on the minimum detectable DM subhalo mass, as this is the quantity most closely tied to the matter power spectrum.
The subhalo mass refers to the total mass assuming that the subhalo is modeled as a Plummer sphere.
We consider the detectability of a single strong impact, motivated by previous studies estimating that there are $O(1)$ impacts leading to a detectable gap over the lifetime of the stream~\citep{Erkal_2016,menker2024,Adams:2024zhi}.
We systematically explore how stream properties affect the minimum detectable subhalo mass in this impact, and apply our resulting model to currently-known stellar streams of the Milky Way.

Streams in the Milky Way can be characterized by a number of properties, and are each on a unique orbit. Rather than modeling all streams in detail, we will consider an idealized model of a circular stream with the observer at the Galactic Center. The streams are characterized by intrinsic properties, in particular
stream width in angle $\sigma_\theta$, total number of stars per unit angle $\lambda$, distance to the Galactic Center $r_0$ and length in angle $l$. 
When the subhalo is modeled as a Plummer potential, there is a simple analytic approximation for the effect of a subhalo impact on the circular stream~\citep{Erkal_2015}. The advantage of this approach is that it enables rapid simulation of impacted streams, evaluation of likelihoods, and extensive exploration over subhalo and stream properties. It also provides a clean way to illustrate the basic dependencies. 

The detectability of subhalos depends of course on the available observations. We will evaluate four observational scenarios: a) idealized ${\emph{Gaia}}$-era data, b) ${\emph{Gaia}}$ data combined with spectroscopic data from DESI, c) LSST single epoch sensitivity with spectroscopic data from Via and d) LSST 10 year sensitivity with spectroscopic data from Via. The LSST 10 year sensitivity data is expected to include 10 times more stars at a distance of 10 kpc compared to  ${\emph{Gaia}}$ (see Fig.~\ref{fig:no_density}). Via will improve the accuracy of radial velocity by a factor of 10 compared to DESI (see top left panel of Fig.~\ref{fig:vrad}).

With this, we build up a general analytic formula for minimum detectable subhalo mass as a function of stream properties, and for different observational scenarios. 
We apply this model to the known catalog of stellar streams and provide a ranking of the most promising candidates. We emphasize this ranking is not the complete picture for the most promising streams, since detection prospects also depend on the expected rate of subhalo impacts as well as baryonic backgrounds. However, our study provides a foundation for considering the population of detectable impacts in Milky Way stellar streams.

In Sec.~\ref{sec:circular}, we describe the setup for modeling subhalo impacts on a circular stream, which relies on a combination of the analytic model and simulated streams to obtain stream dispersion. In Sec.~\ref{sec:observations}, we discuss how observational errors and limits are accounted for, introducing a survey-effective error that folds in all effects. We identify four benchmark observational scenarios. The statistical framework is presented in Sec.~\ref{sec:statistics}, along with the results for minimum detectable subhalo mass. The expected precision on subhalo mass is also given. Our main result for detectability as a function of stream properties is given in Sec.~\ref{sec:minimum_mass_streams}. In Sec.~\ref{sec:stream_catalog}, we apply our result to streams of the Milky Way.
A discussion of how our results would change in more realistic stream models can be found in Sec.~\ref{sec:discussion}, and conclusions in Sec.~\ref{sec:conclusion}.

\section{Subhalo impacts on a circular stream}
\label{sec:circular}

As a starting point, we consider an idealized model of a stream on a circular orbit about the Galactic Center, moving in the $xy$ plane. The progenitor of the stream is moving with a velocity $V_c$ at a distance $r_0$ from the Galactic Center. We assume a logarithmic potential for the galaxy, which implies a circular velocity independent of $r_0$ that we set as $V_c=220~\mathrm{km/s}$. We will consider $r_0$ in the range 5-40 kpc. To further keep things simple, we assume the observer is located at the Galactic Center. In such a setup, the effect of a DM subhalo impact on the stream can be modeled analytically. We can therefore use this model to rapidly simulate data as well as to perform fast likelihood calculations. 

This approach ignores the effect of non-circular orbits of realistic streams and the fact that observers are actually located at the Sun rather than the Galactic Center. However, it enables us to account for a range of stream characteristics such as its length, density, thickness, and distance to the observer. We can then apply our results to known streams in the Milky Way, as a first estimate of their sensitivity to DM subhalos.

We first review the analytic model for subhalo impacts below. The analytic description does not account for the dispersion of the stream, so we will use simulated streams to develop a model of the position and velocity dispersion along the stream.  We then describe how we combine these ingredients for mock data, and validate our method with simulated subhalo impacts.

\subsection{Analytic model for subhalo impact}
\label{sec:analytic}

\begin{figure*}[t]
\centering
\includegraphics[width=0.85\columnwidth]{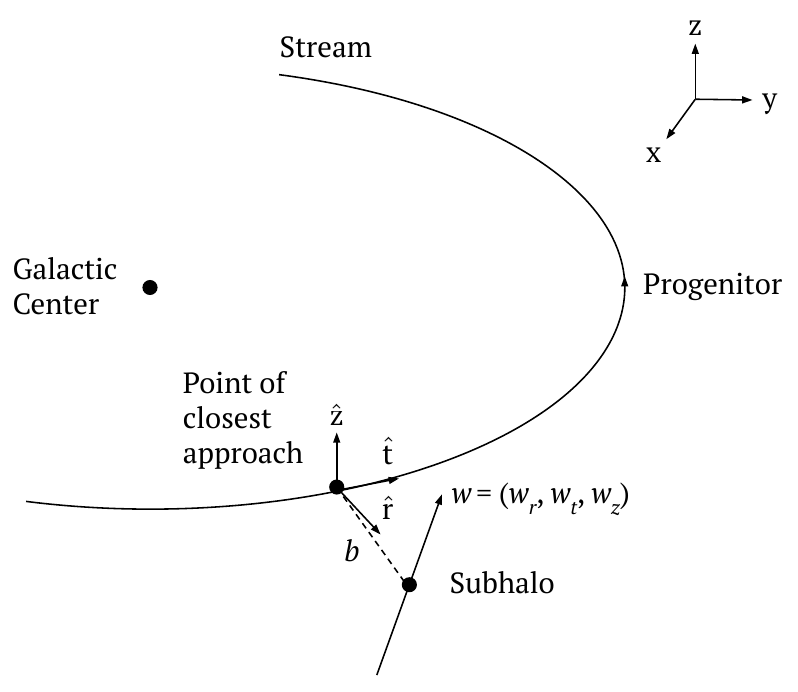}
\hspace{0.5cm}
\includegraphics[width=1.1\columnwidth]{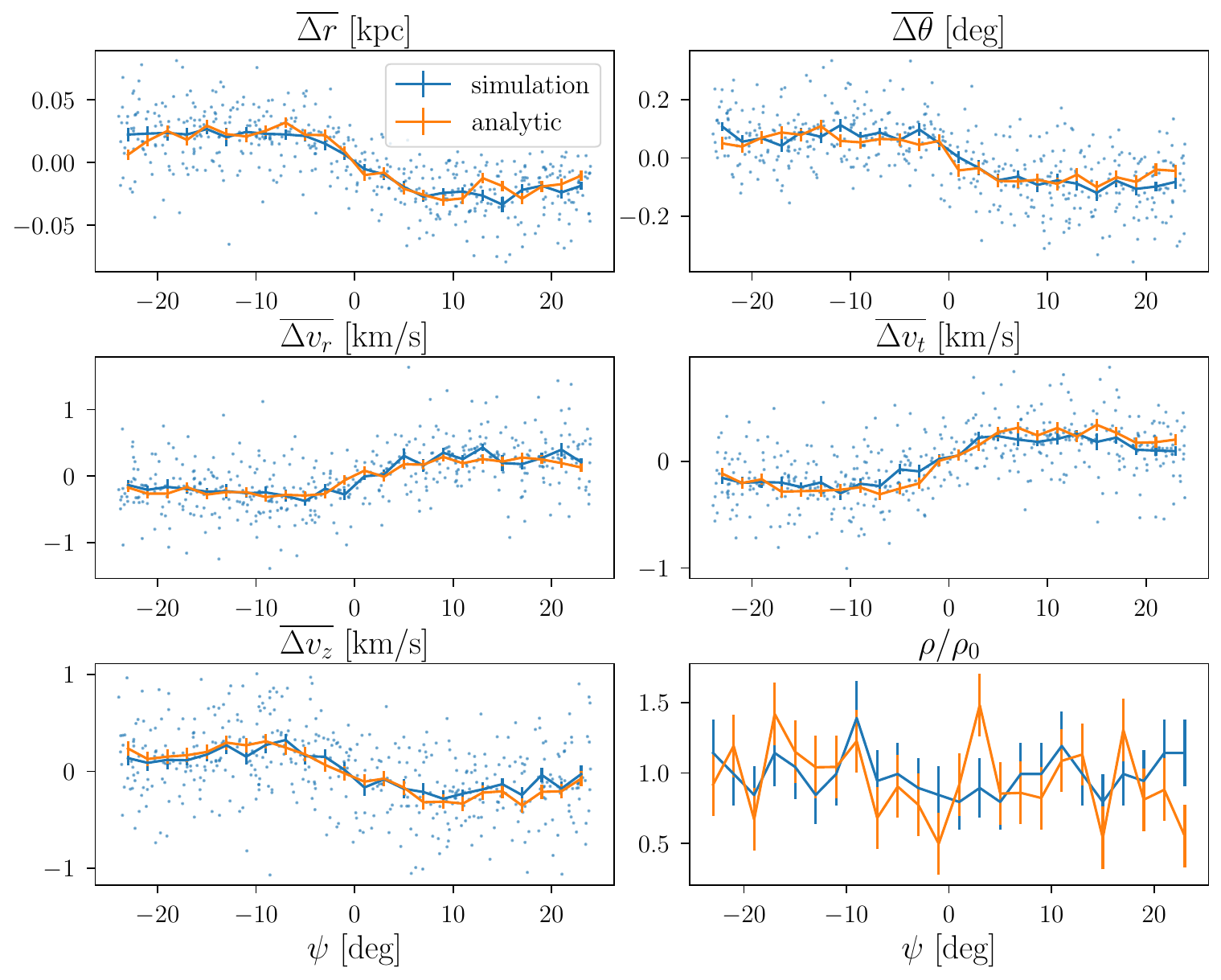}
\caption{\textbf{Left:} The geometry of a subhalo passing by a stellar stream. The stream is moving in a circular orbit in the $xy$ plane of the galaxy. At every point on the stream, the stars are moving along the tangential $\hat t$ direction. At the point of closest approach, the subhalo passes by the stream with distance $b$, and velocity $(w_r, w_t, w_z)$ in the Galactic frame. \textbf{Right:} A comparison between the observables and uncertainties from simulated perturbed streams (blue) versus using the analytic model combined with random Gaussian noises generated based on uncertainties of simulated unperturbed streams (orange). The blue dots show the simulated stars in the perturbed stream.
The stream is generated with $\sigma_\theta=0.2^\circ$, $r_0=10~\mathrm{kpc}$ and $\lambda=10~\mathrm{deg}^{-1}$. The subhalo impact is parametrized by $M_{\mathrm{sh}} = 10^{7.5}~\msun$, $r_s = 0.91~\mathrm{kpc}$, $b=0~\mathrm{kpc}$, $t=315~\mathrm{Myr}$, $w_r=0~\mathrm{km/s}$, $w_t=0~\mathrm{km/s}$ and $w_z=180~\mathrm{km/s}$.}
\label{fig:subhalo_geometry_and_observables}
\end{figure*}

We make use of the model developed in \cite{Erkal_2015,Erkal_2015_2} to approximate the subhalo impact on stellar streams. In this model, all stream stars are assumed to initially be on the progenitor's orbit, and at some point in time, they instantaneously receive a velocity kick due to the gravitational effect of the passing DM subhalo. Treating these kicks as small perturbations, analytic results for the perturbed orbits of the stream stars can be obtained. This gives a model for the perturbed stream as a function of time after the impact. 

The geometry of the stream and the DM subhalo impact are illustrated in the left panel of Fig.~\ref{fig:subhalo_geometry_and_observables}. The stream moves on a circular orbit with radius $r_0$ while the subhalo, modeled as a Plummer sphere, passes by the stream in an arbitrary direction. The subhalo impact is parameterized by the following:
\begin{itemize}
    \item $M_{\mathrm{sh}}$: total subhalo mass
    \item $r_s$: subhalo scale radius
    \item $b$: impact parameter (distance of the closest approach between the subhalo and the stream)
    \item $t$: time since the moment of closest approach
    \item $\bm{w}=(w_r, w_t, w_z)$: subhalo velocity in the Galactic frame, decomposed as the velocity along the radial direction at the point of closest approach $w_r$, the velocity along the tangential direction at the point of closest approach $w_t$, and the velocity perpendicular to the orbital plane $w_z$
\end{itemize}
The potential of the subhalo is also a possible free parameter but as noted above, the analytic model assumes a Plummer potential.

During the subhalo flyby, the stream is approximated to be moving with a constant velocity along a straight line. We parameterize location along the stream arm at this time by an angle $\psi_0$, where $\psi_0 = 0$ is the point of closest approach. Treating the impact with the impulse approximation, then stars along the stream get velocity kicks given by \citep{Erkal_2015_2}
\begin{align}
    \label{eq:vkr}
    \Delta u_r(\psi_0) &=\frac{2G M_{\mathrm{sh}}}{r_0^2 w_\perp^2 w_\mathrm{rel}}\frac{b w_\mathrm{rel}^2 \frac{w_z}{w_\perp}-\psi_0 r_0 w_\parallel w_r}{\psi_0^2 +\frac{(b^2+r_s^2)w_\mathrm{rel}^2}{r_0^2 w_\perp^2}},\\
    \label{eq:vkt}
        \Delta u_t(\psi_0) &=-\frac{2G M_\mathrm{sh} \psi_0}{w_\mathrm{rel}r_0(\psi_0^2+\frac{(b^2+r_s^2)w_\mathrm{rel}^2}{r_0^2w_\perp^2})}, \\
    \label{eq:vkz}
    \Delta u_z(\psi_0) &=-\frac{2G M_{\mathrm{sh}}}{r_0^2 w_\perp^2 w_\mathrm{rel}}\frac{b w_\mathrm{rel}^2 \frac{w_r}{w_\perp}+\psi_0 r_0 w_\parallel w_z}{\psi_0^2 +\frac{(b^2+r_s^2)w_\mathrm{rel}^2}{r_0^2 w_\perp^2}},
\end{align}
where $\Delta u_r$ and $\Delta u_t$ are velocity kicks in the plane of the stream orbit, along the radial and tangential direction respectively, and $\Delta u_z$ is the velocity kick perpendicular to the plane of the orbit. Note that  $w_\parallel=w_t - V_c$ is the relative stream-subhalo velocity along the stream, $w_\perp = \sqrt{w_r^2+w_z^2}$ is the magnitude of the subhalo velocity perpendicular to the stream, and $w_\mathrm{rel}=\sqrt{w_\parallel^2+w_\perp^2}$ is the magnitude of the relative stream-subhalo  velocity.

Kicks perpendicular to the orbital plane $\Delta u_z$ cause the stars to oscillate with respect to the original plane. Kicks along the radial direction $\Delta u_r$ cause oscillations within the orbital plane, including oscillations in the density. Kicks along the orbit $\Delta u_t$ cause oscillations within the orbital plane as well but more importantly, they change the orbital period of stars, so that stars race ahead or fall behind the impact point. This leads to an underdensity (gap) in the stream that grows with time. 

From these velocity kicks, \cite{Erkal_2015_2} derives analytic formulae for 6 observables today as functions of $\psi_0$:
\begin{itemize}
    \item $\Delta r$: shift in radial distance
    \item $\Delta \theta \equiv \Delta z / r_0$: angular shift along $z$ direction
    \item $\Delta v_r$: velocity perturbation in radial direction
    \item $\Delta v_t$: velocity perturbation in tangential direction
    \item $\Delta v_z$: velocity perturbation in $z$ direction
    \item $\rho/\rho_0$: density ratio, where $\rho$ denotes the number of stars per degree today, while $\rho_0$ denotes the number of stars per degree at the time of impact
\end{itemize}
Here $\psi_0$ is the angle along the stream at the time of impact, while we are interested in these observables as functions of the observed angle along the stream today, which is denoted by $\psi$. We transform the functions of $\psi_0$ above to functions of $\psi$ using the formula for $\psi(\psi_0, t)$ in \cite{Erkal_2015_2}. 

Though this analytic model is efficient in estimating the observables, it has some limitations: a) It ignores the internal dispersion and the deviation of the stream stars from the progenitor's orbit; b) It assumes that the impacted stream is moving with a constant velocity along a straight line, which is true only if
\begin{equation}
\label{eq:analytic_assumption}
    \frac{w_\mathrm{rel}}{w_\perp}\sqrt{b^2+r_s^2} \ll r_0~~\text{and}~~\frac{V_c}{w_\perp}\sqrt{b^2+r_s^2}\ll r_0.
\end{equation}
However, in our work, neither of these limitations is a significant concern. As we will see in Sec.~\ref{sec:simulated_impact}, the dispersion in the stream and the deviation from the progenitor's orbit is small compared to the orbit parameters, $\sigma_v \ll V_c$ and $\sigma_r \ll r_0$.  
Furthermore, we will use default values of $b=0~\mathrm{kpc}$ and $w_\mathrm{rel}\sim w_\perp \sim V_c \sim 200~\mathrm{km/s}$ throughout our analysis to comply with the conditions in Eq.~\ref{eq:analytic_assumption}.
Even when considering deviations from these default values in Sec.~\ref{sec:nuisance}, we can still explore a range of parameters while requiring them to satisfy Eq.~\ref{eq:analytic_assumption}.

\subsection{Simulated streams and stream characteristics}
\label{sec:simulated_stream}

In the analytic model, the stream is on a single orbit, which therefore assumes an infinitely thin stream.
For streams with a non-zero dispersion, we use this analytic model to predict the expected mean value of the observables along the stream.
We also need to know the uncertainties for the 6 observables. In this section, we deal with the statistical uncertainty, which is controlled by the internal dispersion of the stream and number density along the stream. For this, we generate simulated streams and calculate their internal dispersion. 

In order to introduce uncertainties for observables, we split the stream into evenly sized angular bins in $\psi$. We choose to use 2 degree bins throughout the paper. This ensures we have as many stars per bin as possible while not losing the perturbation features across bins.
To keep the notation concise, we will use $X$ to denote the shift in any of the position or velocity quantities $\Delta r$, $\Delta \theta$, $\Delta v_r$, $\Delta v_t$ and $\Delta v_z$.
For a given angular bin along $\psi$ with $n$ stars in the bin, the observable is $\overline{X} = \sum_i X_i/n$ with $X_i$ being the individual measurements.
All observed quantities are defined with respect to the unperturbed stream (e.g. $\Delta r_i = r_i - r_0$).
The uncertainty on the mean value of $X$ is given by
\begin{equation}
\label{eq:error_kin}
    \sigma_{\overline{X}}=\frac{\sigma_{X}}{\sqrt{n}}
\end{equation}
where $\sigma_{X}$ is the dispersion of stars in the stream, which can be obtained from the simulated streams. Note that we have not included any observational errors here and will introduce them in more detail in Sec.~\ref{sec:observations}. In addition, to generalize these observables to arbitrary streams, one also has to account for uncertainties on the baseline unperturbed model achieved from fitting the stream orbit; we assume that such uncertainties are small and do not include them.

Finally, for the observable density ratio $\rho/\rho_0$, the Poissonian uncertainty is simply 
\begin{equation}
\label{eq:error_den}
    \sigma_{\rho/\rho_0}=\frac{1}{\sqrt{n}}.
\end{equation}
Again, $\rho_0$ is the density from the unperturbed analytic model, which is a constant across bins and does not contribute to the uncertainty.

To obtain the internal dispersion $\sigma_{X}$, we simulate streams using the particle spray technique from \cite{Fardal_2015} as implemented in \texttt{gala} \citep{gala,adrian_price_whelan_2020_4159870}. The basic idea is to control how the progenitor releases stars and do orbit integration for all the released stars under the host potential. Fortunately, we only need to perform the simulation for a given smooth (unperturbed) stream once. We extract the internal dispersion from the smooth stream for all position and velocity observables. In particular, we generate one stream arm of $90\degree$ length, and use the region between [8\degree, 52\degree] assuming the progenitor is located at 0\degree. This avoids the portions around the progenitor and near the stream ends, resulting in a fairly constant dispersion in the selected region. We use the average dispersion over all the bins in this region for $\sigma_X$.

With the above results, we can generate mock data by sampling the observables in each $\psi$ bin. The mean is given by the analytic model and the uncertainties are given by Eq.~\ref{eq:error_kin} and~\ref{eq:error_den}.
Observational errors can also be added to this procedure, and we will consider them in detail in Sec.~\ref{sec:observations}. Note that this process assumes that the dispersion along the stream is not substantially modified by the presence of the subhalo, which we will validate later in Sec.~\ref{sec:simulated_impact}.

\begin{figure*}[t]
\centering
\includegraphics[width=0.66\columnwidth]{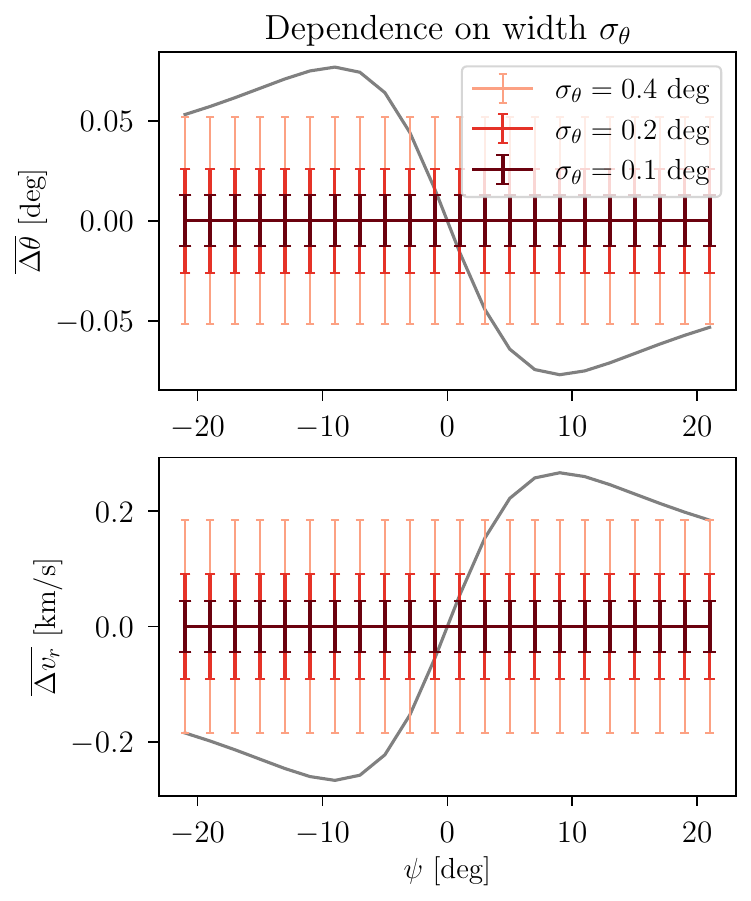}
\hspace{0.01cm}
\includegraphics[width=0.66\columnwidth]{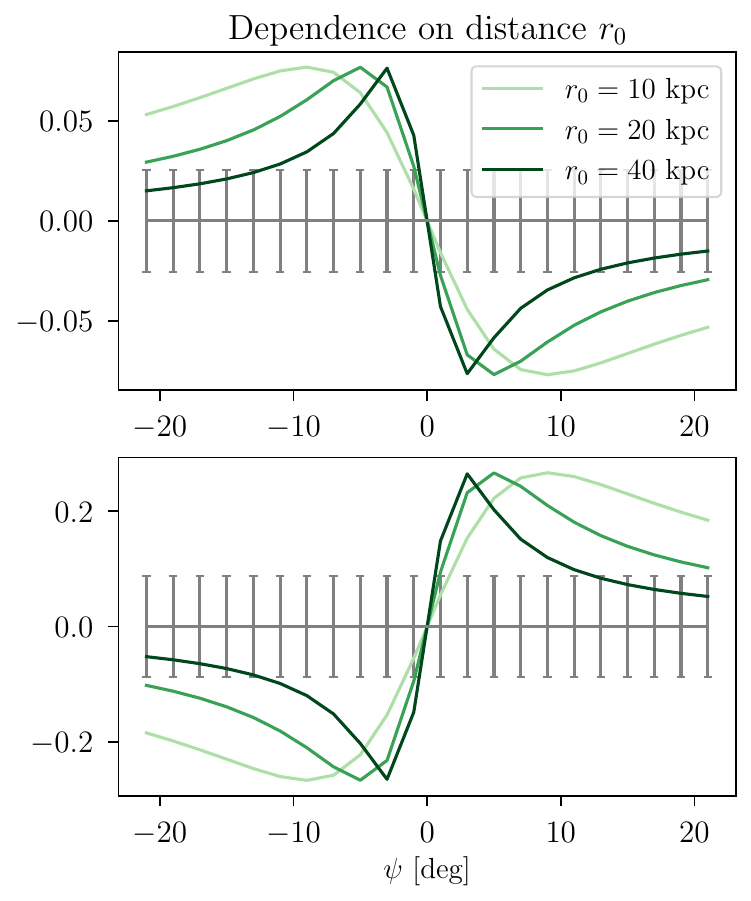}
\hspace{0.01cm}
\includegraphics[width=0.66\columnwidth]{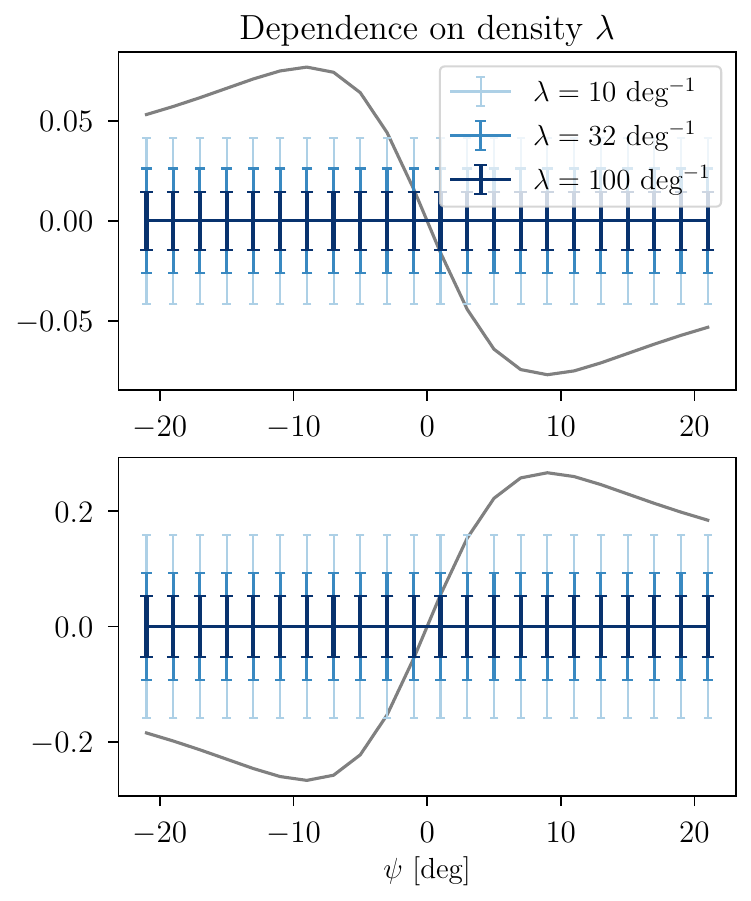}
\caption{Dependence of some key observables, $\overline{\Delta \theta}$ and $\overline{\Delta v_r}$, on stream properties. The subhalo impact is the same as the right panel of Fig.~\ref{fig:subhalo_geometry_and_observables}. In each column, we vary one stream property and fix the others to demonstrate how a given property affects either the size of the error bars or the shape of the signal. The default stream properties are $\sigma_\theta = 0.2$ deg, $r_0=10~\mathrm{kpc}$ and $\lambda=32~\mathrm{deg}^{-1}$. Signals here are only calculated from the analytic model and do not account for the uncertainties. Error bars here are only due to stream dispersion. When observational errors are included (Sec.~\ref{sec:observations}), changing $r_0$ will also affect the error bars.}
\label{fig:diff_properties}
\end{figure*}

Our goal is to characterize the detectability of a subhalo impact in different streams, which is ultimately determined by the errors on the mean observables. In the discussion below, we identify a set of parameters controlling the dispersion and number density of the stream. We first review how these quantities are set in the approach of \cite{Fardal_2015}. The dispersion in the radial and $z$ positions are proportional to the tidal radius,
\begin{equation}
\label{eq:rz_on_rtidal}
    \sigma_{r} \propto (\sigma_{z}=r_0\sigma_\theta) \propto r_{\mathrm{tidal}},
\end{equation}
and the dispersion in the radial, tangential and $z$ velocities are proportional to the circular velocity and the ratio of tidal radius to the orbital radius,
\begin{equation}
\label{eq:vrtz_on_rtidal}
    \sigma_{v_r} \propto \sigma_{v_t} \propto \sigma_{v_z} \propto V_c \frac{r_{\mathrm{tidal}}}{r_0},
\end{equation}
where the tidal radius is~\citep{1962AJ.....67..471K}
\begin{equation}
    r_{\mathrm{tidal}}=\left(\frac{M_{\mathrm{prog}}}{fM(<r_0)}\right)^{1/3}r_0\propto\left(\frac{M_{\mathrm{prog}}}{r_0}\right)^{1/3}r_0.
\end{equation}
Here $M_\mathrm{prog}$ is the progenitor mass, $M(<r_0)$ is the enclosed mass for the host halo at a distance $r_0$, and $f$ is a constant depending on the host potential. We have used $f=2$ and $M(<r_0) \propto  r_0$ for a logarithmic potential.

Plugging $r_{\mathrm{tidal}}$ back into Eq. \ref{eq:rz_on_rtidal} and \ref{eq:vrtz_on_rtidal}, we have
\begin{equation}
    \sigma_{r} \propto \left(\frac{M_{\mathrm{prog}}}{r_0}\right)^{1/3}r_0, 
\end{equation}
\begin{equation}
\label{eq:width_and_mprog}
    \sigma_{\theta} \propto \sigma_{v_r} \propto \sigma_{v_t} \propto \sigma_{v_z} \propto \left(\frac{M_{\mathrm{prog}}}{r_0}\right)^{1/3} .
\end{equation}
Here we have suppressed the constants of proportionality, which are determined by $V_c = 220$ km/s and $O(1)$ constants specified in a given particle spray model. So fundamentally, the internal dispersions of the streams are controlled by the progenitor mass $M_\mathrm{prog}$ and the distance to the Galactic Center $r_0$. At a given $r_0$, we can trade the progenitor mass $M_\mathrm{prog}$ for the angular width of the stream $\sigma_\theta$, as this quantity is more closely related to observed stream properties.

From Eq.~\ref{eq:error_kin} and Eq.~\ref{eq:error_den}, the uncertainty also depends on the number of stars per bin $n$, or essentially the number of stars per unit angle $\lambda$. Following~\cite{Fardal_2015}, the stream density in the central part of the stream is controlled by the total number of stream stars $N_\mathrm{stars}$ as,
\begin{equation}
    \label{eq:lambda}
    \lambda\approx\frac{4}{3}\frac{N_\mathrm{stars}}{l},
\end{equation}
where $l$ is the angular length of the stream. The factor of $4/3$ is to account for the fact that the stream density becomes lower near the two ends, so the density $\lambda$ is correspondingly higher for the central part. For a given stream, $N_\mathrm{stars}$ can be determined from the stellar mass $M_\mathrm{stellar}$ and the mass distribution of stars from an initial mass function. We emphasize that generally the stellar mass $M_\mathrm{stellar}$ is an independent quantity and lower than the progenitor mass $M_{\rm prog}$, which we are using to set the stream dispersion. This could be due to the presence of dark matter for heavier stream progenitors, additional physics such as cluster dynamics or baryonic interactions that increases dispersion, as well as the fact that we only typically observe a part of the stream consisting of more recently stripped stars, while older parts of the stream have been lost.

In addition to uncertainties, the detectability also depends on the signals (observables) of a subhalo impact. From Eqs.~\ref{eq:vkr}-\ref{eq:vkz}, we see that the observables depend on the stream's distance to the Galactic Center $r_0$. The length of the stream is also important for the full signal to be present, since the impact is most relevant over a region of length $\sim w_\mathrm{rel} \sqrt{b^2 + r_s^2} / w_\perp$.  From~\cite{Fardal_2015}, the angular length of the stream $l$ is controlled by the age of the stream $T_\mathrm{form}$ and the velocity dispersion in the tangential direction $\sigma_{v_t}$:
\begin{equation}
\label{eq:length_and_age}
    l \propto \frac{T_\mathrm{form} \sigma_{v_t}}{r_0} \propto  \frac{T_\mathrm{form} \sigma_{\theta}}{r_0}.
\end{equation}
We again choose to characterize the stream by $l$ rather than $T_{\rm form}$ as $l$ is a direct observable.

\begin{table}[t!]
\begin{center}
\renewcommand{\arraystretch}{2}
\begin{tabular}{ |c|c|c|} 
\hline
 & \textbf{Description} & \textbf{Values}  \\ 
\hline
 \ $\sigma_\theta$ \ & \makecell{stream width}  & $0.1^\circ-1.6 ^\circ$  \\ 
\hline
 \ $r_0$ \ & \makecell{distance to the Galactic Center}  & $5-40$ kpc  \\ 
\hline
 \ $\lambda$ \ & \makecell{number of stars per unit angle }  & $10-320~\mathrm{deg}^{-1}$  \\ 
\hline
 \ $l$ \ & \makecell{stream length in angle}  & $10^\circ-100^\circ$  \\ 
 \hline
\end{tabular}
\caption{Intrinsic stream properties used to assess detectability. The last column gives the range of values that we focus on in this work.  $\sigma_\theta$ and $\lambda$ can also be traded for progenitor mass $M_{\rm prog}$ and stellar mass $M_{\rm stellar}$.}
\label{table:stream_props}
\end{center}
\end{table}

Now we can collect all the relevant stream properties that affect the detectability. Tab.~\ref{table:stream_props} summarizes the quantities used to characterize streams as well as a range of values considered here. We have selected the quantities that are most closely identified with measured stream properties. In the simulated streams, we can vary progenitor mass $M_\mathrm{prog}$, the age of the stream $T_\mathrm{form}$, distance to the Galactic Center $r_0$, and total number of stream stars $N_\mathrm{stars}$ to achieve any combination of the parameters in Tab.~\ref{table:stream_props}. For instance, at a fixed $r_0$, we adjust $M_\mathrm{prog}$ according to Eq.~\ref{eq:width_and_mprog} to vary $\sigma_\theta$ and adjust $T_\mathrm{form}$ according to Eq.~\ref{eq:length_and_age} to vary $l$. $M_{\rm stellar}$ then finally determines $\lambda$.

\subsection{Simulated impacts}
\label{sec:simulated_impact}

Combining the analytic model for the 6 observables and the dispersion from the simulated stream, we can assemble everything and generate the stream data for given subhalo impact parameters and stream properties.

To generate the impacted stream data, we compute the 6 observables at the central value of $\psi$ for each bin using the analytic model. Then for each value in each bin, we add a randomly sampled Gaussian noise based on Eq. \ref{eq:error_kin} and \ref{eq:error_den}. For simplicity, we assume the number of stars per bin, $n$, to be the average value across bins and thus a constant. For $n \gg 1$, the error introduced by this approximation is negligible.
The right panel of Fig. \ref{fig:subhalo_geometry_and_observables} shows a comparison between the observables and uncertainties from a simulated perturbed stream (blue) versus the observables from analytic model plus random Gaussian noises generated based on the dispersion of smooth stream simulation (orange). The simulated perturbed streams are obtained using \texttt{gala} by injecting the subhalo potential for a short period of time ($\pm$ 20 Myr) around the time of impact. We can see that the uncertainties from smooth stream simulation and perturbed stream simulation are very similar. Also, the differences in the observables between the simulation and analytic model are comparable to the error bars. This validates our method to rapidly generate observables for a perturbed stream from the analytic model. Note also that we have not included any observational error in this figure; including such errors will render some observables more useful than others as we will see later.

To give an intuitive sense about how stream properties affect detection of a subhalo impact, in Fig.~\ref{fig:diff_properties} we plot the signals and error bars for two main observables, $\overline{\Delta \theta}$ and $\overline{\Delta v_r}$. In each column, we vary one stream property and fix all the others. For the two observables considered, the stream width $\sigma_\theta$ and density $\lambda$ impact the size of the error bars, while the distance to the Galactic Center $r_0$ impacts the shape of the signal. This illustrates how the significance of the detection of the signal will change depending on the stream.


\section{Observational scenarios}
\label{sec:observations}

In this section, we describe how observational errors are accounted for, and present several benchmark observational scenarios. Each observational scenario consists of a combination of surveys or measurements for various observables. We will account for two main effects on stream observations. First, surveys have well-defined cut-offs in brightness which means only a fraction of the total stars are observable. In addition, we account for observational errors as a function of magnitude.

We first consider the distribution of magnitudes of stars in a typical cluster, which in turn depends on the mass distribution. We assume that the stars in the cluster are distributed according to the Chabrier power-law initial mass function (IMF)~\citep{Chabrier:2001dc}. 
Then the mass distribution is further convolved with a MIST isochrone~\citep{Dotter_2016, Choi_2016} with age = 10 Gyr and [Fe/H] $= -2$, corresponding to old, metal-poor stars. This gives the distribution of the magnitudes of stars in various wavelength filters. We implement this using the package \texttt{imf}\footnote{\url{https://github.com/keflavich/imf}} for the initial mass function and package \texttt{minimint}~\citep{minimint} for the MIST isochrone.

Given the distribution of magnitudes, we can generalize Eqs.~\ref{eq:error_kin}-\ref{eq:error_den}, which only account for internal stream dispersion, to include measurement uncertainty and survey limits. As discussed below in Sec.~\ref{sec:observational_errors}, for stream density ratio $\rho/\rho_0$ and position $\overline{\Delta \theta}$, observational errors can be neglected, and the effect can simply be accounted for by the fraction of observable stars, $n_{\rm obs}/n$. For the other kinematic quantities $\overline X$, we consider both the observational errors and the fraction of observable stars.

We now discuss how the observational errors are incorporated into each observable. Throughout this section, we keep the age and metallicity fixed as these are higher order effects in estimating detectability.

\begin{figure}[t]
\centering
\includegraphics[width=\columnwidth]{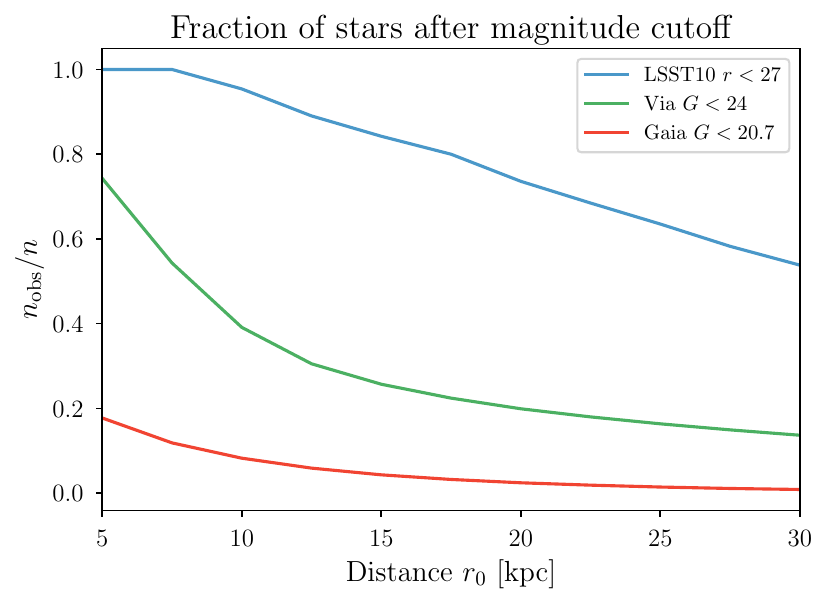}
\caption{The fraction of observable stream stars, $n_\mathrm{obs}/n$, as a function of the stream distance $r_0$ and for various magnitude cutoffs in different surveys. LSST10 is the 10-year sensitivity of LSST.}
\label{fig:no_density}
\end{figure}

\subsection{Observational errors}
\label{sec:observational_errors}

\subsubsection{Number density} 

We first consider the error on measurements of the density ratio $\sigma_{\rho/\rho_0}$. In the ideal case when observational errors are not present, the error follows Eq.~\ref{eq:error_den}. 
Since stars with magnitudes above a certain cutoff are not observable, we modify this to:
\begin{equation}
\label{eq:error_den_with_obs}
    \sigma_{\rho/\rho_0}^2=\frac{1}{n_\mathrm{obs}}=\frac{1}{n}\frac{n}{n_\mathrm{obs}}.
\end{equation}
Here $n_\mathrm{obs}$ is the number of observable stars per bin below a certain magnitude cutoff, while $n=\lambda \cdot (\psi~\mathrm{bin~size})$ is the number of actual stars per bin, observable or not. We have written the error in this way to split it into two factors.
The first factor, $1/n$, is the error in the limit of perfect observations. The second factor, $n/n_\mathrm{obs}$, only depends on the observational scenario, or more explicitly, the magnitude cutoff we use. Since observational errors on the angular positions of stars are typically much smaller than our default bin size of 2 degrees, we assume that stars can be perfectly assigned to a bin.

The magnitude cutoff can be made based on (1) kinematic (proper motions) identification of stream stars, (2) spectroscopic identification or (3) photometric measurements only. 
In the first case, we consider kinematic information from the \Gaia\ data ~\citep{GaiaEDR3} (with magnitude $G < 20.7$).
In the second case, we consider spectroscopy from the Via project (with maximum magnitude $G < 24$). We assume these provide high efficiency of stream star selection with negligible contamination of non-stream stars.

In the last case, LSST~\citep{lsstsciencebook2009} is expected to have a magnitude limit of $r < 27$ (10-year sensitivity).  Without kinematic information, it is possible that there may be a contaminating background of non-stream objects, particularly from mis-identified galaxies and to a lesser extent, the stellar halo~\citep{Nidever_2017}.
\cite{pearson2024forecastingpopulationglobularcluster} suggested that one or several selection techniques (e.g. isochrone-based selection, color-color selection, using proper motions, etc.) could be used to help recover the streams.
In this paper, we are modeling observables of stream stars only. We will neglect contamination in this study and take $r < 27$ as the best-case scenario for photometric measurements.

Fig.~\ref{fig:no_density} shows the fraction of stars remaining after magnitude cutoffs, $n_\mathrm{obs}/n$, as a function of the distance to the Galactic Center $r_0$ and for different surveys. This quantity will also play an important role in the other observables below.

\subsubsection{Positions}

In considering the error for the observable $\overline{\Delta \theta}$, we use the same magnitude cutoff as the density observable in the previous section. So the percentage of observable stars $n_\mathrm{obs}/n$ in Fig.~\ref{fig:no_density} applies here as well. Among the observable stars, the observational errors for star positions on the sky are negligible compared to the stream width. For instance, the position uncertainty with LSST is $<$ 0.1 arcsec for observations down to an apparent r-band magnitude of $r=24$. This contribution to the error, $\sigma_{\theta,\mathrm{obs}} = 0.1~\mathrm{arcsec} = 2.78~\times 10^{-5}~\text{deg}$, is far less than the stream width, $\sigma_{\theta,\text{disp}} \sim 0.1 - 2 $ deg. We will therefore only account for the reduction in the number of stars from the magnitude cutoff, and not consider the observational errors on those stars. The error for $\overline{\Delta \theta}$ can then be written as:
\begin{equation}
\label{eq:error_theta}
    \sigma_{\overline{\theta}}^2=\frac{\sigma_{\theta,\mathrm{disp}}^2}{n_\mathrm{obs}}=\frac{\sigma_{\theta,\mathrm{disp}}^2}{n}\frac{n}{n_\mathrm{obs}}.
\end{equation}

We do not use the measurements of the radial distances to the stars, as the observational errors on them are expected to be too large to be competitive in detecting subhalo impacts~\citep{GaiaEDR3,lsstsciencebook2009}.

\begin{figure*}[t]
\centering
\includegraphics[width=\columnwidth]{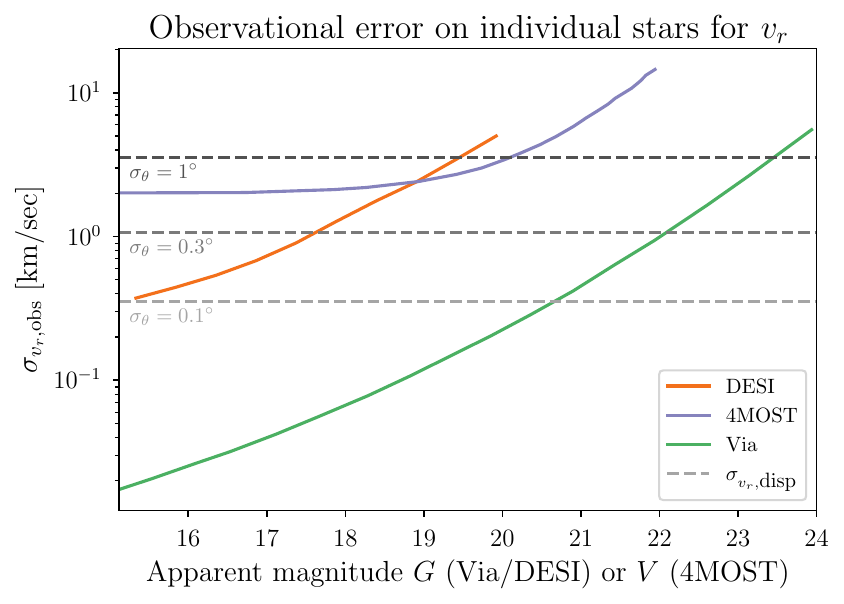}
\includegraphics[width=0.98\columnwidth]{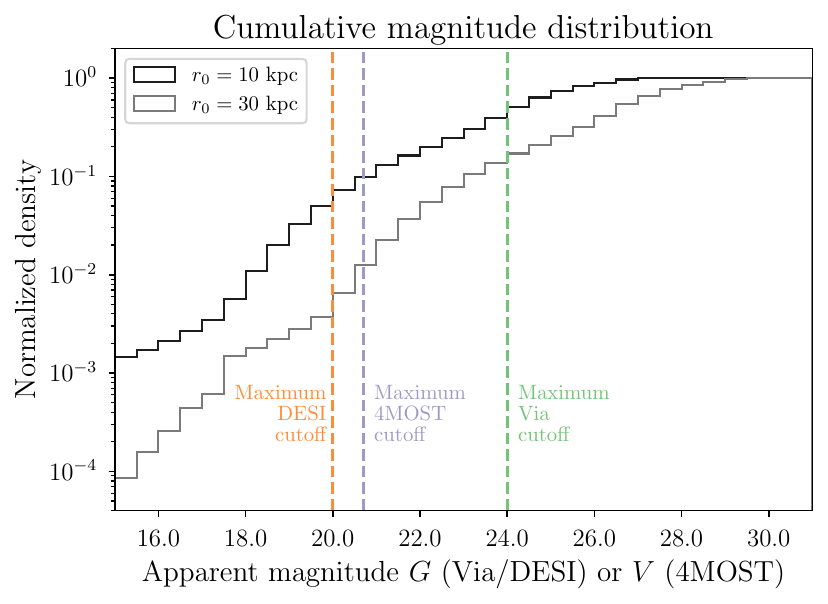}
\vspace{0.4cm}\\
\includegraphics[width=0.98\columnwidth]{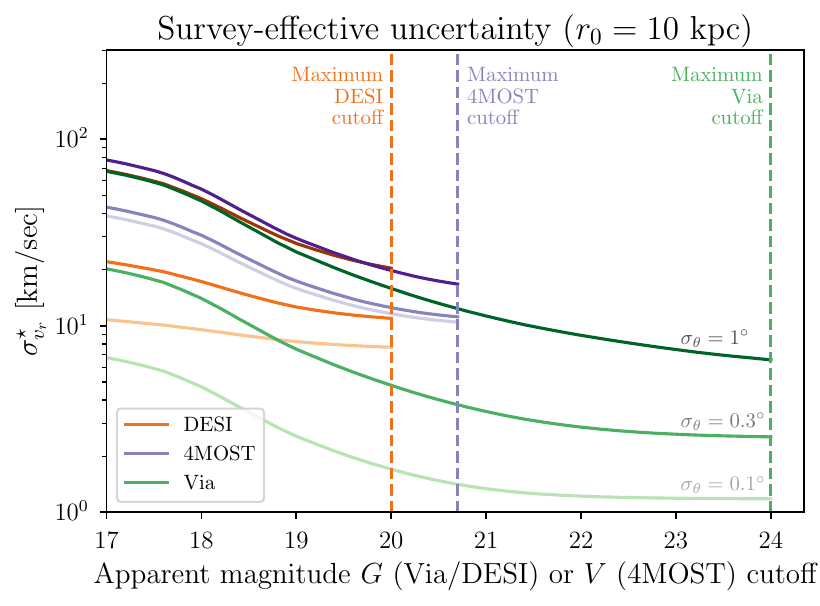}
\includegraphics[width=0.98\columnwidth]{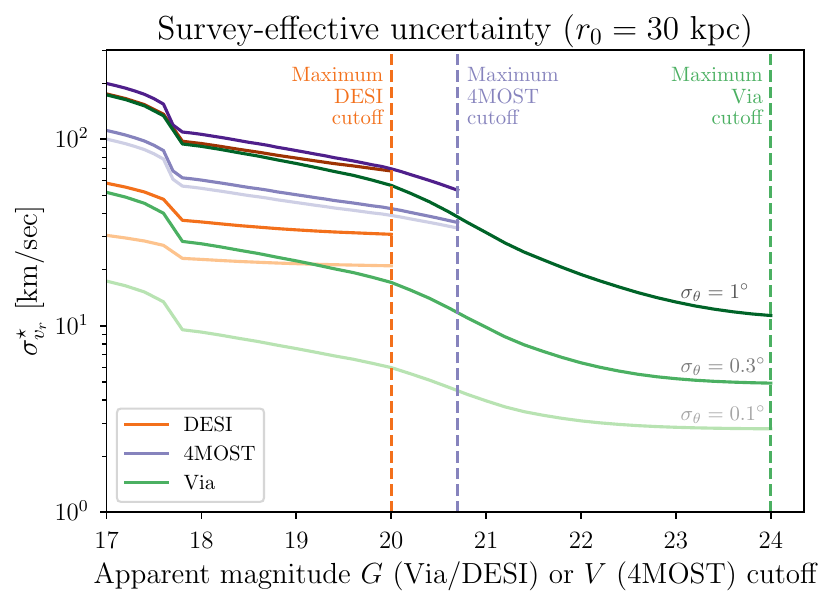}
\caption{\textbf{Top left:} Observational errors on radial velocity as a function of the apparent magnitude G for the Via survey (green) and the DESI survey (orange) and as a function of apparent magnitude V for 4MOST (purple). \textbf{Top right:} The cumulative apparent magnitude distribution for Via (green), DESI (orange) and 4MOST (purple) at different stream distances $r_0$. \textbf{Bottom left:} The survey-effective uncertainty $\sigma_{v_r}^{\star}$ as defined in Eq.~\ref{eq:error_with_obs} and calculated according to Eq.~\ref{eq:error_kin_with_obs_star} as a function of magnitude cutoff for Via (green), DESI(orange) and 4MOST (purple) at distance of $r_0=10$ kpc. The error for different widths of streams are shown in different levels of darkness. The survey-effective uncertainty $\sigma_{v_r}^\star$ is normalized to the total intrinsic number of stream stars and should not be interpreted as the uncertainty of an individual observed star. As the magnitude cutoff increases, a survey observes more stars and $\sigma_{v_r}^\star$ decreases. However, observing additional stars with very large individual uncertainties eventually provides diminishing returns, leading to the plateau at the end. \textbf{Bottom right:} The same thing as bottom left but at $r_0=30$ kpc. In this case, there is a rapid decrease in $\sigma^\star_{v_r}$ at $G \approx 18$, corresponding to the increased number of stars when the red clump is accessible (also visible in the top right figure).}
\label{fig:vrad}
\end{figure*}

\subsubsection{Radial velocity}

For the velocity observables $X=v_r, v_t, v_z$, the observational errors are no longer negligible compared to the internal dispersion.
Furthermore, compared to the statistical error arising solely from stream dispersion, a difference in this case is that observational errors vary significantly from star to star depending on stellar properties, primarily the magnitude. Rather than using a uniformly weighted average $\overline X = \sum_i X_i/n $ to estimate the mean, we will use an inverse-variance weighted estimator when taking both dispersion and observational errors into account:
\begin{equation}
    \overline{X}=\frac{\sum_i X_i / (\sigma_{X,\mathrm{disp}}^2 + \sigma_{X_i,\mathrm{obs}}^2)}{\sum_i 1/ (\sigma_{X,\mathrm{disp}}^2 + \sigma_{X_i,\mathrm{obs}}^2)}
\end{equation}
This inverse-variance weighted average is the minimum-variance estimator of the mean, with variance:
\begin{align}
   \sigma^2_{\overline X}
    &=\frac{1}{n_\mathrm{obs}}\frac{1}{\langle1/(\sigma_{X,\mathrm{disp}}^2+\sigma_{X,\mathrm{obs}}^2)\rangle_{\mathrm{bin}}}
\end{align}
Here the angular brackets refers to an average over observable stars in a given bin. Next, we approximate the average over stars in a given bin with an average over all observable stars in the cluster:
\begin{align}
    \sigma^2_{\overline X} & \approx\frac{1}{n_\mathrm{obs}}\frac{1}{\langle1/(\sigma_{X,\mathrm{disp}}^2+\sigma_{X,\mathrm{obs}}^2)\rangle_{\mathrm{cluster}}}\\
    \label{eq:iw_error}
    &=\frac{1}{n}\frac{n}{n_\mathrm{obs}}\frac{1}{\langle1/(\sigma_{X,\mathrm{disp}}^2+\sigma_{X,\mathrm{obs}}^2)\rangle_{\mathrm{cluster}}}.
\end{align}
This therefore neglects fluctuations in the variance across bins and is a reasonable approximation for $n \gg 1$.
In the last line, we split the error into three parts. The first part, $1/n$, depends only on the total stellar density. The second part, $n/n_\mathrm{obs}$, is the fraction of observable stars as shown in Fig.~\ref{fig:no_density}, which encodes information about the magnitude cutoff and stream distance. The third part, $\langle1/(\sigma_{X,\mathrm{disp}}^2+\sigma_{X,\mathrm{obs}}^2)\rangle_{\mathrm{cluster}}$, depends on the internal dispersion and the distribution of observational errors.

Now we define a survey-effective uncertainty $\sigma_X^\star$ which captures all survey and measurement effects, while being independent of the total stellar density (observable or not) of the stream:
\begin{equation}
    \label{eq:error_kin_with_obs_star}
    \sigma_{X}^{\star 2} \equiv \frac{n}{n_\mathrm{obs}}\frac{1}{\langle1/(\sigma_{X,\mathrm{disp}}^2+\sigma_{X,\mathrm{obs}}^2)\rangle_\mathrm{cluster}}.
\end{equation}
The quantity $\sigma_{X}^{\star 2}$ folds in multiple effects: the fraction of observable stars, internal dispersion of the stream, and measurement uncertainties in the observations. These combine in a non-linear way to give the total error, so we have found it useful to capture all effects in a single quantity. This way, the dependence on the survey properties, including magnitude cut, is entirely captured in $\sigma_X^\star$. As the magnitude cut is increased, there are more observable stars, leading to decreasing $\sigma_X^\star$. However, this effect eventually saturates if the additional observed stars have large measurement uncertainties.

Then the total uncertainty on the mean observable is
\begin{equation}
\label{eq:error_with_obs}
    \sigma_{\overline X}^2 \equiv \frac{\sigma_{X}^{\star 2}}{n},
\end{equation}
where $n$ is the total number of stars per bin, observable or not.  We emphasize that $\sigma_{X}^{\star}$ is not the uncertainty of an individual observed star. Rather, the dependence of $\sigma_{X}^{\star}$ on magnitude cutoffs and distance to the stream provides a useful way to compare different surveys, independent of a stream's stellar density.

\begin{figure*}[t]
\centering
\includegraphics[width=\columnwidth]{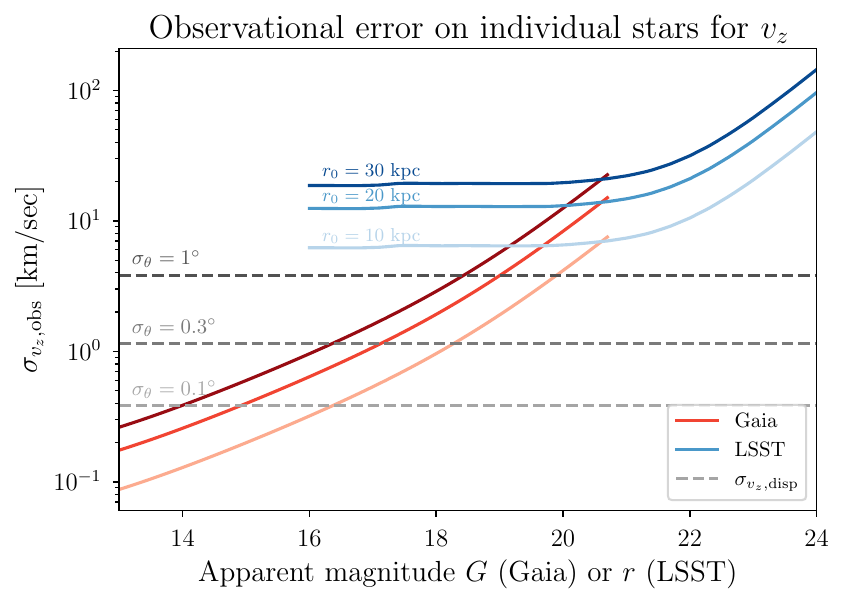}
\includegraphics[width=0.98\columnwidth]{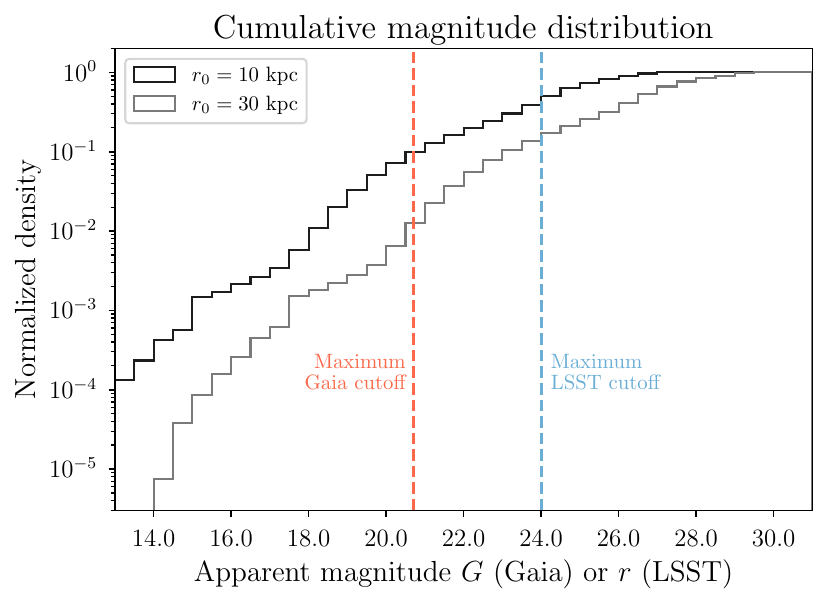}
\vspace{0.4cm}\\
\includegraphics[width=0.98\columnwidth]{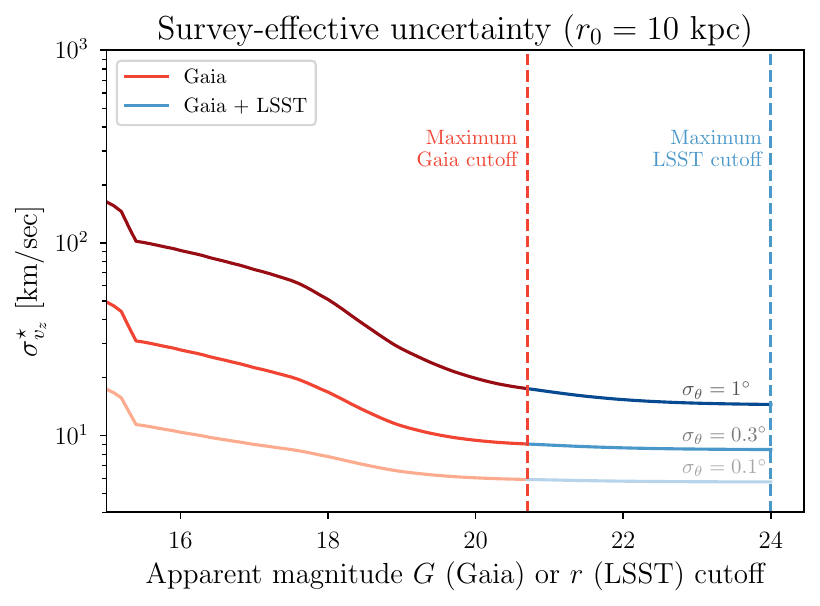}
\includegraphics[width=0.98\columnwidth]{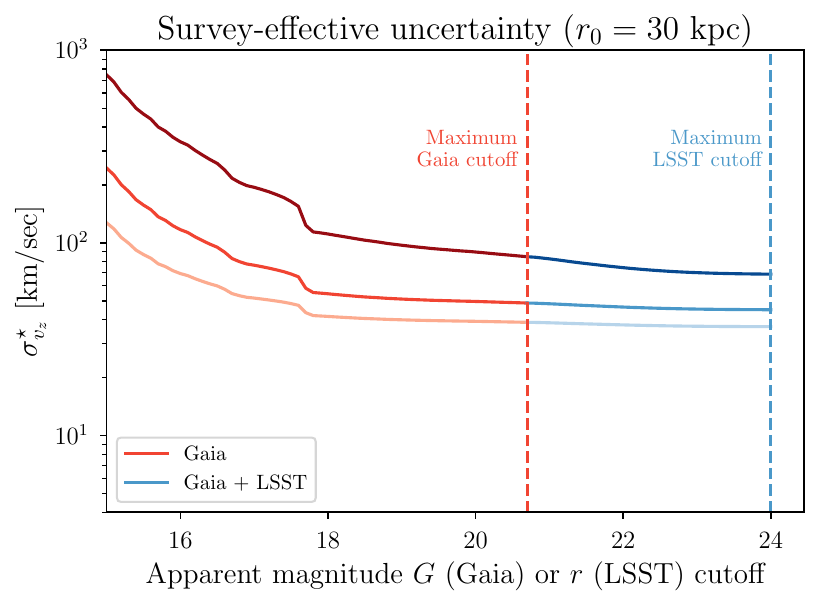}
\caption{\textbf{Top left:} Observational errors on $v_z$ as a function of the apparent magnitude G for \Gaia\ (red) and as a function of the apparent magnitude r for LSST (blue) for different stream distances $r_0$. \textbf{Top right:} The cumulative apparent magnitude distribution for \Gaia\ and LSST at different stream distances $r_0$. \textbf{Bottom left:} The survey-effective uncertainty $\sigma_{v_z}^{\star}$ as defined in Eq.~\ref{eq:error_with_obs} and calculated according to Eq.~\ref{eq:error_kin_with_obs_star} as a function of magnitude cutoff for \Gaia\ (red) and \Gaia\ + LSST (blue), at distance of $r_0=10$ kpc. The errors for different widths of streams are shown in different levels of darkness. The survey-effective uncertainty $\sigma_{v_z}^\star$ is normalized to the total intrinsic number of stream stars and should not be interpreted as the uncertainty of an individual observed star. As the magnitude cutoff increases, a survey observes more stars and $\sigma_{v_z}^\star$ decreases. However, observing additional stars with very large individual uncertainties eventually provides diminishing returns, leading to the plateau at the end. \textbf{Bottom right:} Same as bottom left but at $r_0=30$ kpc. In the bottom panels, there is a rapid decrease in $\sigma^\star_{v_z}$ at $G \approx 15$ (left) and $G \approx 18$ (right), corresponding to the increased number of stars when the red clump is accessible (also visible in the top right figure).}
\label{fig:vz}
\end{figure*}

Now we discuss how to calculate the survey-effective uncertainty for $X=v_r$. Radial velocities are assumed to be measured by one of the 4MOST survey~\citep{de_Jong_2012}, the DESI survey~\citep{Cooper_2023}, or the Via project. The observational errors on individual stars as a function of apparent magnitude is shown in the top left panel of Fig.~\ref{fig:vrad} for each survey. These also depend on the metallicity of the stars, and can change up to a factor of 2 with a change in metallicity [Fe/H] of 1. For 4MOST~\citep{de_Jong_2012} and Via~\citep{Via_errors}, we use the error estimate for the metallicity of [Fe/H]=$-2$. For DESI~\citep{2024MNRAS.533.1012K}, we download the dataset from DESI Data website\footnote{\url{https://data.desi.lbl.gov/doc/releases/edr/vac/mws/}}, and extract the mean errors from stars with metallicity [Fe/H] between -2.5 and -1.5. In the same panel, we also plot the internal dispersion in radial velocity for different widths of streams as gray dashed lines. This can be obtained from the simulated streams introduced in Sec.~\ref{sec:simulated_stream}.

In the top right panel of Fig.~\ref{fig:vrad}, we show the cumulative apparent magnitude (G magnitude for DESI/Via and V magnitude for 4MOST) distribution of stars at different stream distances $r_0$, and compare with the survey magnitude cutoffs (dashed vertical lines). This distribution is obtained under the assumption of Chabrier IMF and MIST isochrones as mentioned at the beginning of this section. Note that this plot is independent of stream density and thus works for streams of any stellar mass and length.

Given the stream dispersion, observational errors at each magnitude (upper left panel of Fig.~\ref{fig:vrad}), and the distribution of magnitudes (upper right panel of Fig.~\ref{fig:vrad}), we can calculate the survey-effective uncertainty in Eq.~\ref{eq:error_kin_with_obs_star} for different surveys. The bottom panels of Fig.~\ref{fig:vrad} show the resulting $\sigma_{v_r}^{\star}$ as a function of the magnitude cutoff for 4MOST (purple), DESI (orange) and Via (green). 

For each survey, to illustrate the dependence on stream width $\sigma_\theta$, we show three lines representing $\sigma_\theta=0.1^\circ$, $\sigma_\theta=0.3^\circ$ and $\sigma_\theta=1^\circ$ respectively. To show the variation with distance to the stream $r_0$, we split the bottom panel into two subplots, $r_0=10$ kpc (bottom left) and $r_0=30$ kpc (bottom right).   We can see that in all cases,  $\sigma_{v_r}^\star$ reaches a plateau before the maximum cutoff magnitude for 4MOST, DESI or Via. This indicates that it is possible to achieve similar performance even if these surveys do not measure all stars in the stream down to the maximum magnitude cutoff. 
The faint end of the stellar distribution is only marginally helpful in detecting subhalos. This point is important because the ability to observe all stars down to the maximum possible cutoff is limited by observing time. For instance, Via plans to target all candidate members identified based on photometry and astrometry down to $G < 21$, and is not guaranteed to observe all stars down to $G < 24$. Nevertheless, we will still use the maximum possible cutoff magnitude and therefore maximum number of stars to minimize the errors. 

\subsubsection{Proper motions}

\begin{figure*}[t!]
\centering
\includegraphics[width=0.98\linewidth]{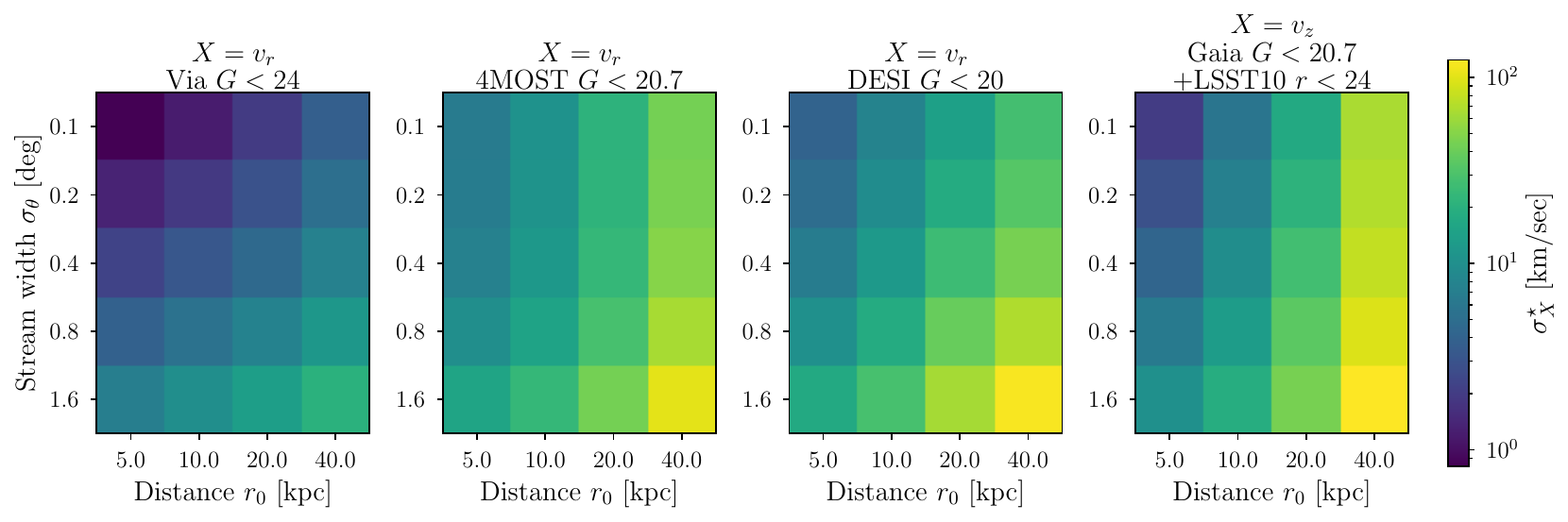}
\caption{The survey-effective uncertainty $\sigma_X^{\star}$, as defined in Eq.~\ref{eq:error_kin_with_obs_star}, for $X=v_r$ with Via, 4MOST, and DESI, as well as $X=v_z$ with \Gaia\ and LSST. The uncertainty is shown as a function of stream distance $r_0$ and stream width $\sigma_\theta$, where the stream width also determines velocity dispersion $\sigma_\theta \propto \sigma_{v_r} \propto \sigma_{v_z}$.}
\label{fig:heat_map}
\end{figure*}

For proper motions, the observational errors are also not negligible compared to internal dispersion, and we follow the same framework as for the radial velocity to compute the survey-effective uncertainty in Eq.~\ref{eq:error_with_obs}.

We assume proper motions to be measured by \Gaia\ and LSST~\citep{lsstsciencebook2009} (10 year sensitivity). For \Gaia, the error in proper motion is given by\footnote{\url{https://www.cosmos.esa.int/web/gaia/science-performance}},
\begin{align}
\label{eq:pm_error}
\sigma_{\mu,\text{obs}} = c_\mu &T_\text{factor}(40 + 800 z +30 z^2)^{0.5} \nonumber\\ &\upmu\text{as/year},
\end{align}
where $T_\text{factor}=0.527$, $c_\mu$ is 0.29 (0.25) for $\mu_{\alpha *}$ ($\mu_\delta$), and $z$ is given as a function of magnitude G as,
\begin{equation}
z=\text{Max}[10^{0.4(13-15)},10^{0.4(G-15)}].
\end{equation}
These errors correspond to \Gaia\ DR5, and we assume the proper motion data to be available for $G <20.7$.

Assuming the same geometric setup of streams as in Sec.~\ref{sec:circular}, the observables $v_z$ and $v_t$ defined in Sec.~\ref{sec:analytic} are given by $v_z = \mu_\delta \times r_0$ and $v_t = \mu_{\alpha *} \times r_0$, where $r_0$ is the star's distance to the Galactic Center. Dropping the subscripts on $\mu$ for simplicity, the errors in the velocities have two contributions of the form $\sigma_{\mu,\text{obs}} \times r_0$ and $\sigma_{r_0,\text{obs}} \times \mu$.
Distances to individual stars typically have large errors, but the distance of the stream track $r_0$ can be extracted more reliably by orbit fitting. We therefore neglect the contribution from $\sigma_{r_0,\text{obs}}$ 
and the total error for $v_t$ or $v_z$ is estimated as $\sigma_{\mu,\text{obs}} \times r_0$.

The top left panel of Fig.~\ref{fig:vz} shows the observational error on $v_z$ for an individual star as a function of magnitude ($G$ magnitude for \Gaia\ and $r$ magnitude for LSST) for distances of 10, 20, and 30 kpc. In the same panel, we also compare with the internal dispersion of $v_z$ for different widths of streams in gray dashed lines. The top right panel of Fig.~\ref{fig:vz} shows the cumulative apparent magnitude distribution at different stream distances $r_0$, compared with the maximum magnitude cutoff for \Gaia\ and LSST proper motions (dashed vertical lines).

With information from both of the top panels of Fig.~\ref{fig:vz}, we calculate the survey-effective uncertainty in Eq.~\ref{eq:error_kin_with_obs_star} for $X=v_z$ under assumptions of different magnitude cutoffs. In the bottom panels of Fig.~\ref{fig:vz}, we show $ \sigma_{v_z}^{\star}$ as a function of magnitude cutoff for \Gaia\ (red) and \Gaia\ + LSST (blue). In the scenario labeled ``Gaia + LSST'', we use  \Gaia\ for stars with magnitude $G < 20.7$ and use LSST for stars with magnitude $G > 20.7$, since the observational errors for  \Gaia\ are smaller than LSST at magnitudes $G<20.7$, as can be seen in the top left panel.
We again show three lines for each case, corresponding to different choices for internal dispersion, which are parameterized by stream widths ($\sigma_\theta=0.1^\circ$, $\sigma_\theta=0.3^\circ$, $\sigma_\theta=1^\circ$). 
Again we see that in all cases,  $\sigma_{v_z}^\star$ starts to flatten before the maximum magnitude cutoff $G<20.7$, with the additional LSST data only giving mild improvement in increasing detectability. Nevertheless, we will still use the cutoff magnitude determined by each survey limit to maximize the number of observable stars per bin.

The process to calculate errors for $v_t$ is the same as $v_z$. The difference in the observational errors comes from the $c_\mu$ factor (0.25 for $v_z$ and 0.29 for $v_t$) in Eq.~\ref{eq:pm_error} for \Gaia, and slightly different internal dispersions. Both are very small effects, so we will calculate errors for $v_t$ explicitly but we do not show the results in the paper.

\subsubsection{Summary}
\label{sec:obs_summary}

The discussion in this section gives us the ingredients to calculate total errors for the 5 observables ($\rho/\rho_0$, $\overline{\Delta \theta}$, $\overline{\Delta v_r}$, $\overline{\Delta v_t}$, and $\overline{\Delta v_z}$) we use for subhalo impacts. The total errors are given by Eq.~\ref{eq:error_den_with_obs} for $\rho/\rho_0$, Eq.~\ref{eq:error_theta} for $\overline{\Delta \theta}$, and Eq.~\ref{eq:error_with_obs} for $\overline{\Delta v_r}$, $\overline{\Delta v_t}$ and $\overline{\Delta v_z}$, with the survey-effective uncertainty $\sigma_{X}^{\star 2}$ given by Eq.~\ref{eq:error_kin_with_obs_star}. These equations in turn use the number of stars per bin $n = \lambda \cdot (\psi {\rm bin size})$, and the fraction of observable stars $n_\mathrm{obs}/n$ under different magnitude cutoffs as summarized in Fig.~\ref{fig:no_density}. 

The survey-effective uncertainty $\sigma_X^{\star}$ depends also on stream width $\sigma_\theta$ and the stream distance $r_0$. Some results for $\sigma_X^{\star}$ are already shown in the bottom panels of Fig.~\ref{fig:vrad} and Fig.~\ref{fig:vz}. In order to perform tests of detectability in all of the stream parameter space, we calculate a denser grid of $\sigma_X^{\star}$. The results are shown in Fig.~\ref{fig:heat_map} for different observables and surveys, with an overall trend of smaller errors for colder and nearby streams. The effective errors on proper motions for \Gaia\ are similar to \Gaia+LSST (see the bottom left panel of Fig.~\ref{fig:vz}) and thus not shown. Also, the effective errors for $v_t$ are similar to $v_z$ and thus not shown.

\subsection{Benchmark scenarios}

Now we consider a few concrete observational scenarios, consisting of different combinations of measurements, and summarize them in Tab.~\ref{table:scenarios}.
The first scenario ``Gaia'' is close to data we have now and serves as a baseline. The second scenario ``Gaia + DESI'' corresponds to the case where DESI data is available and tells what we gain from the line-of-sight measurements. We don't include any scenario using 4MOST data, since the effective errors from 4MOST and DESI are very similar (as shown in Fig.~\ref{fig:heat_map}) and thus the ``DESI + Gaia'' scenario is also representative for the ``4MOST + Gaia'' scenario. The third scenario ``Via + Gaia + LSST'' assumes Via data is available and that position information is based on LSST single epoch sensitivity. Here the selection of stars in LSST data is assumed to be based on the spectroscopy with Via, so the LSST data is using the magnitude cutoff from Via. The last scenario ``Via + Gaia + LSST10'' assumes we have position and proper motion with 10-year LSST data.
We use these four scenarios in the rest of the paper to analyze the detectability of subhalo impacts.

\begin{table}[t]
\centering
\renewcommand{\arraystretch}{2}
\begin{tabular}{ |c|c|c|c|} 
\hline
 & \textbf{Positions} & \makecell{\textbf{Radial}\\\textbf{Velocity}} & \makecell{\textbf{Proper}\\\textbf{Motions}} \\ 
   \hline
 Gaia & \makecell{Gaia\\$G<20.7$} & not used & \makecell{Gaia \\$G<20.7$} \\ 
  \hline
 DESI+Gaia & \makecell{Gaia\\$G<20.7$} & \makecell{DESI \\$G<20$} & \makecell{Gaia \\$G<20.7$} \\ 
 \hline
 Via+Gaia+LSST & \makecell{LSST\\$G<24$ (Via)} & \makecell{Via \\$G<24$} & \makecell{Gaia \\$G<20.7$} \\ 
 \hline
 Via+Gaia+LSST10 & \makecell{LSST10\\$r<27$} & \makecell{Via \\$G<24$} & \makecell{Gaia\\$G<20.7$\\LSST10\\$r<24$} \\ 
 \hline
\end{tabular}
\caption{Benchmark observational scenarios considered.}
\label{table:scenarios}
\end{table}

\section{Detectability of DM subhalos }
\label{sec:statistics}

\begin{table*}[t!]
\centering
\renewcommand{\arraystretch}{2}
\begin{tabular}{ |c|c|c|c|} 
\hline
 & \textbf{Description}
 & \textbf{Default value}
 & \textbf{Bounds for optimization} \\ 
 \hline
 \ $M_{\rm sh}$ \ & Subhalo mass  & -- & $[0, 10^9]~\msun$ \\ 
 \hline
 \ $r_s$ \ & Scale radius & \makecell{Plummer sphere, Eq.~\ref{eq:mass_radius}} & -- \\ 
 \hline
 \ $b$ \ & Impact parameter & \makecell{0} & [-5, 5] kpc \\
  \hline
 \ $t$ \ & Time since flyby & \makecell{\(\displaystyle315~\mathrm{Myr} \times \frac{r_0}{10~\mathrm{kpc}}\) }
 & \(\displaystyle[0, 700]~\mathrm{Myr} \times \frac{r_0}{10~\mathrm{kpc}}\)\\ 
 \hline
 \ $\bm{w}=(w_r, w_t, w_z)$ \ & Subhalo velocity & \makecell{180 km/s in $\hat z$ direction} & [-500, 500] km/s for each component \\ 
 \hline
 \ $\psi_\mathrm{impact}$ \ & Impact location & $0^\circ$ on a $[-22^\circ, 22^\circ]$ region & $[-4^\circ, 4^\circ]$ \\ 
 \hline
\end{tabular}
\caption{Subhalo impact parameters used in the likelihood Eq.~\ref{eq:likelihood}. The middle column gives the default value used to generate mock data and assess detectability of a subhalo with mass $M_{\rm sh}$. The dependence of results when the parameter is varied from the default value is discussed in Sec.~\ref{sec:nuisance} and shown in Fig.~\ref{fig:min_msh_nuisance}. The last column shows the bounds used by \texttt{scipy.optimize} when we maximize the likelihood function over these parameters.
}
\label{table:default_impact}
\end{table*}

With the method in Sec.~\ref{sec:circular} to rapidly generate observables with errors for an impacted stream, we are ready to establish  detectability of subhalos in stellar streams under the different observational scenarios introduced in Sec.~\ref{sec:observations}. Our primary parameter of interest is the subhalo mass $M_{\rm sh}$ as this is most closely related to underlying DM particle physics. 

First, we define a Gaussian likelihood function
\begin{equation}
    L(M_{\mathrm{sh}},~\bm \theta) = \prod_{i, j}
    \frac{1}{\sqrt{2 \pi \sigma_{\nu_{ij}}^2}}
    \exp\left[-\frac{(\nu_{ij}(M_{\mathrm{sh}},~\bm \theta) - \nu_{ij}^{\mathrm{dat}})^2}{2\sigma_{\nu_{ij}}^2}\right]
    \label{eq:likelihood}
\end{equation}
where $\nu_{ij}$ denotes the value of the $i$th observable for the $j$th $\psi$ bin along the stream. We use a bin size of 2 degrees by default.
We generate the mock impact data, $\nu_{ij}^{\rm dat}$, as described in Sec.~\ref{sec:simulated_impact}, along with the total error in each bin $\sigma_{\nu_{ij}}$ as summarized in Sec.~\ref{sec:obs_summary}.  Note that the error $\sigma_{\nu_{ij}}$ is computed assuming the number of stars per bin $n$ is constant across bins, so that the error $\sigma_{\nu_{ij}}$ is also constant across bins. 

The expected value of the observables, $\nu_{ij}(M_{\mathrm{sh}},~\bm \theta)$, is given by the analytic model and computed as a function of $M_{\rm sh}$ and nuisance parameters $\bm\theta=(b,~t,~w_r,~w_t,~w_z, ~\psi_\mathrm{impact})$, which we will profile over. Note that the scale radius of the subhalo, $r_s$, is not considered a nuisance parameter here, since we fix its dependence on $M_{\mathrm{sh}}$ based on the CDM mass-radius relation for a Plummer potential \citep{Erkal_2016}:
\begin{equation}
\label{eq:mass_radius}
    r_s=\left(\frac{M_{\mathrm{sh}}}{10^8~\msun}\right)^{0.5} \times 1.62~\mathrm{kpc}.
\end{equation}
Without this assumption, there would be an exact degeneracy between subhalo mass $M_\mathrm{sh}$ and relative velocity $\bm{w}_\mathrm{rel}=(w_r, w_\parallel, w_z)$ in the analytic model, where taking $M_\mathrm{sh} \to \lambda M_\mathrm{sh}$ and $\bm{w}_\mathrm{rel} \to \lambda \bm{w}_\mathrm{rel}$ leaves the velocity kicks Eqs.~\ref{eq:vkr}-\ref{eq:vkz} unchanged (see also section 5.1 of \cite{Erkal_2015_2}). Taking $r_s$ to be a function of $M_\mathrm{sh}$ breaks the degeneracy, and is essential for us to maximize the likelihood function over its parameters. In addition, we also introduce a new nuisance parameter $\psi_\mathrm{impact}$, which allows the impact location of the model to vary along the stream arm.
Using Eq.~\ref{eq:likelihood}, we can then define test statistics for an impact to be distinguishable from a smooth stream, as well as confidence intervals on $M_{\rm sh}$ for a detected impact. 

Since the detectability of a subhalo impact depends on the nuisance parameters, we will set default values for these parameters when considering mock data. These are given in Tab.~\ref{table:default_impact}. We set an impact parameter $b = 0$ corresponding to a direct impact as this is the best possible case, although the results do not vary significantly when $b$ is smaller or at similar scale as the scale radius $r_s$. 
We set the subhalo velocity $w = \sqrt{w_r^2+w_t^2+w_z^2} = 180$ km/s as this is a typical DM velocity in the halo, and set the direction to be in the $\hat z$ direction. Note that here we are setting the DM velocity in the halo frame so that we can assign a typical value. The relative velocity in the frame of the stream, $w_\mathrm{rel} = \sqrt{w_r^2+(w_t-V_c)^2+w_z^2}$, depends on the DM direction relative to the stream. For time $t$, we pick a default value that is representative of detectability for most values of $t$. The value of $t$ is scaled with stream distance $r_0$ in order for the observables at different $r_0$ to have the same orbital phase, yielding more comparable perturbation profiles. Lastly, $\psi_\mathrm{impact}=0^\circ$ is defined to be at the center of the impact in the data, and we consider a region of $44\degree$ ($[-22^\circ, 22^\circ]$) about the impact center in the likelihood. We find that this angular range is large enough to include the full signal for any impact from subhalos of masses up to $10^9 \msun$. In Sec.~\ref{sec:min_stream_length}, we will explore the effect on detectability when we shrink the region considered, thus revealing the dependence on stream length. Variations with other default parameter choices will also be explored later in Sec.~\ref{sec:nuisance}.

The likelihood defined above neglects the contribution of potential systematics arising from stream dynamics, effects of baryonic perturbers, and observational systematics. This therefore represents an idealized scenario for detection of an impact, similar to the assumption of $b=0$ for the mock data. The aim of this work is to create a simple metric to identify promising streams for subhalo impact detection. This then warrants a more detailed follow up study of promising streams taking into account the specific systematic effects that impact them, and the population of subhalo impacts they experience.

In Sec.~\ref{sec:min_msh}, we first discuss the minimum detectable subhalo mass for different stream widths $\sigma_\theta$, distances to the Galactic Center $r_0$, stream densities $\lambda$, and observational scenarios. In Sec.~\ref{sec:minimum_mass_streams}, we condense the results of Sec.~\ref{sec:min_msh} into a formula which gives the minimum detectable subhalo mass as a function of stream properties $\sigma_\theta$, $r_0$ and $\lambda$. This key result allows us to quickly assess detectability in observed Milky Way streams.
In Sec.~\ref{sec:msh_confidence}, we consider impacts well above the detectability threshold, and evaluate confidence intervals for the detected mass under different observational scenarios. In Sec.~\ref{sec:min_stream_length}, we also add a constraint on the length of streams in order for subhalos to be detectable.
And finally in Sec.~\ref{sec:nuisance}, we discuss how the subhalo detectability is affected by different nuisance parameters for the subhalo impact (e.g. time since flyby $t$ and the direction of subhalo velocity $\hat w$).

\subsection{Minimum detectable subhalo mass}
\label{sec:min_msh}

Following~\cite{Cowan_2011}, the test statistic for discovery of a signal is defined in terms of the log likelihood ratio:
\begin{equation}
\label{eq:q0}
    q_0 = \begin{cases}
    2\ln \frac{L(\hat M_\mathrm{sh},~\hat{\bm\theta})}{L(0)}& \hat{M}_\mathrm{sh}\geq 0\\
    0 & \hat{M}_\mathrm{sh} < 0\\
    \end{cases}
\end{equation}
where $\hat M_\mathrm{sh}$ and $\hat{\bm\theta}$ denote the value of $M_\mathrm{sh}$ and $\bm\theta$ that maximizes $L$. $L(0)$ is the likelihood evaluated assuming no impact ($M_\mathrm{sh}=0$) where we have omitted the nuisance parameters $\bm\theta$, and simply use $\nu_{ij} = 0$ for position/velocity observables and $\nu_{ij}=1$ for $\rho/\rho_0$ in Eq.~\ref{eq:likelihood}. The test statistic is defined in the above way to exclude unphysical negative values of the best fit subhalo mass, $\hat M_{\rm sh} < 0$.

\begin{figure}[t]
\centering
\includegraphics[width=\columnwidth]{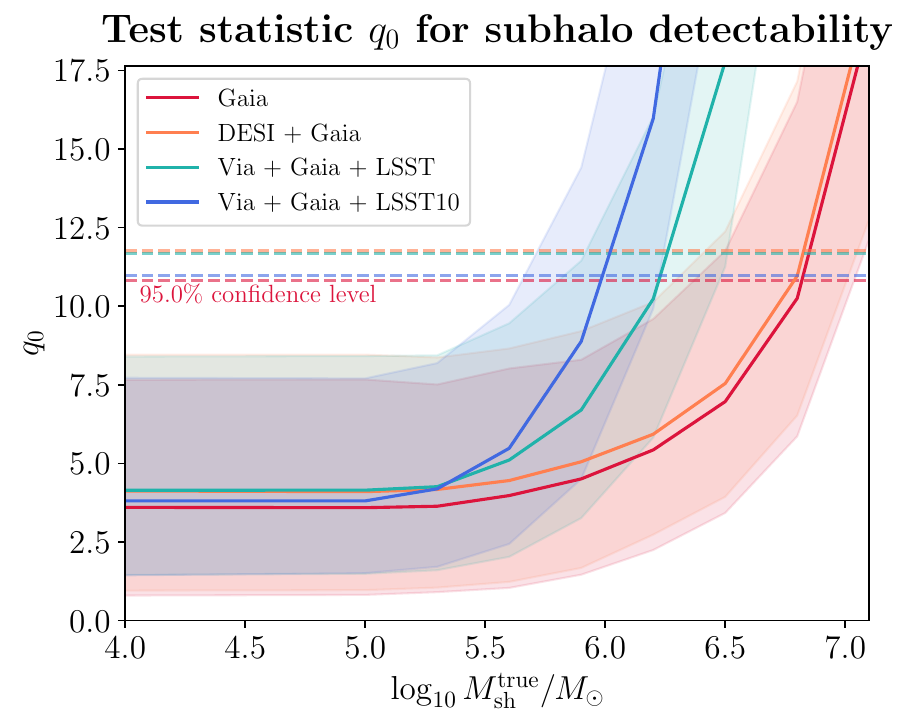}
\caption{Test statistic $q_0$ for subhalo detectability (Eq.~\ref{eq:q0}) as a function of true subhalo mass $M_\mathrm{sh}^\mathrm{true}$. Here we consider a stream similar to GD-1 ($\sigma_\theta=0.2^\circ$, $r_0=10~\mathrm{kpc}$, $\lambda=100~\mathrm{deg}^{-1}$) for all four observational scenarios. The solid lines indicate the median value of $q_0$ for each truth value of the subhalo mass, $M_{\mathrm{sh}}^{\mathrm{true}}$, and the bands indicate the $\pm1\sigma$ region. The dashed lines indicate the threshold to reject the null hypothesis of no impact at 95\% CL.}
\label{fig:ts_for_signal}
\end{figure}

We can then compute the test statistic $q_0$ for mock data generated using a true value of the subhalo mass $M_{\mathrm{sh}}^{\mathrm{true}}$, the default values of nuisance parameters mentioned before, and errors as summarized in Sec.~\ref{sec:obs_summary}. Fig.~\ref{fig:ts_for_signal} shows the distribution of $q_0$ as a function of $M_{\mathrm{sh}}^{\mathrm{true}}$, for a stream similar to GD-1 ($\sigma_\theta=0.2^\circ$, $r_0=10~\mathrm{kpc}$, $\lambda=100~\mathrm{deg}^{-1}$) and different observational scenarios. The solid line is the median $q_0$ value for each $M_{\mathrm{sh}}^{\mathrm{true}}$, while the shaded area is the $\pm 1\sigma$ variation. These are obtained from 500 realizations of mock data at each $M_{\mathrm{sh}}^{\mathrm{true}}$. For each dataset, we evaluate $q_0$ by maximizing the likelihood function over $M_\mathrm{sh}$ and $\bm\theta$. In particular, we use the Nelder\textrm{-}Mead~\citep{nedler_mead} method implemented in \texttt{scipy.optimize.minimize}, and the boundaries for each parameter are specified in Tab.~\ref{table:default_impact}. We emphasize that the analytic model is crucial for us to perform the computation-heavy optimization on hundreds of datasets. Plots of the $q_0$ distribution at specific $M_{\mathrm{sh}}^{\mathrm{true}}$ values, as well as further discussion of the distribution shape can be found in App.~\ref{app:q0_dist}.

\begin{figure*}[t]
\centering
\includegraphics[width=\textwidth]{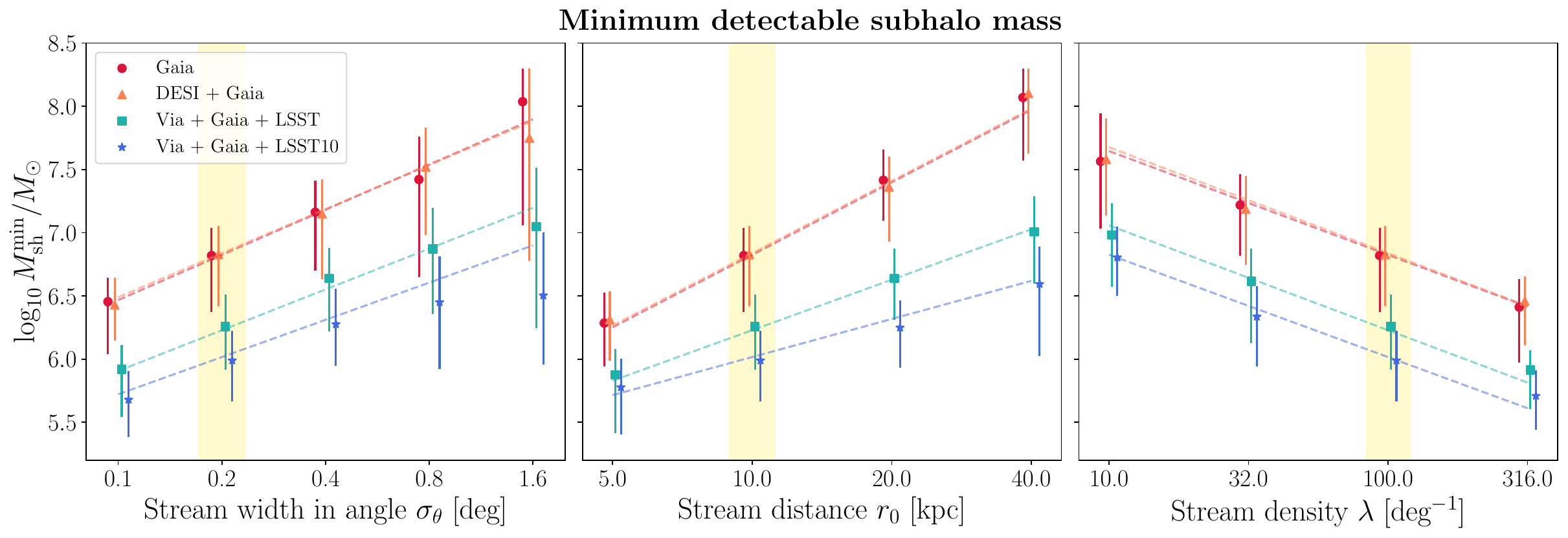}
\caption{\textbf{Left:} Minimum detectable subhalo mass at 95\% confidence level as a function of stream width. \textbf{Middle:} same  but as a function of stream distance. \textbf{Right:} same but as a function of stream density. In all panels, the fixed properties always use values similar to GD-1 ($\sigma_\theta=0.2^\circ$, $r_0=10~\mathrm{kpc}$ and $\lambda=$100 deg$^{-1}$ as highlighted in yellow). The data points indicate the median values. The error bars indicate the $1\sigma$ bounds. The dashed lines are the results from the fitting formula in Eq.~\ref{eq:fitting} with the coefficients in Tab.~\ref{table:fitting}.}
\label{fig:min_msh}
\end{figure*}

In order to obtain the minimum detectable subhalo mass at 95\% confidence level, we use the probability distribution for $q_0$ when the data $\nu_{ij}^\mathrm{dat}$ is generated from $M_{\mathrm{sh}}^{\mathrm{true}}=0$ (no impact). The threshold for detectability is given by the 95th percentile value $q_0^{95\%} |_{M_{\mathrm{sh}}^{\mathrm{true}}=0}$, shown as the dashed lines in Fig.~\ref{fig:ts_for_signal}. For each observational scenario, the intersection of $q_0^{95\%} |_{M_{\mathrm{sh}}^{\mathrm{true}}=0}$ with the distribution of $q_0$ values for nonzero $M_{\mathrm{sh}}^{\mathrm{true}}$ gives the median and $\pm 1 \sigma$ band for the minimum detectable subhalo mass at 95\% CL.

Repeating this procedure for different streams, we can obtain the dependence of subhalo detectability on stream properties. 
Fig.~\ref{fig:min_msh} shows the minimum detectable subhalo mass as functions of stream width $\sigma_\theta$ (left), stream distance $r_0$ (middle) and stream density $\lambda$ (right) respectively. In each panel, we vary one stream property and keep the other two fixed. The fixed properties always take the values similar to GD-1 ($\sigma_\theta=0.2^\circ$, $r_0=10~\mathrm{kpc}$ and $\lambda=$100 deg$^{-1}$). The points in the plots correspond to the median while the error bars correspond to $\pm 1 \sigma$.  It can be seen that the minimum detectable mass increases as the stream width $\sigma_\theta$ increases, as the distance to the Galactic Center $r_0$ increases, or as the stream density $\lambda$ decreases. In addition, improvements in observational scenarios can significantly improve detectability.

\begin{figure}[h!]
\centering
\includegraphics[width=\columnwidth]{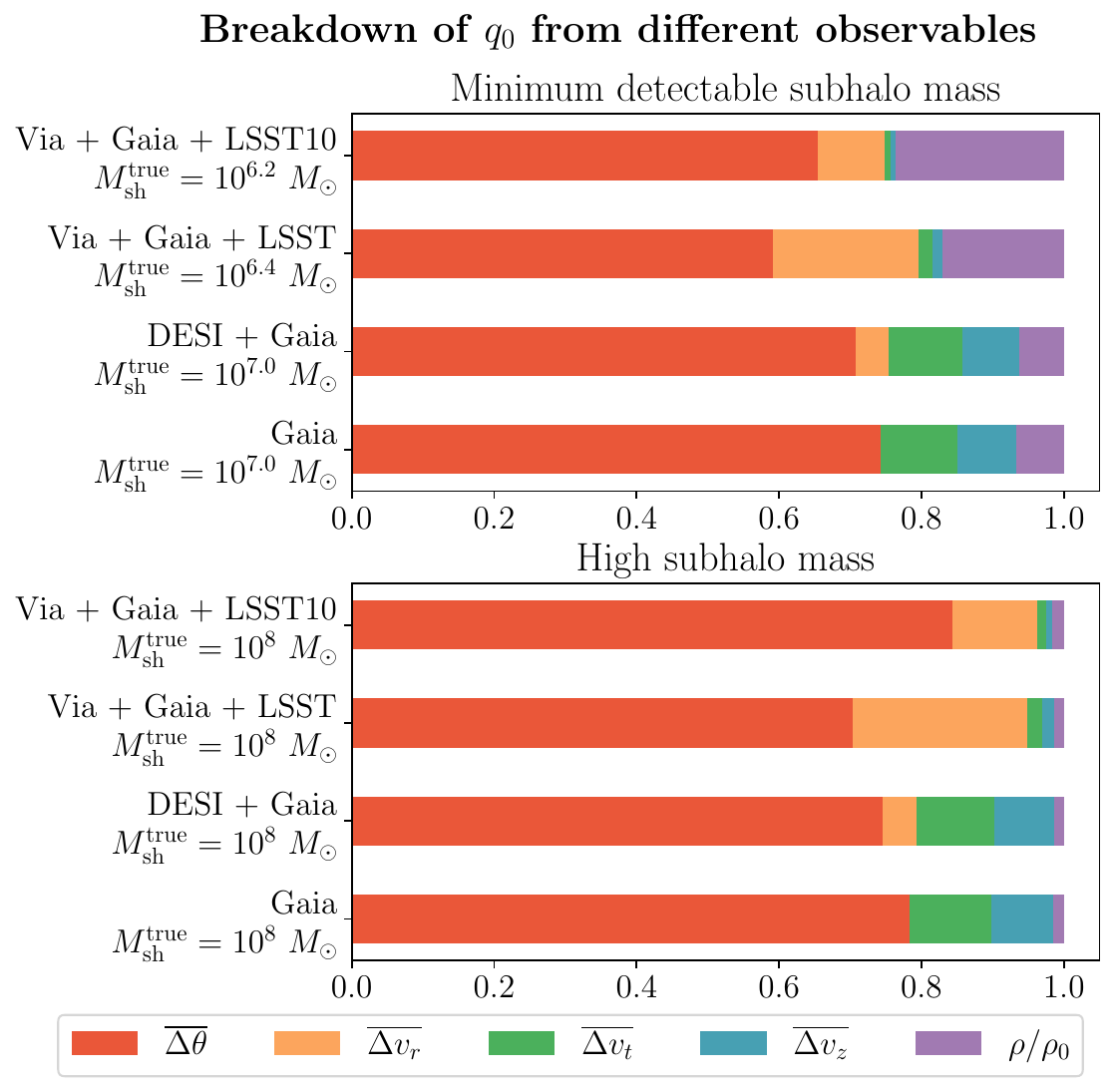}
\caption{Decomposition of the test statistic $q_0$ into contributions from different observables. We assume a cold stream with $\sigma_\theta=0.2^\circ$, $r_0=10~\mathrm{kpc}$, $\lambda=100~\mathrm{deg}^{-1}$ and $q_0$ is evaluated on Asimov data sets, with no errors added, to show a representative breakdown. \textbf{Top:} We assume the minimum detectable subhalo mass in each observational scenario. \textbf{Bottom:}  High subhalo mass $M_\mathrm{sh}=10^8~\msun$ for all scenarios. In both panels, the total $q_0$ is normalized to 1 so that breakdowns are more comparable across different scenarios.}
\label{fig:chi2_dist}
\end{figure}

To see how different types of observations contribute in detecting a subhalo, we show the breakdown of $q_0$ from different observables in Fig.~\ref{fig:chi2_dist}. Considering that the observables can fluctuate for different datasets even at the same set of parameters, we calculate the breakdown of $q_0$ on the representative ``Asimov data set''. The ``Asimov data set'' is defined in \cite{Cowan_2011} such that when one evaluates the estimators for all parameters, one obtains the true parameter values. In particular, here the Asimov data set contains the expected values from the analytic model without any added noise, $ \nu_{ij}^{\mathrm{dat}} = \nu_{ij}(M_{\rm sh}^{\rm true},~\bm\theta^{\rm true})$.

\begin{figure*}[p]
\centering
\includegraphics[width=\textwidth]{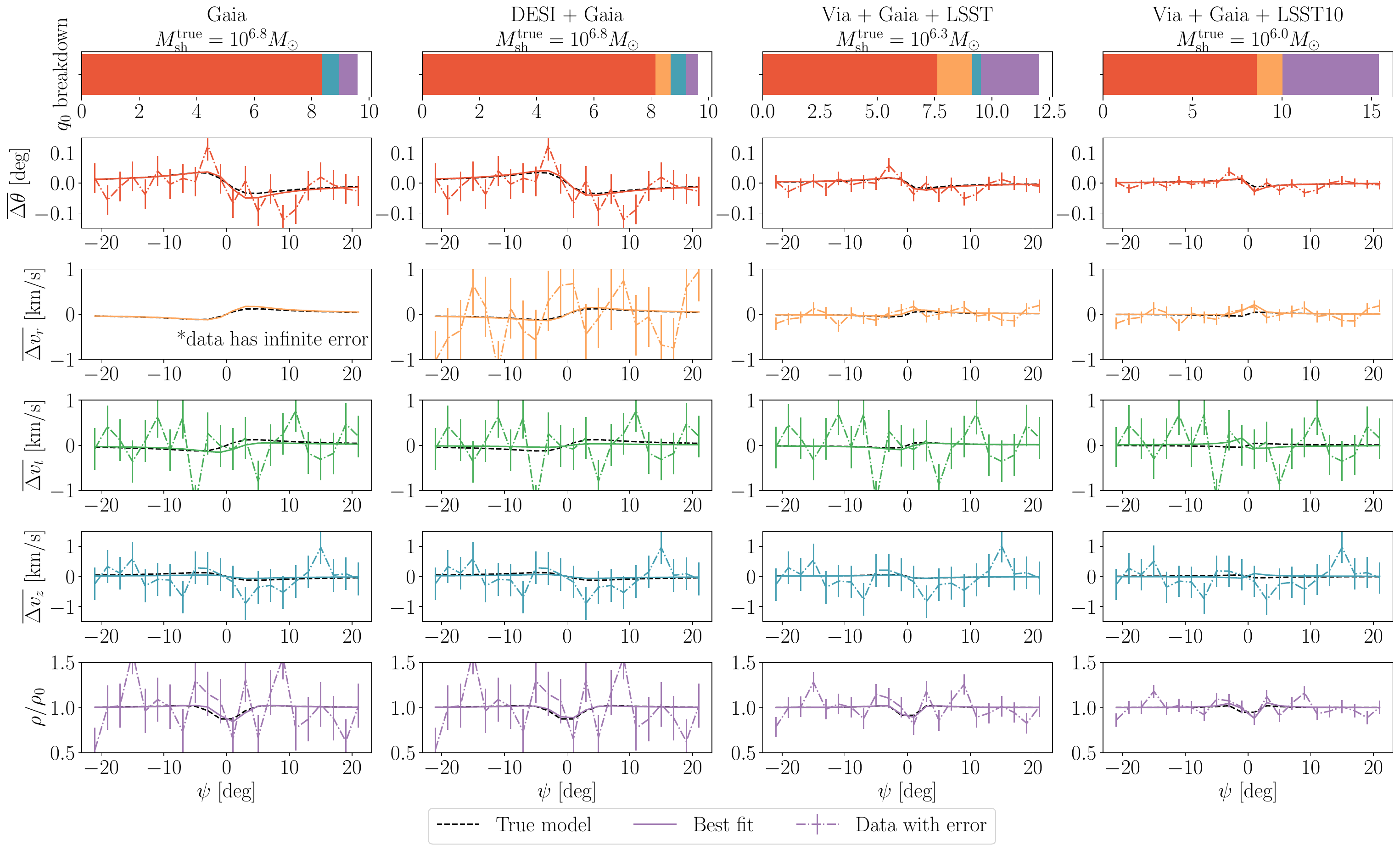}
\par\bigskip
\includegraphics[width=\textwidth]{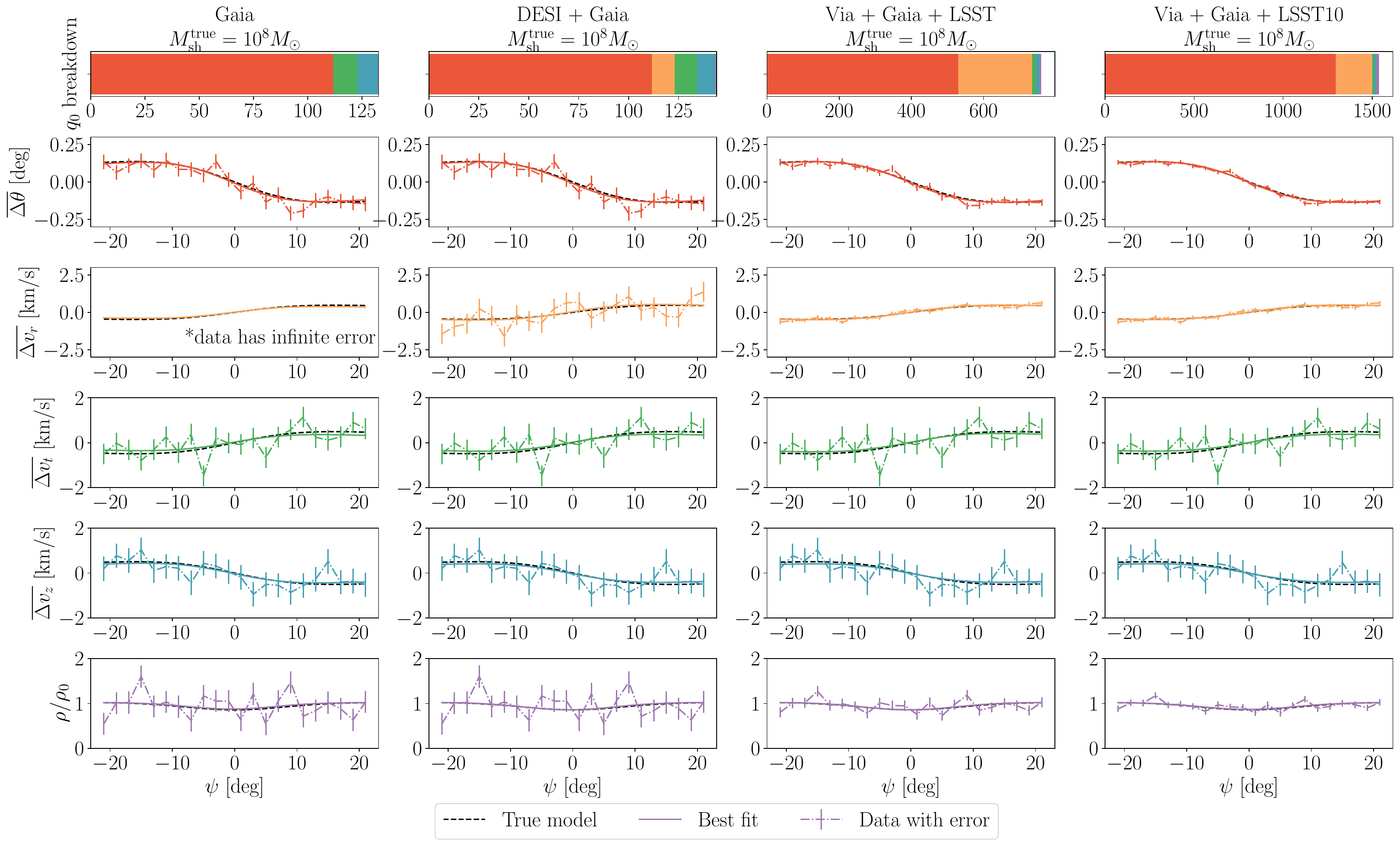}
\caption{Simulated data (dash-dot), where error bars contain both internal dispersion and observational errors.  The data is compared to the best fit model (solid) and true model or Asimov data (dashed). \textbf{Top:} Data generated from minimum detectable subhalo mass for that scenario. \textbf{Bottom:} Data generated from a high subhalo mass $M_\mathrm{sh}=10^8~\msun$.
}
\label{fig:observables_with_errors}
\end{figure*}

In the top panel of Fig.~\ref{fig:chi2_dist}, the true subhalo mass is the minimum detectable subhalo mass for each scenario, meaning the subhalo is just barely detected. In the bottom panel, the true subhalo mass is $10^8~\msun$, indicating a strong impact. In both cases and for all observational scenarios, the position information $\overline{\Delta \theta}$ (red) and $\rho/\rho_0$ (purple) play a major role in detecting a subhalo impact, consistently contributing more than 70\% to $q_0$. This is related to the choice of our default subhalo velocity direction $\hat z$. The radial velocity $\overline{\Delta v_r}$ will contribute much more in other cases, such as when the subhalo has non-zero velocity along the radial direction $w_r$. We discuss this in more detail in Sec.~\ref{sec:nuisance} around Fig.~\ref{fig:q0_breakdown_w_dir}. On the other hand, the density observable $\rho/\rho_0$ (purple) contributes much less in the case of a strong detection. Even though the total density fluctuation grows with $M_{\mathrm{sh}}$, the gap becomes much more spread out in angle, diluting the effect per bin.

Next, Fig.~\ref{fig:observables_with_errors} shows example mock data sets (dash-dot line) along with the best fit (solid line) and true models (dashed line). We again consider both the minimum detectable subhalo mass (top) and a strong impact with $M_{\rm sh} = 10^8 \msun$ (bottom) for each observational scenario and the default stream similar to GD-1 ($\sigma_\theta=0.2^\circ$, $r_0=10~\mathrm{kpc}$ and $\lambda=$100 deg$^{-1}$). The true model refers to the analytic model generated with the true subhalo impact parameters, and is equivalent to the Asimov data. When generating the example datasets, we use the same random seed to sample the Gaussian noise for different observational scenarios, so they are more comparable. We also show the breakdown of $q_0$ for these example datasets, which differs somewhat from that of Fig.~\ref{fig:chi2_dist} due to statistical fluctuations.  The bottom panel of Fig.~\ref{fig:observables_with_errors} illustrates the importance of the $\overline{\Delta \theta}$ observable for strong impacts, as well as how the density perturbation becomes much more spread out in angle.

\subsection{Fitting function for minimum detectable subhalo mass}
\label{sec:minimum_mass_streams}

We now obtain a simple fitting function that estimates the minimum detectable subhalo mass $M_{\mathrm{sh}}^{\mathrm{min}}$ based on stream properties like stream width $\sigma_\theta$, distance to the Galactic Center $r_0$ and stream density $\lambda$.

For each observational scenario, we calculate the median value of $M_{\mathrm{sh}}^{\mathrm{min}}$ on all combinations of stream width $\sigma_\theta \in [0.1^\circ, 0.2^\circ, 0.4^\circ, 0.8^\circ, 1.6^\circ]$, distance to the Galactic Center $r_0 \in [5, 10, 20, 40]$ kpc and stream density $\lambda \in [10, 32, 100, 316]~\mathrm{deg}^{-1}$. We use \texttt{scipy.optimize.curve\_fit} to fit  the $5\times4\times4=80$ data points into the following ansatz:
\begin{equation}
\label{eq:fitting}
     M_{\mathrm{sh}}^{\mathrm{min}}= \left(\frac{\sigma_\theta}{\mathrm{deg}}\right)^{c_{\sigma_\theta}}
    \left(\frac{r_0}{\mathrm{kpc}}\right)^{c_{r_0}} \left(\frac{\lambda}{\mathrm{deg}^{-1}}\right)^{c_\lambda}  10^{c_\mathrm{base}}~\msun
\end{equation}

The best fit coefficients for different observational scenarios and their errors are listed in Tab.~\ref{table:fitting}. The fitting errors are at most 12\% on the coefficients and the typical errors on $\log_{10} M_\mathrm{sh}^\mathrm{min}$ are even smaller. The results of this fit are shown as dashed lines in Fig.~\ref{fig:min_msh} where they can be compared with the actual results on the mock datasets. We observe that the fits agree very well with the data points,  except when the stream is extremely thick ($\sigma_\theta=1.6^\circ$).

\begin{table}[t]
\centering
\renewcommand{\arraystretch}{1.5}
\begin{tabular}{ccccc} 
& $c_{\sigma_\theta}$ & $c_{r_0}$ & $c_\lambda$ & $c_\mathrm{base}$ \\ 
  \hline
Gaia&$1.19^{\pm4\%}$ & $1.91^{\pm3\%}$ & $-0.82^{\pm4\%}$ & $7.39^{\pm1\%}$ \\
DESI+Gaia&$1.16^{\pm4\%}$ & $1.9^{\pm3\%}$ & $-0.84^{\pm4\%}$ & $7.42^{\pm1\%}$ \\
Via+Gaia+LSST&$1.07^{\pm5\%}$ & $1.33^{\pm5\%}$ & $-0.83^{\pm5\%}$ & $7.31^{\pm1\%}$ \\
Via+Gaia+LSST10&$0.98^{\pm9\%}$ & $1.0^{\pm12\%}$ & $-0.81^{\pm8\%}$ & $7.32^{\pm2\%}$ \\
\end{tabular}
\caption{Best fit coefficients for the power-law dependence of $M_{\mathrm{sh}}^{\mathrm{min}}$ on stream properties $\sigma_\theta, r_0$ and $\lambda$ (see Eq.~\ref{eq:fitting}).}
\label{table:fitting}
\end{table}

From Tab.~\ref{table:fitting}, we see that the detectability depends strongly on the stream properties for all observational scenarios. For example, an approximate scaling is given by $M_\mathrm{sh}^\mathrm{min}\propto\sigma_\theta^{1.2} r_0^{1.9} \lambda^{-0.8}$ for the \Gaia\ scenario and $M_\mathrm{sh}^\mathrm{min}\propto\sigma_\theta^{0.98} r_0^{1.0} \lambda^{-0.8}$ for LSST 10 year sensitivity. In particular, the coefficient for the stream distance $r_0$, $c_{r_0}$, has very different values for the observational scenarios. The detectability under the first two scenarios degrades significantly ($c_{r_0}\approx 2$) with the increase of $r_0$, while in the latter two scenarios, the detectability degrades much more mildly ($c_{r_0}\approx 1$) with the increase of $r_0$. This is because the number of observable stars in \Gaia\ decreases much faster with $r_0$ than for LSST 10 year sensitivity (see Fig.~\ref{fig:no_density}). The other coefficients have little or no dependence on observational scenario.

The application of this fit to a specific observed stream may not be that accurate because of the assumptions made in this work, including a) the circular orbit for the stream, b) the default parameters for the subhalo impact, c) the assumption of fixed age and metallicity used in the MIST isochrones, and d) the usage of fixed angular bin size of $2^\circ$ which might not be optimal for farther away streams, where impacts typically have smaller angular size. However it gives us an effective way to quickly identify promising streams with potential for subhalo detection, from a large number of observed streams. We can always perform more accurate tests on the smaller set of promising streams that are selected from this approximate method. 

\subsection{Confidence intervals on subhalo mass}
\label{sec:msh_confidence}

For subhalo mass estimation, following~\cite{Cowan_2011} again, we use the following likelihood ratio as the test statistic:
\begin{equation}
\label{eq:t}
    t(M_{\mathrm{sh}})=2\ln \frac{L(\hat M_\mathrm{sh},~\hat{\bm\theta})}{L(M_\mathrm{sh},~\hat{\bm\theta})}.
\end{equation}
In the denominator, $\hat{\bm\theta}$ refers to the values of the nuisance parameters which maximize the likelihood for that particular value of $M_{\rm sh}$. By definition, $t(M_{\mathrm{sh}})=0$ at the maximum likelihood point. In the large sample limit, the maximum likelihood point will be at $M_\mathrm{sh}^\mathrm{true}$ and the $n\sigma$ confidence interval on $M_\mathrm{sh}^\mathrm{true}$ is obtained by solving for the $M_{\rm sh}$ at which $t(M_{\mathrm{sh}}) = n^2$. 

In this case, we evaluate the expected confidence interval on the Asimov data set (analytic model with no noise) for $M_\mathrm{sh}^\mathrm{true}=10^8~\msun$, rather than generating many realizations of the data. By definition, the maximum likelihood on the Asimov data set will be given by $L(\hat M_\mathrm{sh},~\hat{\bm\theta}) = L(M_\mathrm{sh}^\mathrm{true},~{\bm\theta}^\mathrm{true})$. For strong detections such as the ones expected for $M_\mathrm{sh}^\mathrm{true}=10^8~\msun$, the confidence interval achieved in the Asimov data set approaches the median expected confidence interval from many realizations of datasets. This is further justified in Appendix~\ref{app:q0_dist},
where we show that the median value of the optimal subhalo mass, $\hat M_\mathrm{sh}^\mathrm{median}$, approaches the true subhalo mass when $M_\mathrm{sh}^\mathrm{true}=10^8~\msun$ (see the second and fourth columns of the bottom panel of Fig.~\ref{fig:q0_dist_all_settings}).

\begin{figure}[t]
\centering
\includegraphics[width=\columnwidth]{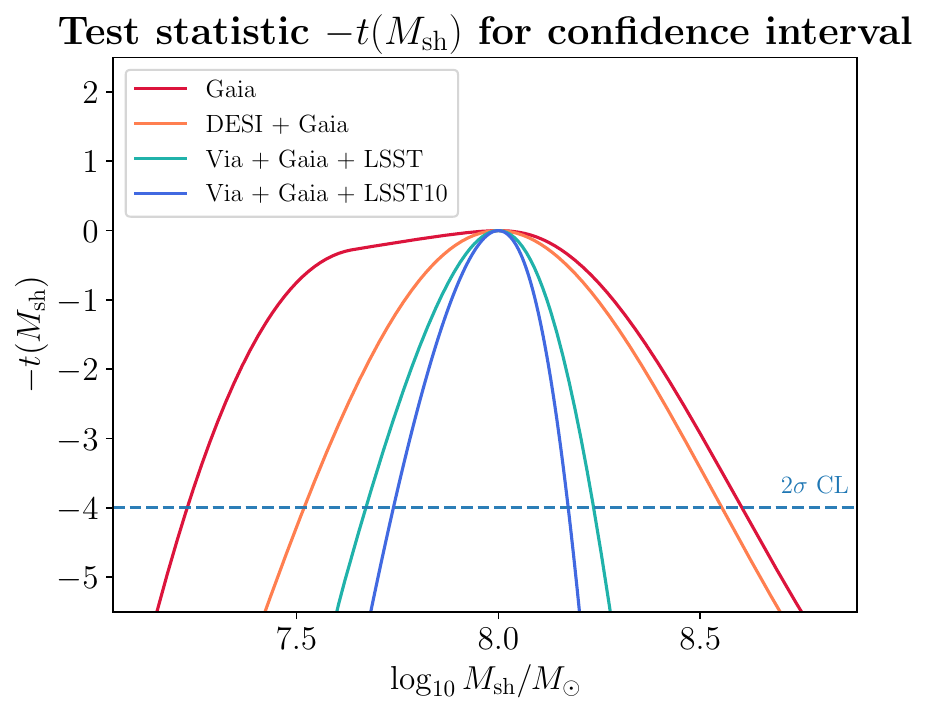}
\caption{Test statistic $-t(M_{\mathrm{sh}})$ used to obtain confidence intervals, as defined in Eq.~\ref{eq:t}. Here $-t(M_{\mathrm{sh}})$ is evaluated on the Asimov data set for a stream with $\sigma_\theta=0.2^\circ$, $r_0=10~\mathrm{kpc}$, $\lambda=100~\mathrm{deg}^{-1}$ and all the four observational scenarios. The dataset is generated from $M_\mathrm{sh}^\mathrm{true}=10^8 \msun$. }
\label{fig:ts_for_mass}
\end{figure}

\begin{figure*}[t]
\centering
\includegraphics[width=\textwidth]{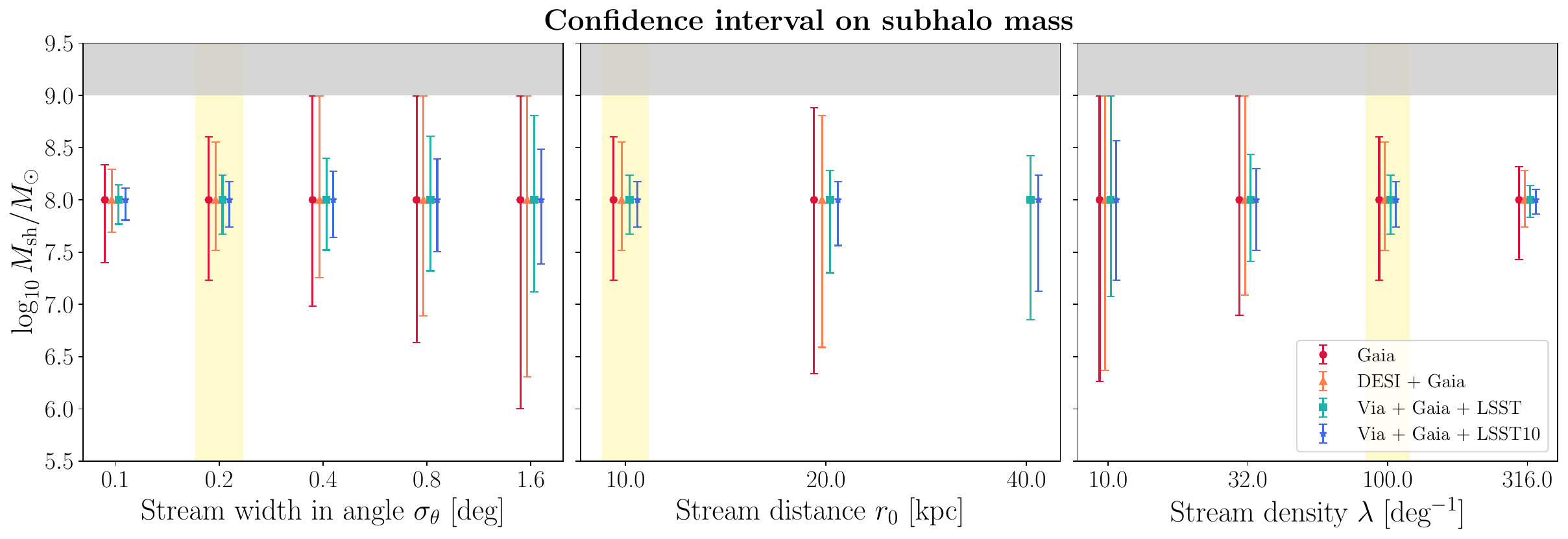}
\caption{\textbf{Left:} $2\sigma$ confidence interval on subhalo mass for a dataset of $M_\mathrm{sh}^\mathrm{true}=10^8~\msun$ as a function of stream width. \textbf{Middle:} same thing but as a function of stream distance. Note that we did not show $r_0 = 5$ kpc here due to a breakdown of the impulse approximation Eq.~\ref{eq:analytic_assumption} at $M_\mathrm{sh}=10^8~M_\odot$ and $r_0 = 5$ kpc. \textbf{Right:} same thing but as a function of stream density. In all panels, the fixed properties always use GD1-like values ($\sigma_\theta=0.2^\circ$, $r_0=10~\mathrm{kpc}$ and $\lambda=$100 deg$^{-1}$ as highlighted in yellow). The region above $10^{9}~\msun$ is shaded because the analytic model breaks down in that regime.}
\label{fig:CL_msh}
\end{figure*}

Fig.~\ref{fig:ts_for_mass} shows $-t(M_{\mathrm{sh}})$ evaluated on the Asimov data set. 
As in the case for $q_0$, we evaluate $t(M_\mathrm{sh})$ using \texttt{scipy.optimize.minimize} to find the maximized likelihood $L(M_\mathrm{sh},~\hat{\bm\theta})$ as a function of $M_\mathrm{sh}$. We determine the $2\sigma$ confidence interval on $M_\mathrm{sh}^\mathrm{true}$ from the intersection of the lines with $-t(M_{\mathrm{sh}}) = -n^2=- 4$ (dashed line). We see that the confidence interval spans further on the lower mass end compared to the higher mass end. This is expected because the subhalo is more difficult to detect at the low subhalo mass end, leading to higher uncertainties and wider confidence intervals. The opposite is true for the high subhalo mass end.

Repeating this process for different streams, we show the dependence of the subhalo mass estimation on stream properties in Fig.~\ref{fig:CL_msh}. In cases where a $M_\mathrm{sh}^\mathrm{true}=10^8~\msun$ subhalo is not detectable, a confidence interval is not shown. In addition, in some cases the confidence interval can extend to very large subhalo mass. However, the condition in  Eq.~\ref{eq:analytic_assumption} required for the validity of the analytic model no longer holds when the subhalo mass is above $\sim 10^{9}~\msun$. We thus do not consider masses above $10^9~\msun$, and shade this region in gray. Also, the confidence intervals for $r_0=5$ kpc streams are not shown as the combination of small $r_0$ and big $M_\mathrm{sh}$ break the assumption in Eq.~\ref{eq:analytic_assumption}.


\begin{figure*}[t]
\centering
\includegraphics[width=\textwidth]{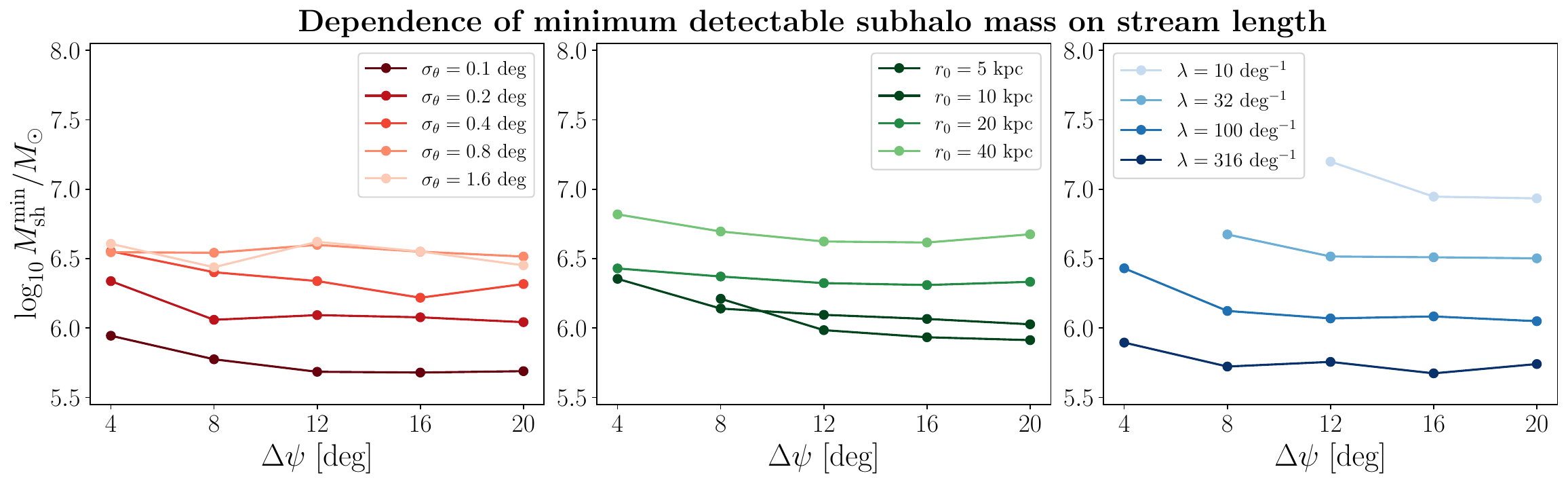}
\caption{Minimum detectable subhalo mass at 95\% confidence level, as a function of angular range used. A missing data point means no subhalo below $10^8~\msun$ could be detected. Only the median value of $M_{\rm sh}^{\rm min}$ is shown. The different lines show variation with stream width (left), stream distance (middle) and stream density (right). In each plot, we vary one stream property and keep the other two fixed at our default values ($\sigma_\theta = 0.2~\mathrm{deg}$, $r_0=10~\mathrm{kpc}$ and $\lambda=100~\mathrm{deg}^{-1}$). All the plots assume the ``Via + Gaia + LSST10'' scenario.}
\label{fig:psi_range}
\end{figure*}

\subsection{Minimal stream length in angle}
\label{sec:min_stream_length}

In addition to the stream properties used in Eq.~\ref{eq:fitting}, the stream length in angle, denoted by $l$, also affects detectability. So far, the default angular region in the likelihood calculations is $[-22^\circ, 22^\circ ]$ around the impact location, for a total range of $\Delta \psi = 44^\circ$ on a single stream arm. This region was selected to be able to capture the full perturbation profile of a fairly massive subhalo as high as $\sim 10^9~\msun$. However, this assumes a stream which is at least twice the angular length, which would impose a strong requirement when considering Milky Way streams.

On the other hand, for the minimum detectable subhalo mass $M_{\rm sh}^{\rm min}$, the requirement on the stream length can be reduced significantly. We now consider variable angular regions of size $[ - \Delta\psi/2, \Delta\psi/2 ]$ with total range $\Delta\psi$. In Fig.~\ref{fig:psi_range}, we plot the median $M_{\rm sh}^{\rm min}$ as a function of $\Delta\psi$ for varying stream properties $\sigma_\theta$, $r_0$, and $\lambda$. For simplicity, we only consider the best observational scenario ``Via + Gaia + LSST10'' here. We follow the same procedure as that used to obtain Fig.~\ref{fig:min_msh}. We only show a data point if the resulting $M_{\rm sh}^{\rm min}$ is smaller than $10^8~\msun$. 

Streams which are shorter will capture a smaller fraction of the full perturbation profile caused by the subhalo, leading to a larger threshold of $M_{\rm sh}$ for detectability. For most streams, particularly the ones with properties resulting in low $M_{\rm sh}^{\rm min}$, the minimum detectable mass flattens out for $\Delta\psi \ge 8^\circ$. This region of $8^\circ$ roughly corresponds to the angular extent of the perturbation for a subhalo impact of $M_{\rm sh} \sim 10^6 M_\odot$ on our fiducial stream (see top panel of Fig.~\ref{fig:observables_with_errors}).
Depending on stream and impact properties, the angular extent of a perturbation can vary: for instance, gap size will increase with time since flyby $t$, while the angular extent of all observables scales inversely with stream distance $r_0$. 

In applying our results to Milky Way streams, we will simply impose a uniform minimum on the total stellar stream length, such that an angular region of $8^\circ$ for the impact can be achieved. Since this must be on one arm of the stream, the total stream length must be at least $16^\circ$. Furthermore, regions around the progenitor or near the two ends of the stream cannot be reliably used for the statistical tests considered in this work.
Accounting for this, we require that the total stellar stream length is at least $20^\circ$. This selects streams with better prospects for subhalo detectability, and also ensures that our results in Sec.~\ref{sec:minimum_mass_streams} can be applied.

\subsection{Dependence on impact properties}
\label{sec:nuisance}

Up to this point, we have been using the default values for the parameters $(b,~t,~w_r,~w_t,~w_z)$ describing the impact. We have also assumed a fixed relationship for the scale radius $r_s(M_{\rm sh})$ as given in Eq.~\ref{eq:mass_radius}. In this subsection, we explore the effects of varying these parameters assuming a GD1-like stream ($\sigma_\theta=0.2^\circ$, $r_0=10$ kpc, $\lambda=100$ deg$^{-1}$), and the best observational scenario ``Via + Gaia + LSST10''.

\begin{figure*}[t]
\centering
\includegraphics[width=\textwidth]{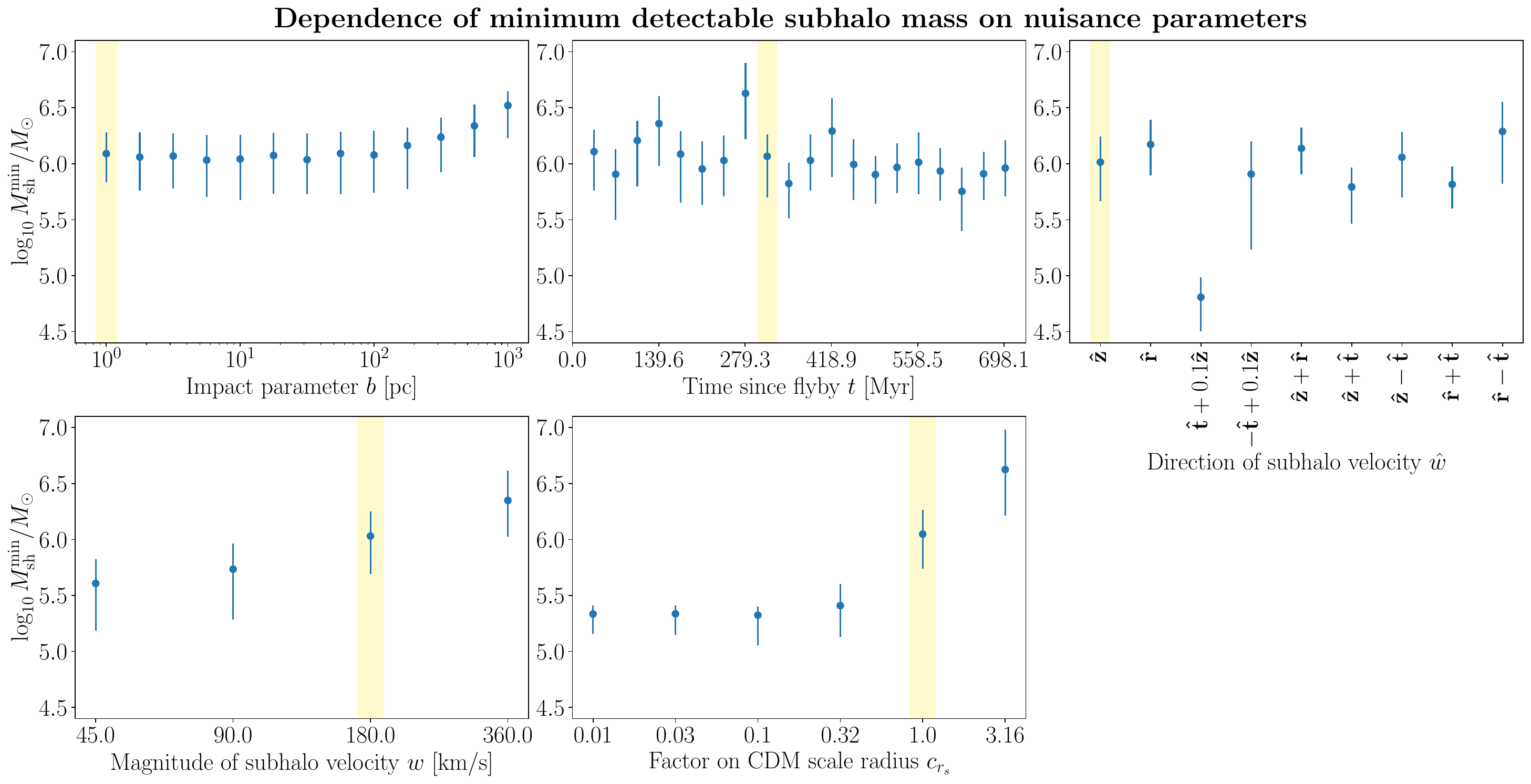}
\caption{Dependence of subhalo detectability on different aspects of the impact. We keep the stream properties fixed at the GD1-like values $\sigma_\theta=0.2^\circ$, $r_0=10~\mathrm{kpc}$, $\lambda=100~\mathrm{deg}^{-1}$, $\sigma_z=35$ pc, and assume the ``Via + Gaia + LSST10'' scenario. In each subpanel, we vary one parameter for the subhalo impact, and keep all the other parameters fixed at the default values. The default values for each  parameter are highlighted in yellow.}
\label{fig:min_msh_nuisance}
\end{figure*}

\begin{itemize}
    \item {\emph{Impact parameter $b$ --}} Our default value of $b=0$ is the optimal case. The results for non-zero $b$ are shown in the top left panel of Fig.~\ref{fig:min_msh_nuisance}. The detectability remains almost the same when the impact parameter $b$ is below 100 pc, and then begins to degrade when $b$ goes beyond 100 pc.
    The turning point is at the same scale as the scale radius of a CDM subhalo of $10^6~\msun$, which is 162 pc according to Eq.~\ref{eq:mass_radius}. This is consistent with our statement that the results do not vary significantly for impact parameter $b \lesssim r_s$, which can also be seen from the dependence of the velocity kicks on $b^2+r_s^2$ in Eqs.~\ref{eq:vkr}-\ref{eq:vkz}.
    Our results are limited to $ b \le 10^3$ pc given the assumptions of the analytic approximation, Eq.~\ref{eq:analytic_assumption}. 
    
    \item {\emph{Time since flyby $t$ --}} We show $M_{\rm sh}^{\rm min}$ as a function of the time since flyby, $t$, in the top middle panel of Fig.~\ref{fig:min_msh_nuisance}. There is a larger variation with $t$ for more recent impacts, especially within a few hundreds of Myr, due to the oscillations in the position and velocity observables over the orbit. The minimum detectable subhalo mass is locally maximized at \(\displaystyle279.3~\mathrm{Myr} \times\tfrac{n}{2}\) ($n=0, 1, 2, \cdots$), where \(\displaystyle279.3~\mathrm{Myr}=\frac{2\pi r_0}{V_c}\) is exactly the oscillation period for $\overline{\Delta \theta}$ in the analytic model. At these times, the most important observable $\overline{\Delta \theta}$ is close to 0, so that the impact is the least detectable. Fig.~\ref{fig:q0_breakdown_zero_z} shows the breakdown of the test statistic at these special times, with no contribution from $\overline{\Delta \theta}$.
    
\begin{figure}[t]
\centering
\includegraphics[width=\columnwidth]{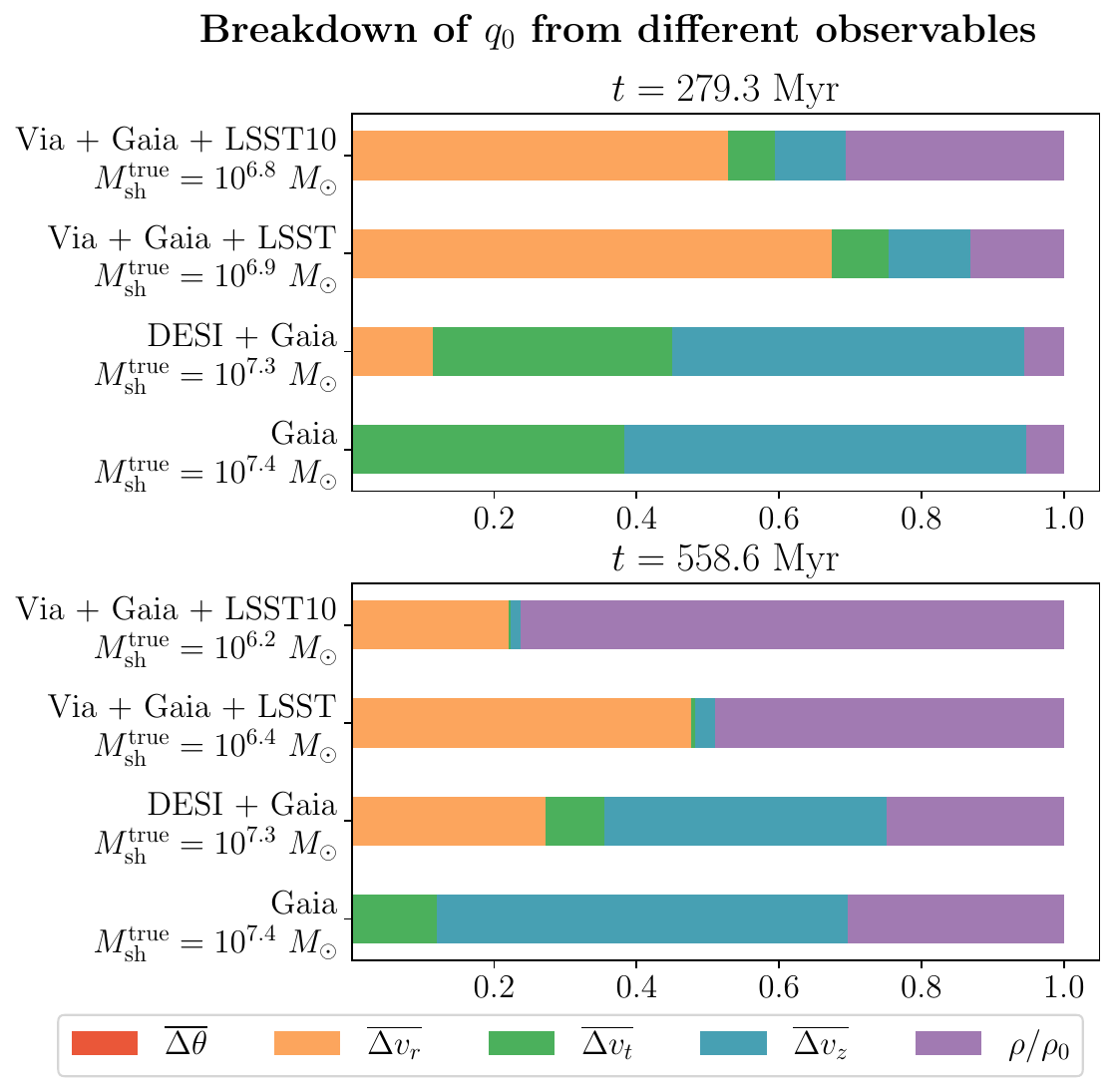}
\caption{Decomposition of $q_0$ contributions from different observables. All settings are the same as the top panel of Fig.~\ref{fig:chi2_dist} except that we consider special times since flyby, $t=279.3$ Myr (\textbf{top}) and $t=558.6$ Myr (\textbf{bottom}), when the perturbation in the $z$ direction is 0. The contribution from $\overline{\Delta \theta}$ vanishes completely in these cases. Subhalo masses are chosen to be the minimum detectable subhalo mass, and $q_0$ is evaluated on the Asimov data set.}
\label{fig:q0_breakdown_zero_z}
\end{figure}

    As $t$ increases, the variations with $t$ become smaller. This is because the size of the gap caused by a subhalo impact grows with time, which makes the density observable $\rho/\rho_0$ more important over other periodic observables for detectability. The gap growth also causes a slight decrease in $M_{\rm sh}^{\rm min}$ as larger gaps are easier to detect. Fig.~\ref{fig:q0_breakdown_zero_z} also shows how the density observable plays a more important role as $t$ grows. Note that we do not consider much older impacts, which may be more difficult to detect. Impacts older than half the stream age tend to be located in the outer parts of the stream, where the stellar density is lower, and the perturbation may also be partially washed out by the stream dispersion. These effects are not captured in the analytic model.

    \item {\emph{Direction of subhalo velocity $\hat w$ --}} In studying the effects of changing the absolute subhalo velocity $\bm{w}=(w_r, w_t, w_z)$, we first keep its magnitude fixed at $w=180$ km/s, and look into the effect of different directions $\hat{w}$.
    Taking symmetry considerations into account, we pick the following subhalo directions as representatives: $\hat z~\mathrm{(default)}, \hat r, \pm\hat t+0.1\hat z, \hat z + \hat r, \hat z\pm\hat{t}$ and $\hat r \pm \hat t$, where $\hat t$ is direction along the stream motion. Note that velocities purely along $\pm\hat t$ break the assumption of the analytic model, Eq.~\ref{eq:analytic_assumption}, so we have added a small perpendicular component $0.1\hat z$ to them. The direction $\pm\hat{t}+0.1\hat z$ is consistent with the impulse approximation, with $ 
    \frac{w_\mathrm{rel}}{w_\perp}\sqrt{b^2+r_s^2} \lesssim 0.3 r_0~~\text{and}~~\frac{V_c}{w_\perp}\sqrt{b^2+r_s^2}\lesssim 0.3 r_0
    $ at the minimum detectable subhalo mass. The results are shown in the top right panel of Fig.~\ref{fig:min_msh_nuisance}. 
    
    For most of the directions, there is only a mild variation in $M_{\rm sh}^{\rm min}$ (within $1\sigma$ of the variation from data errors). The largest variation occurs when $\hat{w}$ is close to the $\hat t$ direction. The subhalo impact is easier to detect when the subhalo is moving in the same direction as the stream stars ($\hat t$), since the relative velocity is smaller and the effective interaction time is longer.
    Aside from this special case, our results in previous subsections for the default $\hat w = \hat z$ direction are representative for general $\hat w$.

\begin{figure}[t]
\centering
\includegraphics[width=\columnwidth]{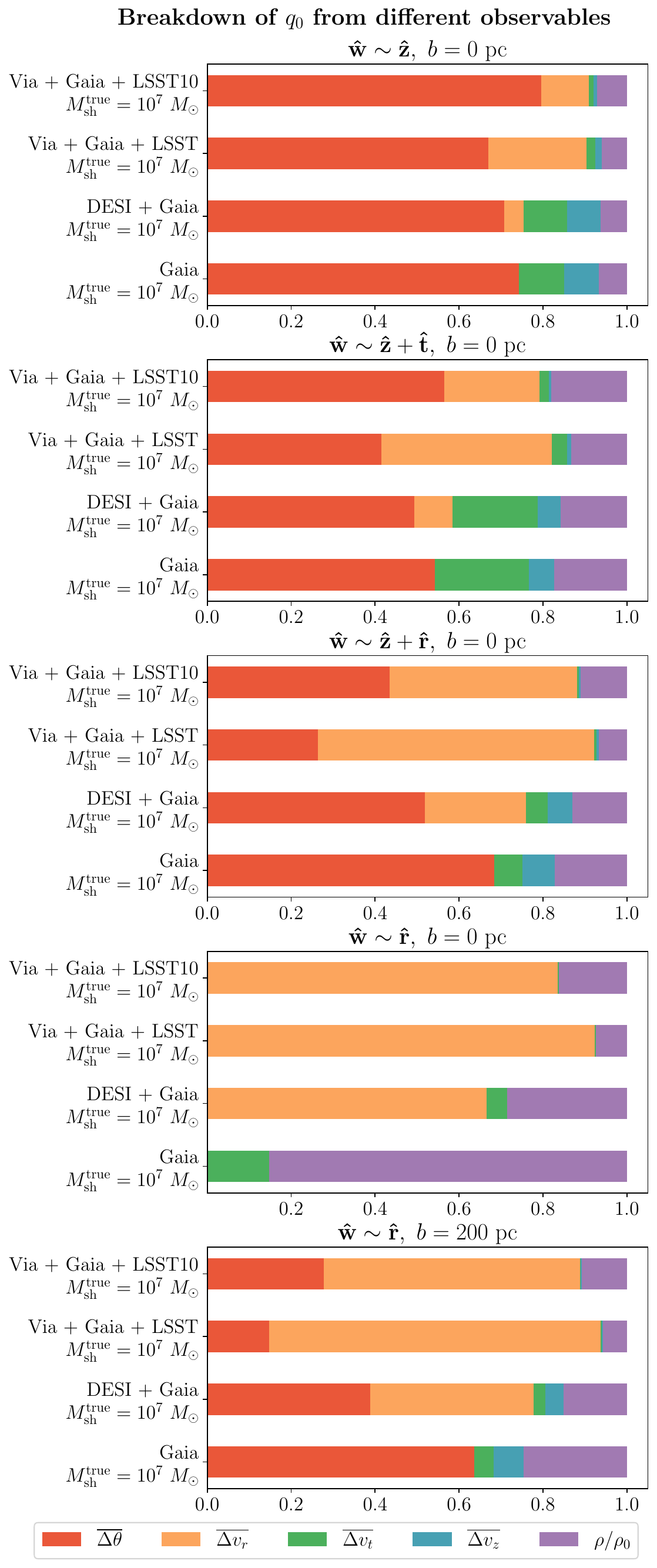}
\caption{Decomposition of $q_0$ contributions from different observables. All settings are the same as the top panel of Fig.~\ref{fig:chi2_dist} except that we consider different directions of subhalo velocity $\hat w$ and nonzero $b$ in some cases. All subhalo masses are chosen to be $10^7~\msun$ and $q_0$ is evaluated on the Asimov data set.}
\label{fig:q0_breakdown_w_dir}
\end{figure}

    Though the minimum detectable subhalo mass does not vary a lot for most of the subhalo directions, it is still interesting to look at the breakdown of test statistics $q_0$ (Eq.~\ref{eq:q0}) from different observables for different subhalo directions, which is shown in Fig.~\ref{fig:q0_breakdown_w_dir}. As we can see, when the subhalo velocity deviates from the default $\hat z$ direction (top panel), the importance of  radial velocity measurements $\overline{\Delta v_r}$ (orange) increases significantly, and the contribution from the position information $\overline{\Delta \theta}$ (red) no longer dominates. The second panel with $\hat w \sim \hat z + \hat t$ shows that the radial velocity is important even when the subhalo velocity has no sub component in $\hat r$ direction. The third panel with $\hat w \sim \hat z + \hat r$ shows that in the best scenario where we have $r<27$ position information and $G<24$ radial velocity information, $\overline{\Delta \theta}$ and $\overline{\Delta v_r}$ are equally important. The fourth panel with $\hat w\sim \hat r$ and $b=0$ shows an extreme case where the position information $\overline{\Delta \theta}$ does not provide any information, though this particular geometry is very unlikely to occur in reality. However, we show a slightly more realistic case in the last panel by increasing the impact parameter $b$ to 200 pc. In this case the contribution from radial velocity still dominates.

    \item {\emph{Magnitude of subhalo velocity $w$ --}} The bottom left panel of Fig.~\ref{fig:min_msh_nuisance} shows how the magnitude of the subhalo velocity $w$ affects the detectability, while keeping the direction fixed to the default direction $\hat z$. The slower the subhalo moves, the stronger the impact, and the easier it is to detect.

    \item {\emph{Subhalo concentration $c_{r_s}$ --}} So far, we have assumed a subhalo mass-radius relation based on modeling CDM subhalos with a Plummer potential, Eq.~\ref{eq:mass_radius}. However, the actual subhalo mass-radius relation could deviate from this with different particle models of DM (e.g. warm dark matter, WDM~\citep{2016MNRAS.455..318B,2016MNRAS.460.1214L}; self-interacting dark matter, SIDM~\citep{2023ApJ...949...67Y} or fuzzy dark matter, FDM~\citep{2017PhRvD..95d3519D,2020ApJ...904..161B,2022MNRAS.511..943C}). We include a coefficient $c_{r_s}$ for the scale radius relation, generalizing  Eq.~\ref{eq:mass_radius} to the form:
    \begin{equation}
    \label{eq:mass_radius_crs}
        r_s=c_{r_s} \left(\frac{M_{\mathrm{sh}}}{10^8~\msun}\right)^{0.5}1.62~\mathrm{kpc}.
    \end{equation}
    The dependence of $M_{\rm sh}^{\rm min}$ on $c_{r_s}$ is shown in the bottom middle panel of Fig.~\ref{fig:min_msh_nuisance}. 
    Note that $c_{r_s}$ is included self-consistently both in generating mock data and evaluating the likelihood. Compact subhalos have a smaller minimum detectable subhalo mass, with a plateau in detectability below $c_{r_s} \sim 0.32$. We see a strong dependence of subhalo detectability on the concentration, however this is primarily because we are assuming zero impact parameter $b=0$. In more realistic cases with $b\gtrsim100$ pc, the dependence on the concentration will be much milder.

\end{itemize}

\section{Application to known streams}
\label{sec:stream_catalog}

Having studied the dependence of subhalo detectability on stream properties, we can now turn to the catalog of currently-identified stellar streams and evaluate their prospects for subhalo detection.

\begin{table*}[p]
  \centering
  {\renewcommand{\arraystretch}{1.2}
  \begin{tabular}{ccccccccccc}

Name & $\sigma_\theta$ & $l$ & $r_h$ & $M_\mathrm{stellar}$ & $\lambda$ & Retro & $M_\mathrm{sh}^\mathrm{min}$ [$\msun$] & $M_\mathrm{sh}^\mathrm{min}$ [$\msun$] & $M_\mathrm{sh}^\mathrm{min}$ [$\msun$] & $M_\mathrm{sh}^\mathrm{min}$ [$\msun$] \\
 & [$^\circ$] & [$^\circ$] & [kpc] & [$\msun$] & [deg$^{-1}$] & /Prograde & Gaia & DESI + Gaia & Via + LSST & Via + LSST10 \\
\hline
\hline
C-12 & 0.51 & 28 & 11.5 & 14000 & 528 & P & 6.75e+06 & 6.58e+06 & 1.40e+06 & 7.75e+05 \\
ATLAS-Aliqa Uma & 0.26 & 41 & 21.4 & 19000 & 490 & P & 1.05e+07 & 1.05e+07 & 1.65e+06 & 7.94e+05 \\
300S & 0.34 & 25 & 15.9 & 7600 & 321 & R & 1.16e+07 & 1.16e+07 & 2.10e+06 & 1.08e+06 \\
NGC 6397 & 0.79 & 32 & 2.5 & 2500 & 83 & P & 2.85e+06 & 2.84e+06 & 1.38e+06 & 1.16e+06 \\
Palomar 5 & 0.54 & 32 & 21.3 & 17000 & 561 & P & 2.23e+07 & 2.15e+07 & 3.20e+06 & 1.45e+06 \\
GD-1 & 0.43 & 119 & 8.0 & 14000 & 124 & R & 9.06e+06 & 9.11e+06 & 2.39e+06 & 1.47e+06 \\
Orphan-Chenab & 1.02 & 137 & 20.7 & 130000 & 1003 & P & 2.79e+07 & 2.62e+07 & 3.77e+06 & 1.64e+06 \\
Ylgr & 0.72 & 49 & 9.5 & 11000 & 237 & R & 1.36e+07 & 1.33e+07 & 3.05e+06 & 1.72e+06 \\
Gaia-6 & 0.4 & 21 & 8.3 & 1800 & 91 & R & 1.16e+07 & 1.17e+07 & 3.02e+06 & 1.84e+06 \\
Kshir & 0.23 & 37 & 10.7 & 2200 & 63 & R & 1.31e+07 & 1.36e+07 & 3.16e+06 & 1.86e+06 \\
C-7 & 0.42 & 34 & 5.8 & 1500 & 47 & R & 1.07e+07 & 1.09e+07 & 3.43e+06 & 2.31e+06 \\
NGC 5466 & 0.23 & 23 & 17.4 & 1900 & 87 & R & 2.53e+07 & 2.60e+07 & 4.59e+06 & 2.31e+06 \\
Gaia-1 & 0.34 & 40 & 5.0 & 1100 & 29 & R & 9.22e+06 & 9.59e+06 & 3.32e+06 & 2.37e+06 \\
Jhelum & 0.65 & 97 & 13.0 & 17000 & 185 & P & 2.69e+07 & 2.65e+07 & 5.09e+06 & 2.60e+06 \\
Leiptr & 0.48 & 73 & 7.1 & 3000 & 43 & R & 1.95e+07 & 1.99e+07 & 5.49e+06 & 3.41e+06 \\
Gaia-8 & 0.83 & 47 & 7.1 & 3600 & 81 & R & 2.24e+07 & 2.23e+07 & 5.90e+06 & 3.52e+06 \\
Gaia-12 & 0.33 & 29 & 11.4 & 1200 & 44 & R & 3.07e+07 & 3.16e+07 & 6.85e+06 & 3.78e+06 \\
SGP-S & 0.21 & 29 & 9.5 & 530 & 19 & P & 2.48e+07 & 2.62e+07 & 6.52e+06 & 3.92e+06 \\
New-25 & 0.88 & 23 & 50.3 & 18000 & 827 & P & 1.49e+08 & 1.40e+08 & 1.22e+07 & 4.02e+06 \\
C-19 & 0.35 & 47 & 18.0 & 3200 & 72 & P & 5.21e+07 & 5.30e+07 & 8.83e+06 & 4.22e+06 \\
Hrid & 1.0 & 77 & 3.2 & 2000 & 27 & R & 1.49e+07 & 1.50e+07 & 6.14e+06 & 4.57e+06 \\
Kwando & 6.58 & 57 & 7.7 & 40000 & 742 & P & 4.96e+07 & 4.45e+07 & 9.62e+06 & 4.81e+06 \\
Omega Centauri & 4.25 & 81 & 5.2 & 20400 & 266 & R & 3.24e+07 & 3.00e+07 & 8.37e+06 & 4.86e+06 \\
C-13 & 0.49 & 23 & 8.4 & 750 & 34 & P & 3.33e+07 & 3.40e+07 & 8.49e+06 & 4.96e+06 \\
Sylgr & 0.99 & 30 & 3.3 & 700 & 25 & P & 1.70e+07 & 1.72e+07 & 6.90e+06 & 5.08e+06 \\
Phlegethon & 0.93 & 77 & 3.4 & 1700 & 23 & R & 1.75e+07 & 1.78e+07 & 7.04e+06 & 5.16e+06 \\
Indus & 1.98 & 90 & 16.6 & 34000 & 399 & P & 8.57e+07 & 8.03e+07 & 1.23e+07 & 5.29e+06 \\
C-24 & 0.58 & 37 & 14.8 & 2700 & 77 & P & 6.18e+07 & 6.19e+07 & 1.11e+07 & 5.38e+06 \\
C-23 & 0.33 & 20 & 8.7 & 310 & 16 & P & 4.10e+07 & 4.29e+07 & 1.08e+07 & 6.37e+06 \\
New-16 & 0.34 & 22 & 5.0 & 170 & 8 & P & 2.61e+07 & 2.77e+07 & 9.49e+06 & 6.61e+06 \\
NGC 3201 & 0.75 & 111 & 4.9 & 2100 & 20 & R & 3.09e+07 & 3.16e+07 & 1.03e+07 & 6.82e+06 \\
New-1 & 0.2 & 31 & 15.5 & 420 & 14 & P & 7.60e+07 & 8.09e+07 & 1.52e+07 & 7.77e+06 \\
Slidr & 1.67 & 35 & 3.1 & 840 & 25 & P & 2.75e+07 & 2.73e+07 & 1.09e+07 & 7.78e+06 \\
NGC 288 & 1.23 & 30 & 11.0 & 2200 & 78 & R & 8.53e+07 & 8.36e+07 & 1.66e+07 & 8.30e+06 \\
Gaia-9 & 1.49 & 36 & 4.2 & 950 & 28 & R & 3.96e+07 & 3.94e+07 & 1.33e+07 & 8.74e+06 \\
M5 & 0.33 & 39 & 14.2 & 710 & 19 & P & 9.15e+07 & 9.55e+07 & 1.81e+07 & 9.16e+06 \\
Gaia-11 & 0.46 & 50 & 12.5 & 1100 & 23 & R & 9.12e+07 & 9.39e+07 & 1.87e+07 & 9.57e+06 \\
C-11 & 1.48 & 33 & 6.9 & 1100 & 35 & P & 8.36e+07 & 8.27e+07 & 2.10e+07 & 1.18e+07 \\
New-7 & 0.76 & 26 & 2.8 & 110 & 4 & P & 3.68e+07 & 3.86e+07 & 1.71e+07 & 1.32e+07 \\
C-9 & 1.71 & 25 & 5.4 & 630 & 27 & P & 7.83e+07 & 7.76e+07 & 2.24e+07 & 1.34e+07 \\
M3 & 0.97 & 76 & 9.8 & 2000 & 28 & P & 1.20e+08 & 1.20e+08 & 2.59e+07 & 1.34e+07 \\
New-10 & 1.0 & 26 & 2.3 & 100 & 4 & P & 3.78e+07 & 3.95e+07 & 1.91e+07 & 1.53e+07 \\
M68 & 1.37 & 105 & 5.4 & 1700 & 17 & P & 8.65e+07 & 8.69e+07 & 2.55e+07 & 1.54e+07 \\
New-13 & 0.75 & 29 & 6.2 & 250 & 9 & R & 9.23e+07 & 9.55e+07 & 2.71e+07 & 1.63e+07 \\
C-25 & 2.13 & 25 & 11.0 & 1300 & 55 & P & 2.17e+08 & 2.10e+08 & 3.99e+07 & 1.87e+07 \\
New-23 & 0.69 & 30 & 4.2 & 120 & 4 & R & 7.51e+07 & 7.91e+07 & 2.80e+07 & 1.91e+07 \\
Wukong & 5.99 & 63 & 18.1 & 15000 & 252 & P & 5.49e+08 & 5.01e+08 & 6.62e+07 & 2.48e+07 \\
New-3 & 2.17 & 71 & 1.7 & 120 & 2 & P & 1.05e+08 & 1.08e+08 & 5.82e+07 & 4.70e+07 \\
New-6 & 2.54 & 80 & 3.1 & 340 & 4 & P & 1.87e+08 & 1.89e+08 & 7.15e+07 & 4.76e+07 \\
New-2 & 3.47 & 61 & 2.2 & 73 & 1 & P & 3.91e+08 & 4.00e+08 & 1.78e+08 & 1.25e+08 \\

  \end{tabular}
  }
  \caption{Milky Way streams and estimated minimum detectable subhalo mass at 95\% CL under different observational scenarios. Stream properties are obtained from the catalog in \cite{bonaca2024stellarstreamsgaiaera}. We only include streams with reported stellar mass, angular length $l > 20^\circ$, and density of observable stars greater than 1 star per $2^\circ$ in Gaia.}
  \label{tab:stream_catalog}
\end{table*}

\begin{figure*}[p]
\centering
\includegraphics[width=\textwidth]{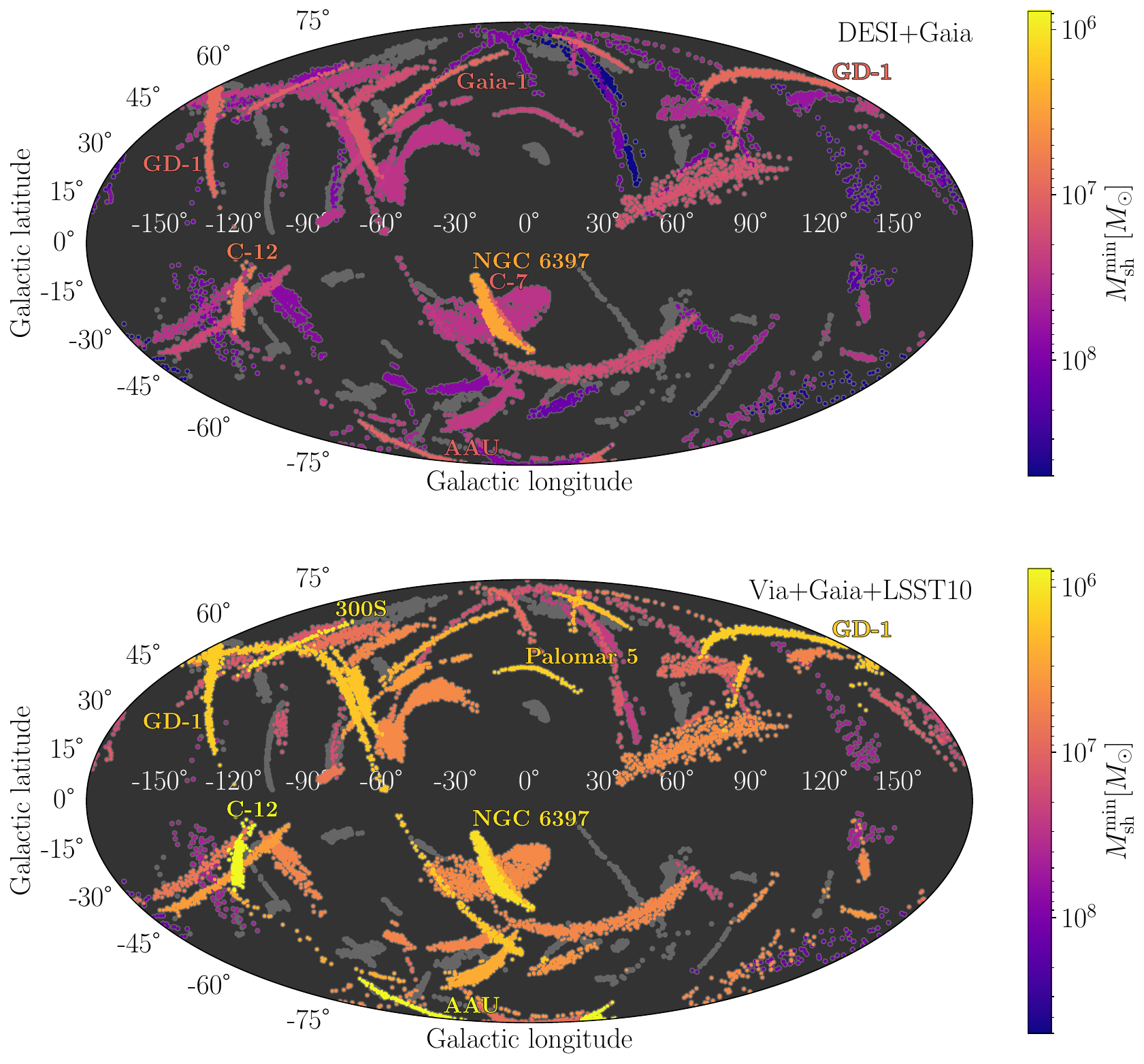}
\caption{Visualized stream catalog in the sky, where color scale indicates the minimum detectable subhalo mass, $M_\mathrm{sh}^\mathrm{min}$, with respect to observational scenario ``DESI+Gaia'' (\textbf{top}) and ``Via+Gaia+LSST10'' (\textbf{bottom}). The six streams with the best $M_\mathrm{sh}^\mathrm{min}$ under each observational scenario are labeled. Streams shown in light gray do not have an estimated $M_\mathrm{sh}^\mathrm{min}$: these have missing $M_{\rm stellar}$, are too short, or the observable number of stars is too low. }
\label{fig:stream_catalog_skymap}
\end{figure*}

We use the catalog of cold stellar streams\footnote{\url{https://github.com/abonaca/streams_overview}} from
\cite{bonaca2024stellarstreamsgaiaera}, which is based on streams currently in the \texttt{galstreams} package~\citep{2023MNRAS.520.5225M}. The catalog contains information on stream width $\sigma_\theta$, stream length $l$, heliocentric distance to the stream $r_h$, and stellar mass $M_\mathrm{stellar}$. Note that we use $r_h$ as the analog of $r_0$ in the analytic model, as this determines the distance to the observer and thus the number of observable stars. Of the 131 total streams in the catalog, we only consider the 86 streams with a reported stellar mass. 
Streams with length $l < 20^\circ$ are also filtered out based on the analysis in Sec.~\ref{sec:min_stream_length}. We estimate the total number of stars per unit angle as $\lambda \approx 4 N_{\rm stars} / 3 l$ (see discussion around Eq.~\ref{eq:lambda}),  where $N_{\rm stars}$ is obtained from $M_\mathrm{stellar}$ using the distribution of stars in the IMF. We also exclude streams whose observable number of stars, $n_\mathrm{obs}$, is less than 1 star per $2^\circ$ for the \Gaia\ magnitude limit of $ G < 20.7$, since our sensitivity estimates break down in that limit. While such streams might have sufficiently large $n_\mathrm{obs}$ for Via or LSST observations, they still generally have worse statistics and therefore are sensitive only to higher subhalo masses.

In total, there are 50 streams remaining that have a reported stellar mass, $l \ge 20^\circ$, and sufficiently high stellar density.
We then use $\sigma_\theta$, $r_0$, $\lambda$ in our fitting function, Eq.~\ref{eq:fitting}, to obtain a minimum detectable subhalo mass $M_\mathrm{sh}^\mathrm{min}$ for each stream under different observational scenarios. 
Tab.~\ref{tab:stream_catalog} lists the streams and their properties, in increasing order of $M_\mathrm{sh}^\mathrm{min}$ with respect to the best observational scenario ``Via + Gaia + LSST10''. We have also included $M_\mathrm{sh}^\mathrm{min}$ for other observational scenarios, which follows a similar (but not identical) ordering as for the best case. For example, streams that are far away from us and  with high stellar density (e.g., New-25, Orphan-Chenab, ATLAS-Aliqua Uma) benefit much more from the improved magnitude limits with LSST.

\begin{figure}[h]
\centering
\includegraphics[width=0.41\textwidth]{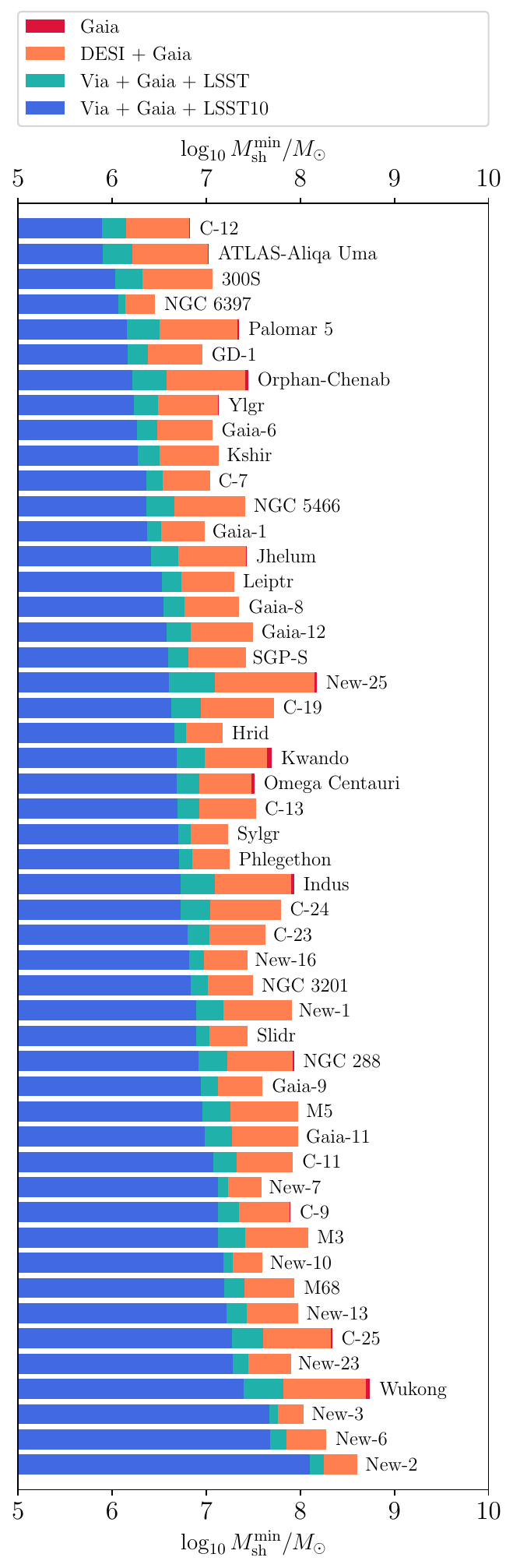}
\caption{Visualized stream catalog ranked by detectability under Via + Gaia + LSST10.}
\label{fig:stream_catalog_ranked}
\end{figure}

\begin{figure*}[p]
\centering
\includegraphics[width=\textwidth]{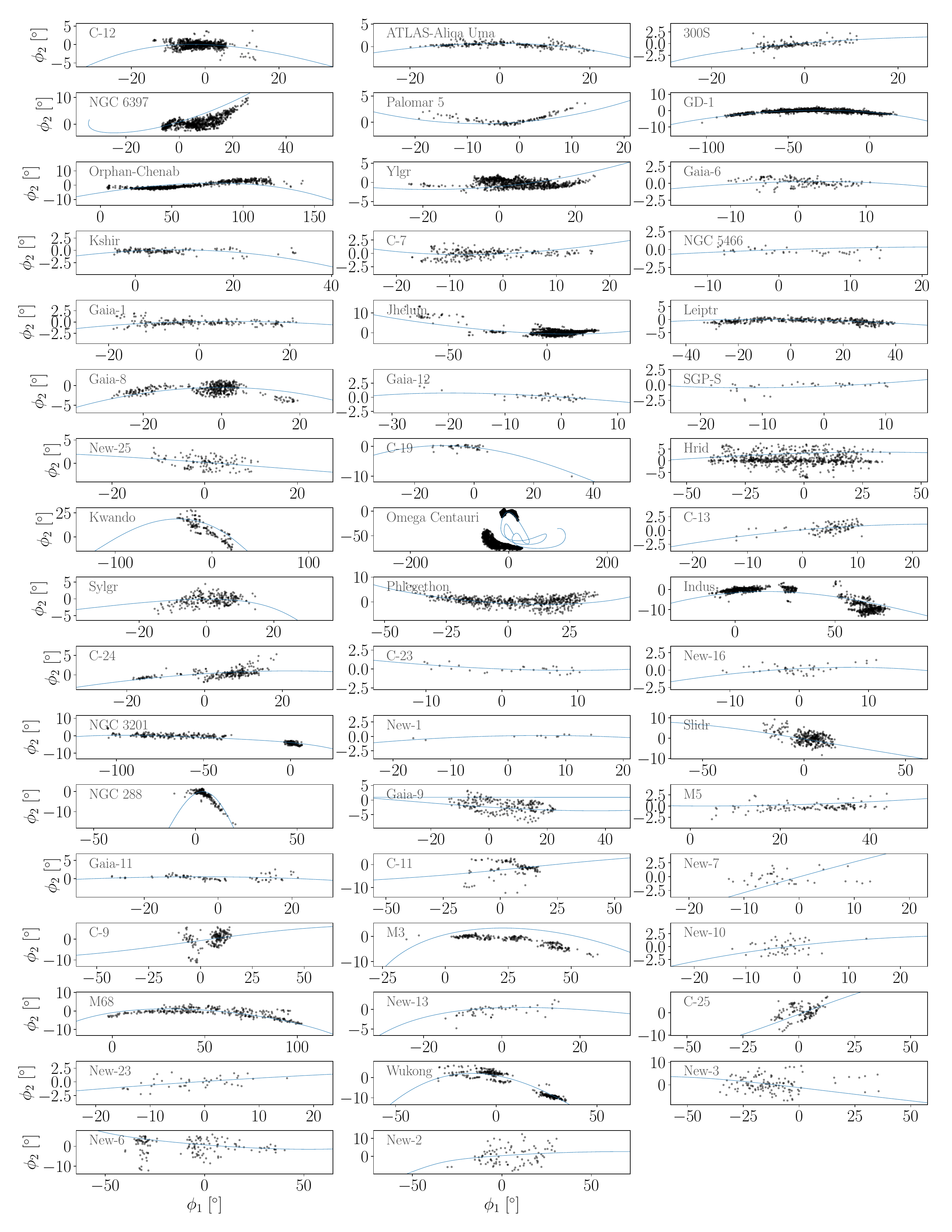}
\caption{Sky plots for individual streams ranked by detectability (left to right, top to bottom) under Via + Gaia + LSST10.}
\label{fig:stream_catalog_ranked_sky}
\end{figure*}

\begin{figure*}[t]
\centering
\includegraphics[width=\textwidth]{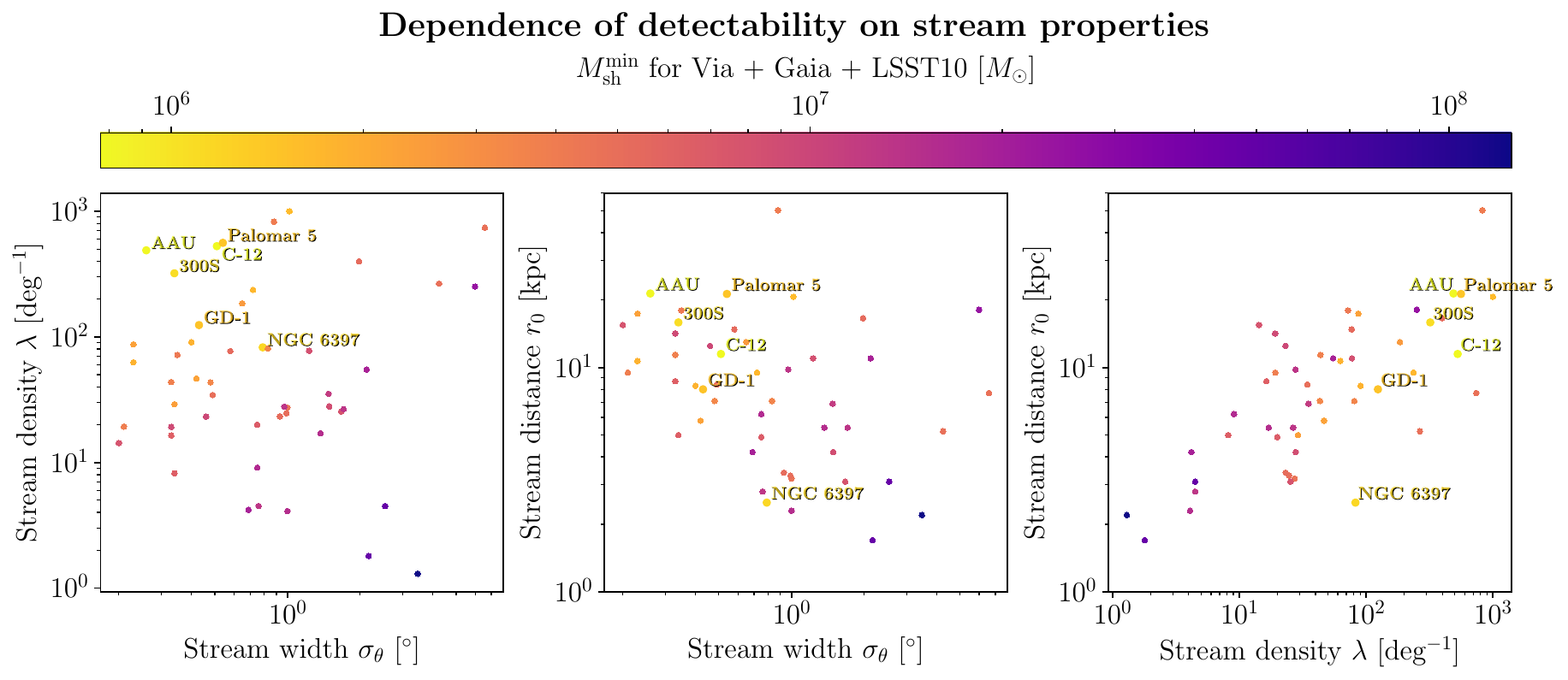}
\caption{Visualized stream catalog, where color scale indicates the minimum detectable subhalo mass, $M_\mathrm{sh}^\mathrm{min}$, with respect to the best observational scenario. The stream properties of stellar density $\lambda$, width $\sigma_\theta$, and distance to observer $r_0$ all play a role in determining $M_\mathrm{sh}^\mathrm{min}$. The six streams with lowest $M_\mathrm{sh}^\mathrm{min}$ are labeled. }
\label{fig:stream_catalog}
\end{figure*}

\begin{figure*}[t]
\centering
\includegraphics[width=\textwidth]{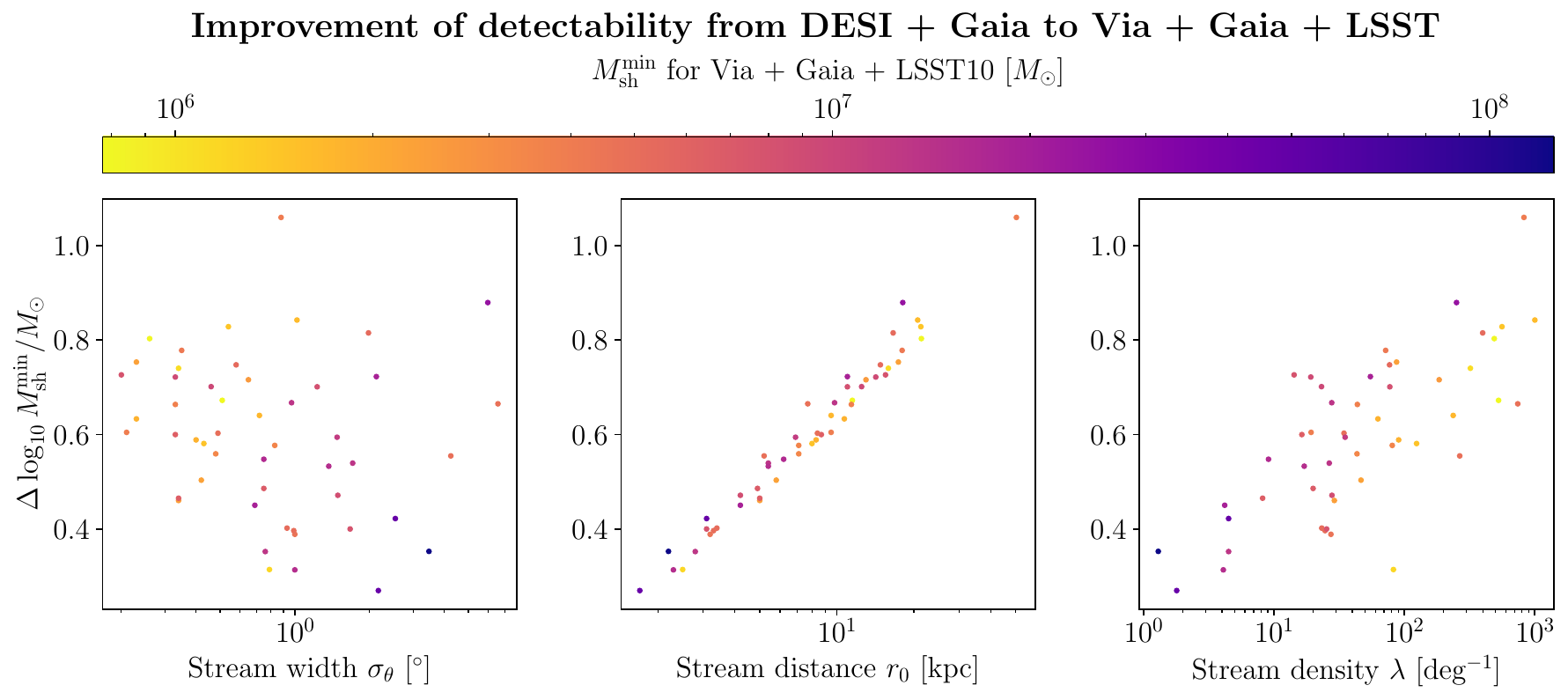}
\caption{Improvement from DESI + Gaia to Via + Gaia + LSST in $\log_{10} M_\mathrm{sh}^\mathrm{min}/\msun$ as functions of stream width (\textbf{left}), stream distance (\textbf{middle}) and stream density (\textbf{right}). The color scale indicates the minimum detectable subhalo mass, $M_\mathrm{sh}^\mathrm{min}$, with respect to the best observational scenario.}
\label{fig:improve_by_via_lsst}
\end{figure*}

We present visualizations of the stream catalog and our results in Figs.~\ref{fig:stream_catalog_skymap}-\ref{fig:improve_by_via_lsst}.  First, Fig.~\ref{fig:stream_catalog_skymap} summarizes our main result, showing all of the streams on the sky with color scale indicating subhalo detectability. The 6 most promising streams with respect to each observational scenario are also indicated.

Fig.~\ref{fig:stream_catalog_ranked} shows $M_\mathrm{sh}^\mathrm{min}$ for all scenarios and streams, where the streams are ranked by the best case observational scenario.
Fig.~\ref{fig:stream_catalog_ranked_sky}, adapted from \cite{bonaca2024stellarstreamsgaiaera}, shows how those streams appear on the sky in the same order (left to right, and top to bottom) as Fig.~\ref{fig:stream_catalog_ranked}. Each subplot shows sky positions of stream members (black points) and stream orbits (blue lines) in individual stream reference frames. Further details on how the stream members and orbit are obtained can be found in Appendix A and B of \cite{bonaca2024stellarstreamsgaiaera}. Generally, the streams with the best prospects appear thin, dense, and long on the sky. However, the sky plots also make it clear that streams must be evaluated on a case-by-case basis, as some streams are not well described by a simple stream orbit and other effects can play an important role.

Some of the streams with promising prospects are likely to be highly impacted by baryonic structures, which will limit their use for subhalo detection. For instance, Palomar 5 has features induced by interactions with the bar~\citep{pearson2017gapslengthasymmetrystellar,2017MNRAS.470...60E,2019MNRAS.484.2009B}.  This is a result of the particular orbit associated with Palomar 5, which is prograde and passes close to the Galactic bar. In general, streams on orbits far from the Galactic Center or on retrograde orbits are less likely to be affected by the Galactic bar and more promising for subhalo detection. In Tab.~\ref{tab:stream_catalog}, we have included a column ``Retrograde/Prograde'' for the orbital direction of the streams, but more study is needed to determine the impact of the bar on individual streams.
Orphan-Chenab appears to be similarly promising in our simple model, but it has features which have been attributed to a close encounter with the Large Magellanic Cloud~\citep{Erkal_2019}. The effects above would have to be accounted for in searching for subhalo signatures. 

Another example where our model cannot capture all effects comes from Jhelum. In terms of subhalo detection, it is ranked relatively highly here, partly due to its narrow width $\sigma_\theta = 0.65^\circ$.  However, there is also a much thicker component associated with Jhelum as well, which may be evidence of some more complicated dynamics~\citep{Bonaca_Jhelum_2019}. Furthermore, while we have assumed a fairly uniform stellar density along the streams, Jhelum illustrates how observed streams can exhibit many more fluctuations due to observational selection criteria in addition to real density variations.

Fig.~\ref{fig:stream_catalog} illustrates the correlations in stream properties and prospects for subhalo detectability, for the best-case observational scenario. All three properties ($\lambda$, $\sigma_\theta$, $r_0$) play a role in determining detectability. The six streams with the best prospects are labeled. While high stellar density $\lambda$ and low width $\sigma_\theta$ are clearly important, there is no single stream property that can be used to assess detectability.

In Fig.~\ref{fig:improve_by_via_lsst}, we plot the improvement in detectability from the ``DESI + \Gaia'' scenario to the ``Via + \Gaia + LSST'' scenario, given in terms of $\Delta \log_{10} M_\mathrm{sh}^\mathrm{min}/\msun$, as functions of stream width, distance and density. The increase of statistics in going from \Gaia-era to LSST-era data brings greater improvement on distant (large $r_0$) and dense (large $\lambda$) streams. Note that the improvement with $\lambda$ appears to largely be from the correlation in $r_0$ and $\lambda$, visible in the top right of Fig.~\ref{fig:stream_catalog}. This ultimately comes from a correlation in stream distance and total stellar mass. This effect may be due to selection effects: at larger distances, larger stellar mass is needed for a sufficient number of observable stars. \cite{pearson2024forecastingpopulationglobularcluster} predicts that at high Galactocentric radius, many lower surface brightness streams exist beyond the current detection limit.

\section{Discussion}
\label{sec:discussion}

In order to assess how subhalo detectability depends on stream properties, we have made simplifying assumptions throughout this work. Below, we briefly discuss how going beyond such assumptions could impact results, and compare with other work.

\subsection{Generalization to realistic streams}
\label{sec:generalization}

{\emph{Circular orbits and observer location --}} 
We expect that some of the largest differences between our analysis and more realistic models  will arise in cases of large variation of heliocentric and/or Galactocentric distance  along the stream orbit. In our study, streams are moving on circular orbits and observers are located at the Galactic Center. Accounting for eccentric orbits and placing the observer at the Sun will change the shape of subhalo-induced perturbations, their time dependence, as well as the relative importance of different observables. 
For a stream with a large variation in heliocentric distance along its length, a subhalo impact will be compressed to a shorter angular arc from the point of view of the observer, leading to a diminished detectability. Subhalo impacts will also be compressed and stretched depending on whether impacts occurred closer to pericenter or apocenter, as well as whether the stream today is observed near pericenter or apocenter. This will lead to much stronger dependence of subhalo detectability on impact time. We expect streams with relatively constant heliocentric and Galactocentric distances along the stream to have the most similar detectability as our idealized study.

Another effect in going from circular orbit to eccentric orbit is the increased fluctuations across angular bins caused by periodic stripping and epicyclic motion~\citep{2012MNRAS.420.2700K}. This introduces additional intrinsic fluctuations across bins, which breaks our assumption of a constant dispersion along the stream. 
We will leave the details of how to incorporate this effect into the statistical test for future work.

Furthermore, throughout the paper, we have been assuming that we know the progenitor orbit perfectly with zero uncertainty. However, the orbit depends on the Milky Way potential, and in practice stream tracks are used to constrain the potential~\citep{KoposovRixHogg2010,Bonaca2014ApJ...795...94B,Sanders2014MNRAS.443..423S}. This introduces additional uncertainties which can reduce the significance of a signal, although this could be mitigated by simultaneous fits to the orbits of multiple streams~\citep{Reino2021MNRAS.502.4170R}. Another issue arises when there has been a massive impact which causes a perturbation extending over the entire stream, which can be challenging to distinguish from an alternative stream orbit. This mostly affects the detectability in more massive subhalo impacts ($\gtrsim 10^8~\msun$ depending on the stream length), but has less of an effect on our results on the minimum detectable subhalo mass.

{\emph{Milky Way potential --}} The actual Milky Way potential is much more complicated than a simple logarithmic potential used here. Baryonic structures like the bar~\citep{pearson2017gapslengthasymmetrystellar}, spiral arms, as well as giant molecular clouds (GMC)~\citep{2016MNRAS.463L..17A} and globular clusters (GC)~\citep{2017MNRAS.470...60E,Ferrone2025A&A...699A.289F}, could potentially cause perturbations in the streams. See also \cite{2019MNRAS.484.2009B}, which considered the effect of all these structures on Pal-5. This will further raise the threshold of detectability, beyond the ideal case shown here. At the same time, kinematic information like radial velocity and proper motions could help distinguish between the impact of baryonic structures and DM subhalos~\citep{Bonaca2020, hilmi2024inferringdarkmattersubhalo,Price_Whelan_2016}.

{\emph{Stream model --}} Our stream model uses position/velocity dispersions assuming a particle spray model, where the parameters governing the dispersions are taken from \cite{Fardal_2015}. We investigated another particle spray model, based on a newer suite of simulations in \cite{chen2024improvedparticlesprayalgorithm}. This model includes initial radial velocity dispersions for the released particles and generates a stream with twice the $\sigma_{v_r}$ and $\sigma_{v_t}$ compared to a stream generated using~\cite{Fardal_2015}, holding the angular width $\sigma_\theta$ fixed. For our default geometry where the subhalo moves along the $\hat z$ direction, this difference does not change the estimated detectability in a significant way, because the shift in the $z$ direction, $\overline{\Delta \theta}$, is the most effective observable in subhalo detection (see Fig.~\ref{fig:chi2_dist}). The dispersion for this observable is governed primarily by the directly observed quantity $\sigma_\theta$. However, for more generic subhalo impact directions like $\hat w\sim\hat z+\hat r$ as in the third panel of Fig.~\ref{fig:q0_breakdown_w_dir}, $\overline{\Delta v_r}$ also plays an important role. In this case, the minimum detectable subhalo mass will be higher by $\sim0.2$ dex when adopting the particle spray model from~\cite{chen2024improvedparticlesprayalgorithm}, due to its higher $\sigma_{v_r}$.

In addition to affecting the dispersions, the dynamics of progenitor disruption could induce perturbations in the streams and further raise the threshold for detectability. However, these effects are expected to mainly be in the region near the progenitor. 

{\emph{Subhalo mass-radius relation and potential --}} As discussed below Eq.~\ref{eq:mass_radius}, there is a degeneracy in subhalo mass and velocity in the analytic model and we used a fixed relationship for $r_s(M_\mathrm{sh})$ to break this degeneracy. This means our results are valid only under the assumption of CDM subhalos following Eq.~\ref{eq:mass_radius}. When the actual mass-radius relation of the subhalo significantly deviates from this relation, the inferred subhalo mass is not valid.

Rather than fixing $r_s(M_\mathrm{sh})$, the degeneracy in mass and velocity can also be broken by using other subhalo potentials (such as Hernquist or NFW) and/or using the method of orbit integration, where we model the perturbed stream using the perturbed orbit of the progenitor (more details are discussed in App.~\ref{sec:oi}). The method of orbit integration does not rely on the impulse approximation, thus effectively breaking the degeneracy of the analytic model. We then evaluated how expected confidence intervals change when the $r_s(M_\mathrm{sh})$ relationship is removed and the method of orbit integration and/or Hernquist potential is used. For the Hernquist potential, the CDM mass-radius relation follows~\citep{Erkal_2016}:
    \begin{equation}
    \label{eq:mass_radius_hernquist}
        r_s=\left(\frac{M_{\mathrm{sh}}}{10^8~\msun}\right)^{0.5} \times 1.05~\mathrm{kpc}.
    \end{equation}

\begin{figure}[t]
\centering
\includegraphics[width=\columnwidth]{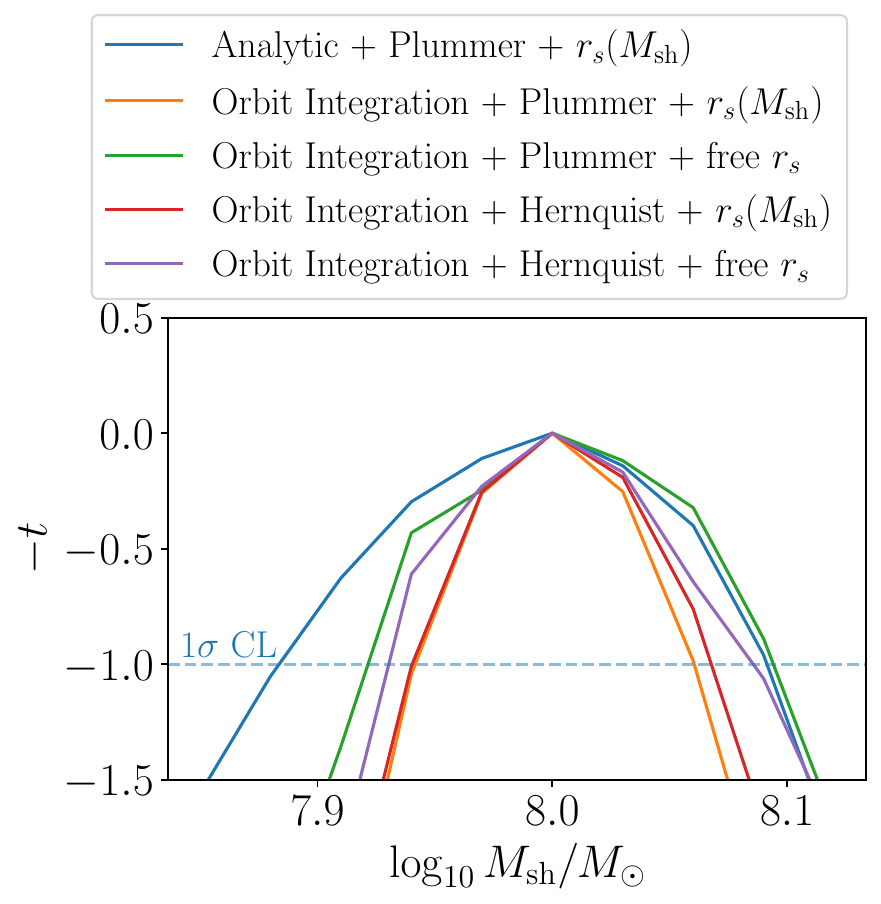}
\caption{Comparison of the test statistic $-t(M_\mathrm{sh})$, as defined in Eq.~\ref{eq:t}, using the analytic model and using an orbit integration method. The dataset is generated from $M_{\mathrm{sh}}^\mathrm{true} = 10^{8}~\msun$. The stream has properties similar to GD-1 ($\sigma_\theta=0.2^\circ$, $r_0=10~\mathrm{kpc}$, $\lambda=100~\mathrm{deg}^{-1}$) and the observational scenario is LSST 10 year sensitivity.}
\label{fig:CI_subhalo_potential}
\end{figure}

Fig.~\ref{fig:CI_subhalo_potential} shows a comparison of the modified approaches with our default approach (Analytic + Plummer + $r_s(M_\mathrm{sh})$). In all cases, we evaluated the test statistic on the Asimov data set. We find that using orbit integration generally results in tighter confidence intervals, even when $r_s$ is a free parameter of the fit. This indicates that our assumption of $r_s(M_\mathrm{sh})$ is not overly restrictive, and in fact the model likely underestimates the detectability compared to orbit integration. The results also show that Plummer and Hernquist potentials give similar signal strengths, as long as consistent $r_s$ values are used (see also the lower right panel of Fig.~\ref{fig:oi}).  Assessing the impact on the minimum detectable subhalo mass is much more expensive computationally, since it requires many more statistical realizations, and therefore we reserve this type of analysis for dedicated studies of individual streams.

More generally, the orbit integration method can be used to account for many of the concerns above. This would allow us to go beyond circular orbits and adopt a more general Milky Way potential, including known structures like the bar, spiral arms, and the Large Magellanic Cloud. It also provides an intermediate approach to modeling perturbed streams, which avoids orbit integration of the entire stream. It is a good approximation for CDM subhalos and again Gaussian noise can be added to account for dispersion. However, given that this approach is still much slower than the analytic model, we reserve this for future work.

\subsection{Comparisons with previous work}
\label{subsec:comparisons}

Our work draws on \cite{Erkal_2015_2}, which derived the analytic model. That work focused on parameter estimation from a subhalo impact rather than the minimum detectable subhalo. Using a Bayesian framework, the 6 observables from the analytic model are used to infer all the parameters of the subhalo impact at the same time. To address the subhalo velocity-mass degeneracy of the analytic model discussed above, they imposed a subhalo velocity prior rather than a mass-radius relationship. The confidence bands are then highly dependent on the velocity prior. Despite this difference as well as slight differences in the observational scenarios, the confidence intervals on subhalo mass estimation are very similar to our results for LSST-era data.

\cite{Erkal_2015_2} also estimated the minimum detectable subhalo mass based on individual observables, as a function of the error of that observable $\sigma_{\overline X}$. They similarly find that the shape of the stream on the sky (or $\overline{\Delta \theta}$) is generally the most discerning observable, with the potential to probe subhalos down to $\sim 10^6 \msun$. Our work provides a more robust result by performing a statistical test considering all 6 observables at the same time, while also accounting for different stream distances, densities, and dispersion. 

More recently, \cite{hilmi2024inferringdarkmattersubhalo} performed a similar analysis of parameter estimation for the stream ATLAS-Aliqa Uma, this time using simulated streams with orbit integration in a more realistic potential. By going beyond the analytic model, the subhalo velocity-mass degeneracy is broken. The method illustrates a more detailed study of a single stream, which is much more computationally intensive but also more realistic. It also provides a useful comparison for our approach based on the analytic model. 

We calculated confidence intervals on $M_\mathrm{sh}$ for a similar set of parameters and observational errors as in Fig. 6 of \cite{hilmi2024inferringdarkmattersubhalo}, except with a circular orbit for the stream. Specifically, we took a subhalo impact with $M_{\mathrm{sh}} = 10^{7}~\msun$, $r_s = 0.3~\mathrm{kpc}$, $b=0.1~\mathrm{kpc}$, $t=250~\mathrm{Myr}$, $w_r=\cos 250^\circ\times 35~\mathrm{km/s}$, $w_\parallel=-10~\mathrm{km/s}$ and $w_z=\sin 250^\circ\times35~\mathrm{km/s}$. The stream properties are assumed to be $\sigma_\theta=0.3^\circ$, $r_0=10~\mathrm{kpc}$, $\lambda=268~\mathrm{deg}^{-1}$ and the observational scenario is the LSST 10 year sensitivity. Using the Asimov data set for the expected confidence interval, we find a $1\sigma$ confidence interval of $[0.96, 1.03]\times 10^7~\msun$ for our analytic model with a fixed $r_s(M_{\rm sh})$ relationship. We also used the orbit integration method discussed in App.~\ref{sec:oi} without any assumed $r_s(M_{\rm sh})$ relationship, finding $[0.97, 1.03]\times 10^7~\msun$ at 1$\sigma$. Both results are similar to their result of $[0.94, 1.02]\times 10^7~\msun$. Despite the differences in orbit, statistical framework, and observational scenarios, this comparison adds confidence that our approach gives a good first estimate for subhalo sensitivity in streams.

Another important work estimating subhalo detection in stellar streams is \cite{drlicawagner2019probing}. Focusing on detectability of a gap, they obtain the minimum detectable subhalo mass as a function of the stream surface brightness, which implicitly decides the stream density. While this work considered different stream distances and surface brightness, it did not consider stream width, which is another important factor in detectability. For a thin stream, the results are similar. Considering a stream with $\mu=31.5$ mag/arcsec$^2$, $\sigma_z=20$ pc  and $r_0=20$ kpc,  the corresponding parameters are $\sigma_\theta=0.06^\circ$ and $\lambda=222$ deg$^{-1}$. Plugging this into our Eq.~\ref{eq:fitting}, we find $M_\mathrm{sh}^\mathrm{min}=3.3 \times 10^5~\msun$ at 95\% CL for LSST 10 year. This is somewhat smaller than their result of $10^6~\msun$. However, that work requires a gap present at $5\sigma$ which is a higher threshold, and does not use other observables.

All of the above have focused on detectability of a single strong impact. Previous studies have estimated the number of strong impacts for a given stream~\citep{Erkal_2016,Barry:2023ksd,menker2024,Adams:2024zhi}, which has typically led to $O(1)$ detectable gaps. For instance, recent works \citep{menker2024,Adams:2024zhi} have evaluated the distribution of gap sizes using semi-analytic models of structure formation. \cite{menker2024} uses the same analytic model of gap evolution for circular orbits, but improves upon it by numerically calculating velocity kicks for more realistic DM subhalo potentials. Meanwhile, \cite{Adams:2024zhi} goes beyond circular orbits and calculates velocity kicks using a 1D model of the stream, and then simulates stronger stream-subhalo encounters using \texttt{gala}. An interesting future direction is to evaluate the number of detectable impacts accounting for the properties of the stream, observables beyond density perturbations, as well as the DM subhalo population in a cosmological setting.

\section{Conclusions}
\label{sec:conclusion}

In this work, we have studied the detectability of DM subhalo impacts on stellar streams, taking into account stream properties such as width, distance, and stream density, as well as different observational scenarios. We have performed our study under the assumption of circular orbits for the stream, using the analytic model from~\cite{Erkal_2015_2} to enable fast simulation of impacts.
Our statistical framework allows us to estimate the minimum detectable subhalo mass and the confidence interval for the detected subhalo mass in a given stream. One of our main results is a fitting formula, Eq.~\ref{eq:fitting}, that estimates the minimum detectable subhalo mass as a function of stream properties and for various observational scenarios. We applied this formula to 50 confirmed streams in the Milky Way with reported stellar mass, sufficient angular length, and sufficient stellar density. This led to a ranked list which identifies promising candidates in terms of DM subhalo impact detectability, as shown in Tab.~\ref{tab:stream_catalog}. The following are our main conclusions:

\begin{itemize}

\item With a baseline observational scenario representing idealized \Gaia\ data, the minimum detectable subhalo mass for a stream like GD-1 (width $\sigma_\theta=0.2^\circ$, distance $r_0=10~\mathrm{kpc}$, stellar number density $\lambda=100~\mathrm{deg}^{-1}$) is $\sim6\times10^6~\msun$ at 95\% confidence level. This threshold mass can be improved by almost an order of magnitude down to $\sim8\times10^5~\msun$ if an optimistic observational scenario with 10-year LSST sensitivity is assumed (Sec.~\ref{sec:min_msh}).

\item In the case of a strong impact by a subhalo of mass $10^8~\msun$, we see a similar improvement in subhalo mass estimation.  The $2\sigma$ confidence interval is 1.5 (0.5) orders of magnitude around the true $M_{\rm sh}$ value for the baseline \Gaia\ (10-year LSST) scenario (Sec.~\ref{sec:msh_confidence}).

\item Detectability depends strongly on stream properties for all observational scenarios. As an example, for the baseline \Gaia\ scenario, an approximate scaling is given by $M_\mathrm{sh}^\mathrm{min}\propto\sigma_\theta^{1.2} r_0^{1.9} \lambda^{-0.8}$. There is a strong dependence on the distance of the stream $r_0$, which governs the number of observable stars. For 10-year LSST sensitivity at a magnitude limit of $r < 27$, the dependence on $r_0$ is mitigated, with $M_\mathrm{sh}^\mathrm{min}\propto\sigma_\theta^{0.98} r_0^{1.0} \lambda^{-0.8}$ (Sec.~\ref{sec:minimum_mass_streams}).

\item Both the shift in position perpendicular to the stream's orbital plane and the radial velocity are important observables to the detectability of an impact (Sec.~\ref{sec:min_msh}). Depending on the direction of the subhalo velocity relative to the stream motion, the importance of one observable could dominate over the other (Fig.~\ref{fig:q0_breakdown_w_dir}).

\item A minimum stream length of $\sim 20$ degrees is required in order to capture the perturbation caused by low-mass ($\sim 10^6~\msun$) subhalos. This guarantees at least $\sim 8$ degrees of angular coverage for the impact analysis  (Sec.~\ref{sec:min_stream_length}), though the exact requirement may vary based on stream and impact properties.

\item The above results are for our default impact configuration with zero impact parameter. Going beyond the default parameters, the minimum detectable subhalo mass increases with subhalo velocity and subhalo radius. The sensitivity is fairly constant for an impact parameter up to several times the stream width, beyond which the sensitivity drops. Dependence on other subhalo properties and configurations is mostly mild (Sec.~\ref{sec:nuisance}).

\item Having applied Eq.~\ref{eq:fitting} to known streams in the Milky Way, we have identified promising candidate streams like C-12, ATLAS-Aliqa Uma, 300S, NGC 6397, Palomar 5 and GD-1 which have the best sensitivity to low mass subhalos (see Tab.~\ref{tab:stream_catalog} and Sec.~\ref{sec:stream_catalog}). While some of the most promising streams have been studied in detail for DM impacts already (Pal-5, GD-1 in particular), there are many others which have received more limited attention. This work adds motivation for more detailed follow up analyses and observations. 
\end{itemize}

Though our results are based on an analytic model,  which has a number of limitations, they compare very well to more sophisticated estimates done on some real streams, as discussed in Sec.~\ref{subsec:comparisons}. A thorough analysis of individual streams is much more expensive, but can take into account the actual stream geometry, the Milky Way potential including baryonic perturbers, the actual position of the observer, and other effects including background sources of perturbations and sources of errors. We briefly discussed potential generalizations in Sec.~\ref{sec:generalization} and the method of orbit integration in App.~\ref{sec:oi}, which can take into account real stream orbits and a more general Milky Way potential, but leave the other details to future work.

Furthermore, we emphasize that the stream ranking on the minimum detectable subhalo mass we calculated in this work is not the complete picture for the most promising streams. The detection prospects also depend on the expected rate of subhalo impacts, which potentially depends on stream length and stream age, as well as baryonic backgrounds. By incorporating all these effects, one would eventually probe the DM subhalo distribution by testing the expected number of detectable impacts against observations. Sensitivity to the distribution will be improved by increasing the effective exposure, and multiple streams can be considered in the analysis together for this purpose. Our work in this paper provides a first step to quickly assess the discovery threshold of subhalo impacts for a general stream in the Milky Way, and the remained steps will be done in future work.


\section*{Acknowledgements}
We are grateful to participants of the KITP program ``Dark Matter Theory, Simulation, and Analysis in the Era of Large Surveys'' for valuable discussions and questions that helped improve this work, and to Ethan Nadler and David Shih for feedback on a draft of this manuscript. We also thank Alex Drlica-Wagner, Adrian Price-Whelan, and group members at UCSD including Bhavya Gupta, Tiffany Liou, Nicole Liu, and Brigette Vazquez-Segovia for helpful discussions.  JL and TL were supported by the US Department of Energy Office of Science under Award No. DE-SC0022104, a Lattimer Research Fellowship, and a Harold and Suzy Ticho Endowed Fellowship. MS's work was supported by the US Department of Energy Office of Science under Award No. DE-SC0022104 and No. DE-SC0009919, the Research Network Quantum Aspects of Spacetime (TURIS), and funded/co-funded by the European Union (ERC, NLO-DM, 101044443).

{\emph{Software:}}
Python \citep{python}, 
numpy \citep{numpy:2020}, scipy \citep{scipy:2020}, 
astropy \citep{astropy_2013, astropy_2018, The_Astropy_Collaboration_2022}, 
jupyter \citep{jupyter}, matplotlib \citep{matplotlib}, gala \citep{gala}. 

\bibliographystyle{mnras}
\bibliography{refs}

\appendix

\section{Distribution of test statistic}
\label{app:q0_dist}

\begin{figure*}[t]
\centering
\includegraphics[width=\textwidth]{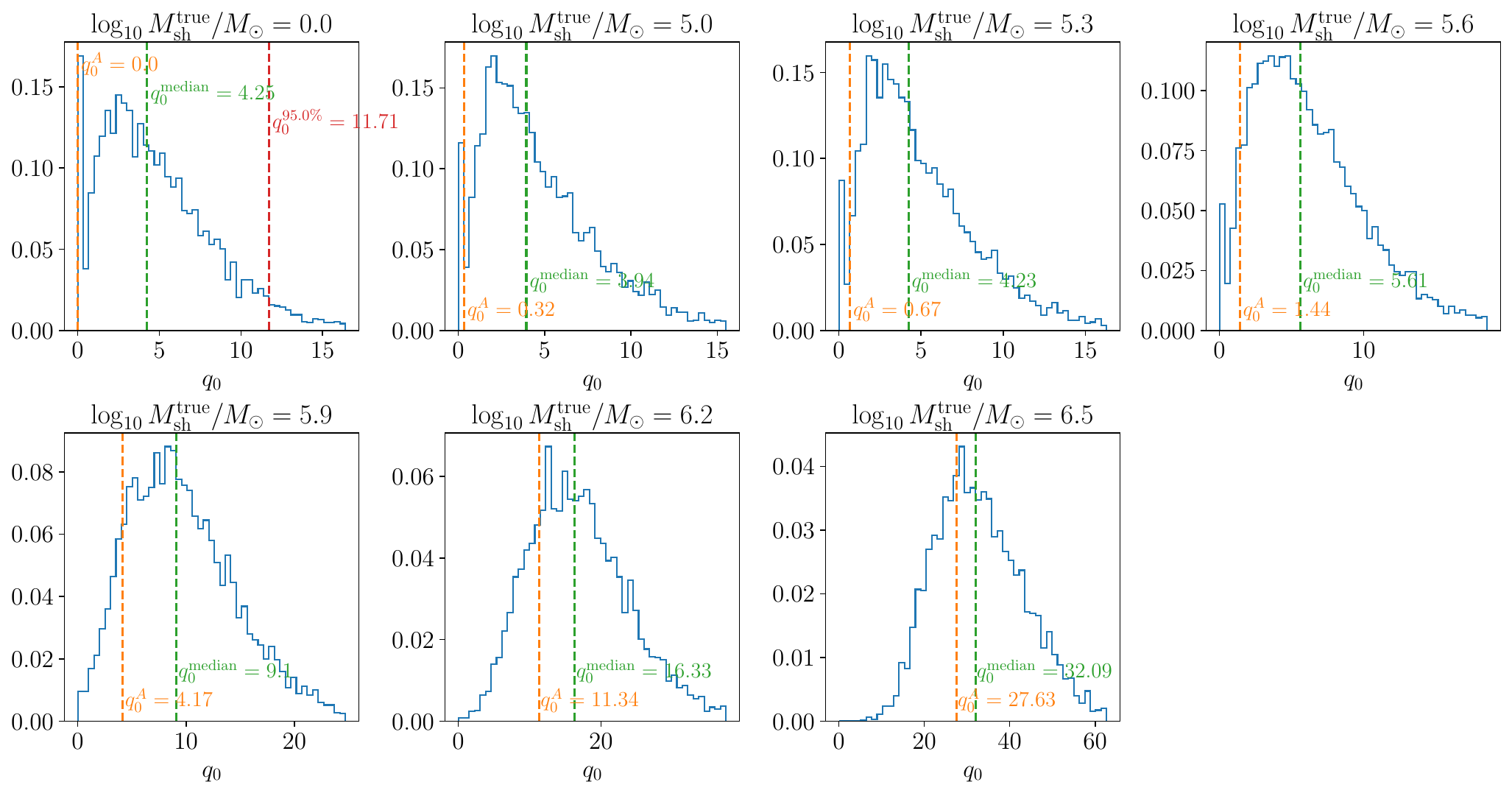}
\caption{$q_0$ distribution for a GD1-like stream ($\sigma_\theta=0.2^\circ$, $r_0=10~\mathrm{kpc}$, $\lambda=100~\mathrm{deg}^{-1}$) under ``Via + Gaia + LSST10'' scenario for different $M_\mathrm{sh}^\mathrm{true}$.}
\label{fig:q0_dist}
\end{figure*}

As discussed in Sec.~\ref{sec:min_msh}, the distribution of $q_0$ is essential for deciding the threshold for a detectable subhalo at 95\% confidence level. It is also needed in computing the median and $1\sigma$ band in Fig.~\ref{fig:ts_for_signal}. Here we pick a stream like GD-1 ($\sigma_\theta=0.2^\circ$, $r_0=10~\mathrm{kpc}$, $\lambda=100~\mathrm{deg}^{-1}$) under the ``Via + Gaia + LSST10'' observational scenario, and provide a more detailed look at the probability distribution of $q_0$ at different subhalo masses $M_\mathrm{sh}^\mathrm{true}$. For each $M_\mathrm{sh}^\mathrm{true}$, we generate 5000 realizations of the datasets, and calculate $q_0$ for each dataset. The resulting distributions of $q_0$ are shown in Fig.~\ref{fig:q0_dist}.

\begin{figure*}[t]
\centering
\includegraphics[width=0.48\textwidth]{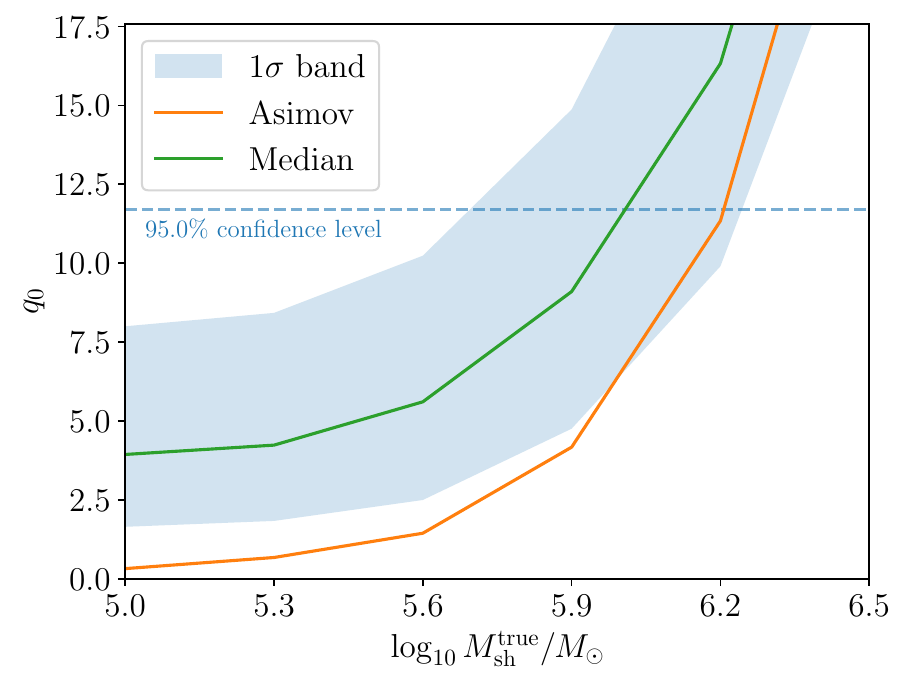}
\caption{The distributions from Fig.~\ref{fig:q0_dist} are condensed into a single plot of $q_0$ as a function of $M_\mathrm{sh}^\mathrm{true}$.}
\label{fig:q0_dist_combined}
\end{figure*}

\begin{figure*}[t]
\centering
\includegraphics[width=0.32\textwidth]{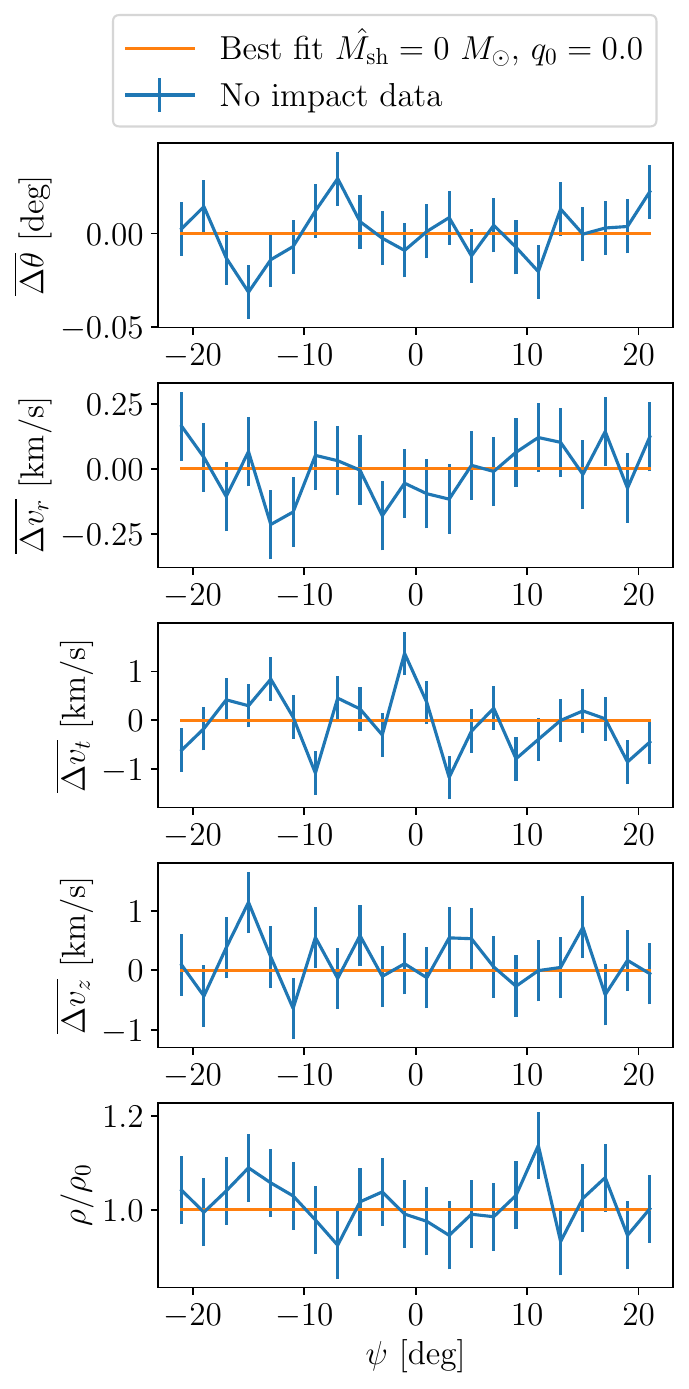}
\includegraphics[width=0.328\textwidth]{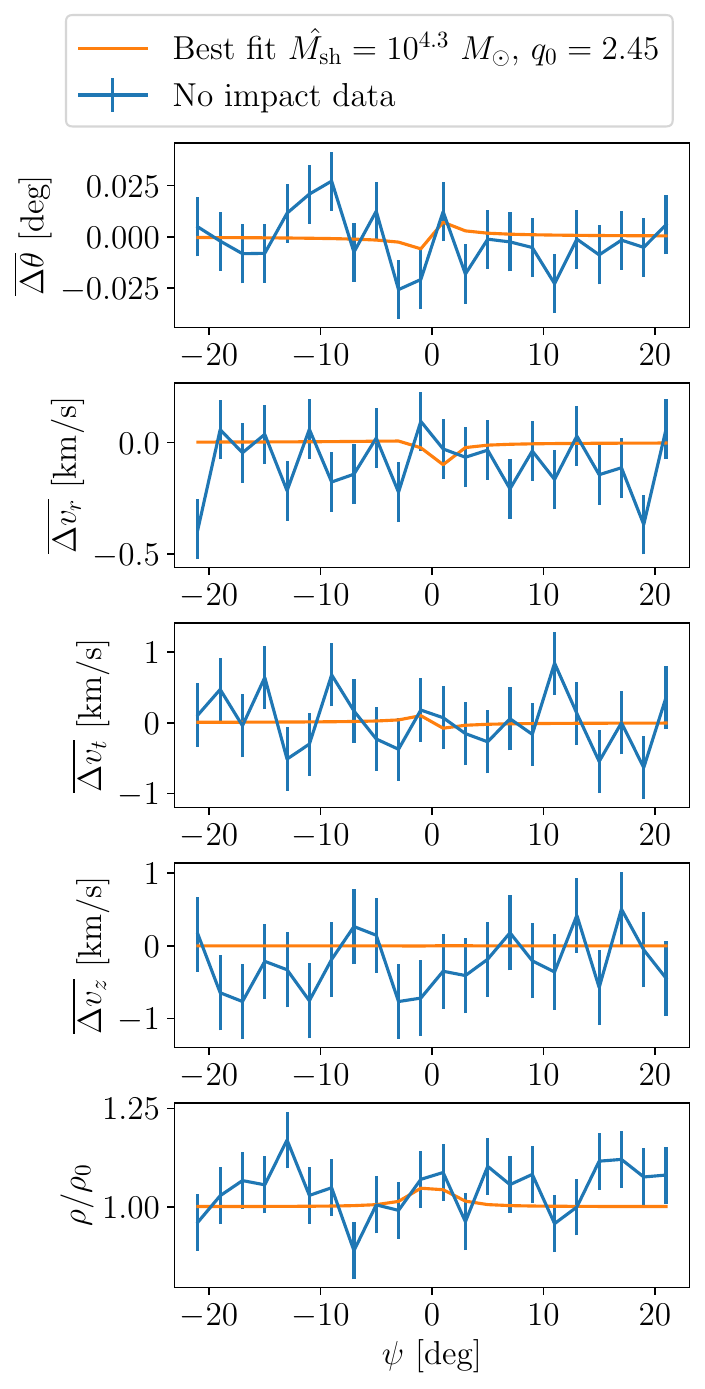}
\includegraphics[width=0.32\textwidth]{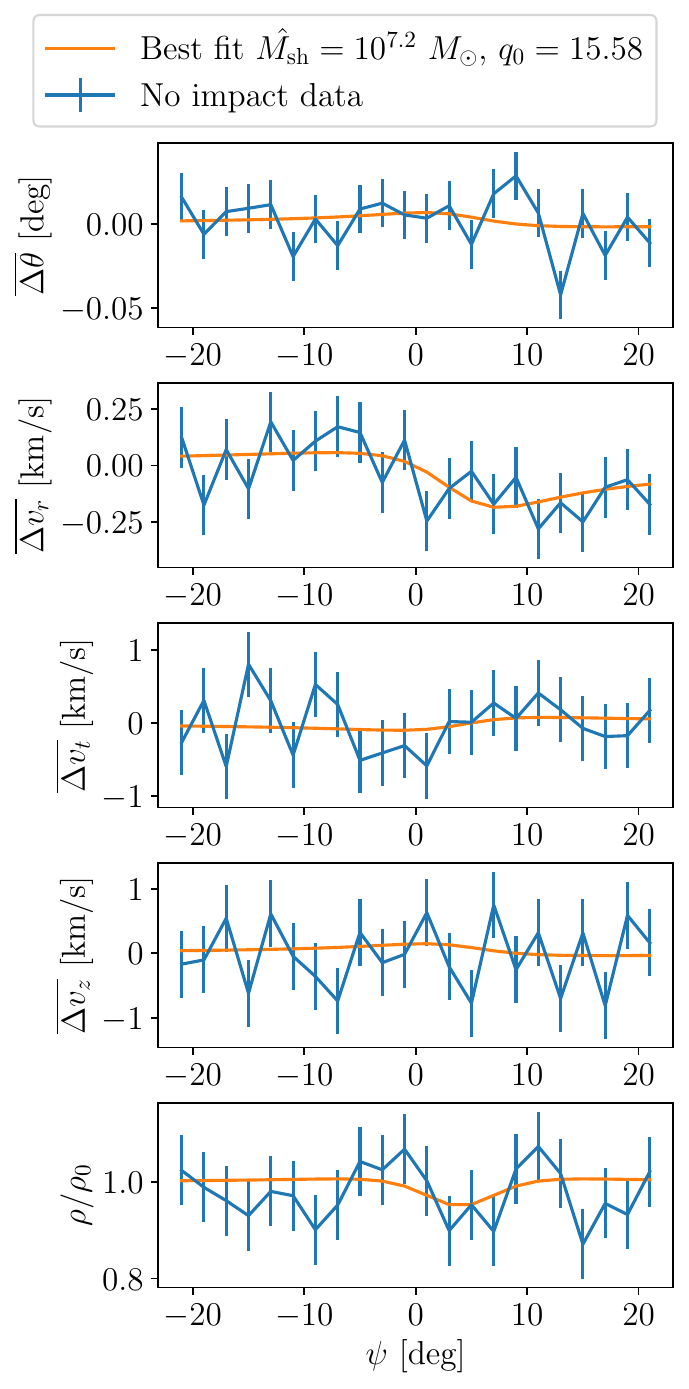}
\caption{Some examples of best fit models for data with no impact ($M_\mathrm{sh}^\mathrm{true}=0$) and Gaussian noise. The simulated data are based on a GD-1 like stream under ``Via + Gaia + LSST10'' scenario.}
\label{fig:best_fit_for_low_mass}
\end{figure*}

The first subplot is for $M_\mathrm{sh}^\mathrm{true}=0$, meaning no subhalo impact. The distribution is a mixture of a delta function at $q_0 = 0$ and a noncentral chi-squared distribution. The 95th percentile value is marked in red dashed line, which determines the $q_0$  detectability threshold at 95\% CL for this particular stream and observational scenario. Another notable feature is the presence of a gap between the Asimov value of $q_0^A$ (orange dashed line) and the median value $q_0^\mathrm{median}$ (green dashed line) at $M_\mathrm{sh}^\mathrm{true}=0$. This is because both positive and negative fluctuations for the no-impact data may look like a signal, leading to a median $q_0 > 0$. In contrast, the Asimov data for no impact just consists of uniform lines for all observables. Fig.~\ref{fig:best_fit_for_low_mass} shows some realizations of the no-impact data including noise (blue) and their best fit models (orange). We show example cases where the best fit is $\hat M_\mathrm{sh} = 0$, but also cases where the best fit is as large as $\hat M_\mathrm{sh} = 10^{7.2} \msun$.

Returning to Fig.~\ref{fig:q0_dist}, the subsequent subplots show the $q_0$ distribution for higher injected subhalo masses. At high masses, the distribution eventually approaches a Gaussian, and $q_0^A$ and $q_0^\mathrm{median}$ become much closer. With these distributions, we can obtain the median and 1$\sigma$ band of $q_0$ as a function $M_\mathrm{sh}^\mathrm{true}$ as shown in Fig.~\ref{fig:q0_dist_combined}. We use the intersection of the band with the $q_0$ threshold at 95\% CL to find the minimum detectable subhalo mass for this stream and observational setting.

\begin{figure*}[h]
\centering
\includegraphics[width=0.9\textwidth]{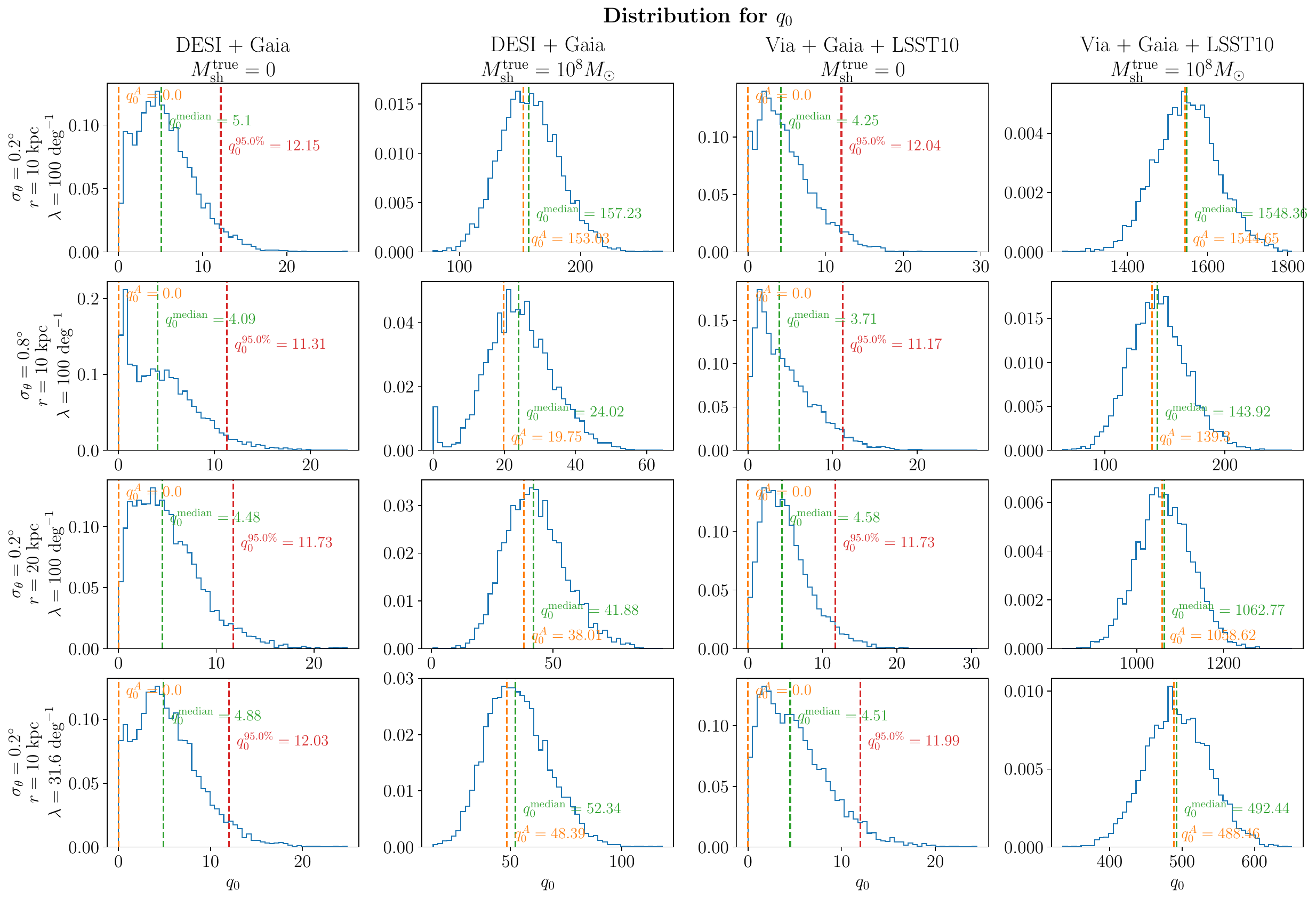}
\includegraphics[width=0.9\textwidth]{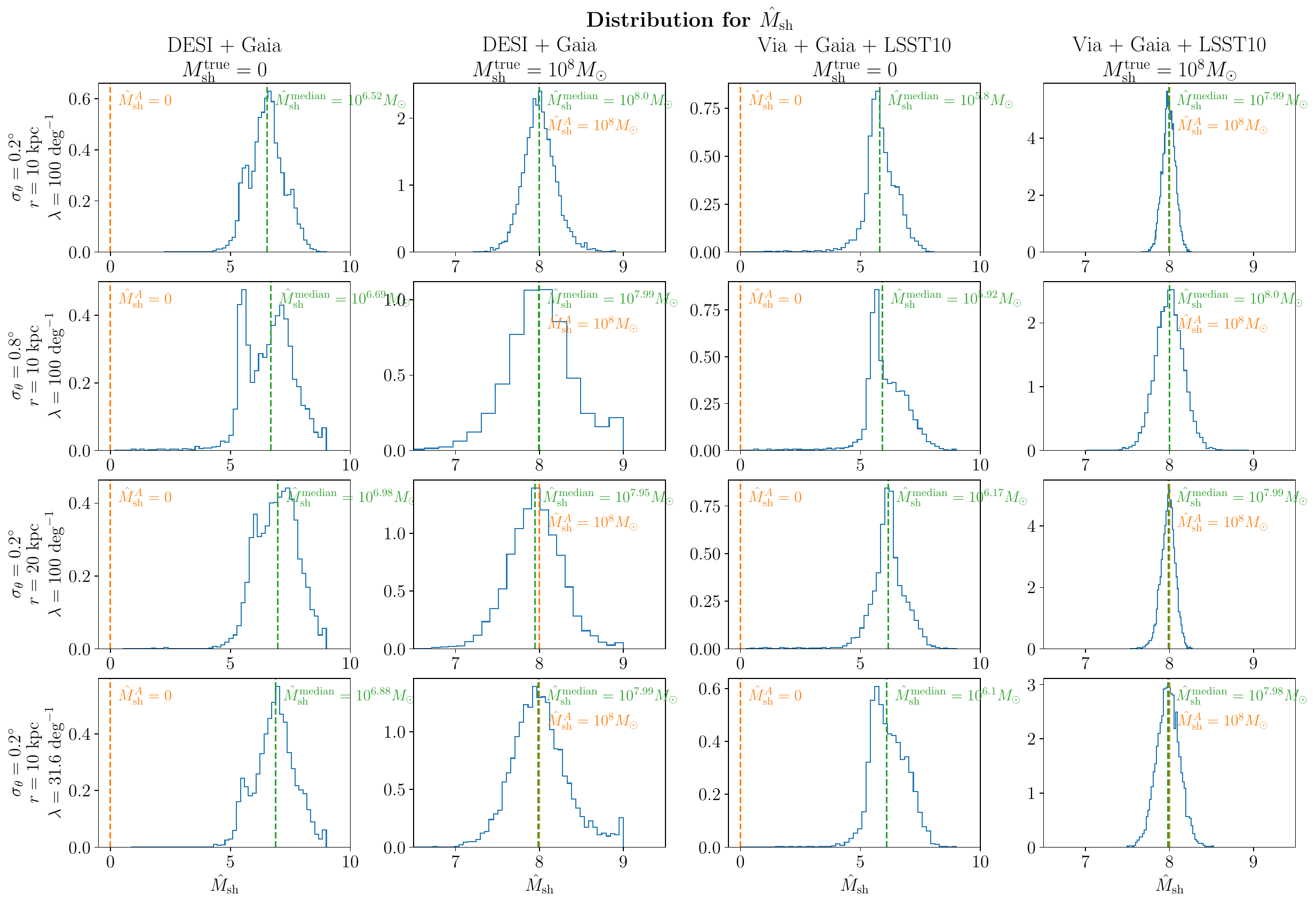}
\caption{\textbf{Top:} $q_0$ distribution for different stream properties and observational scenarios. The first row is for the default GD-1 like stream, while in subsequent rows we vary the thickness (second row), distance (third row), and density (fourth row). The columns correspond to two observational scenarios (DESI + Gaia and Via + Gaia + LSST10) at mass points $M_\mathrm{sh}^\mathrm{true}=0$ (no subhalo impact) and $M_\mathrm{sh}^\mathrm{true}=10^8~\msun$ (high subhalo mass). \textbf{Bottom:} The corresponding distribution for $\hat M_\mathrm{sh}$, the subhalo mass that maximizes the likelihood function, with the same set of settings as top.}
\label{fig:q0_dist_all_settings}
\end{figure*}

We next see how the distribution varies with different stream properties and observational scenarios. To this end, we plot the $q_0$ distribution for two representative mass points ($M_\mathrm{sh}^\mathrm{true}=0$ for no impact and $M_\mathrm{sh}^\mathrm{true}=10^8~\msun$ for large impact) for 8 different settings in the top panel of Fig.~\ref{fig:q0_dist_all_settings}. In the case of $M_\mathrm{sh}^\mathrm{true}=0$ (first and third columns), we see there are slightly different thresholds for $q_0$ at 95\% CL depending on the stream and scenario. 
In the case of $M_\mathrm{sh}^\mathrm{true}=10^8~\msun$ (second and fourth columns), the mean and variance of the distribution depend much more strongly on stream and scenario. However, a generic feature is that the value on Asimov data set $q_0^A$ (orange vertical line) and the median value $q_0^\mathrm{median}$ (green vertical line) become quite similar. This supports our usage of Asimov data set to obtain the median confidence interval for a high subhalo mass in Sec.~\ref{sec:msh_confidence}.

The bottom panel of Fig.~\ref{fig:q0_dist_all_settings} further illustrates the phenomena discussed above. Here we plot the corresponding distribution for the optimal subhalo mass that maximizes the likelihood function, $\hat M_\mathrm{sh}$. In the case of $M_\mathrm{sh}^\mathrm{true}=0$ (first and third columns), we see best fit values clustered around nonzero subhalo mass, due to random fluctuations in the data mimicking a signal. In the case of $M_\mathrm{sh}^\mathrm{true}=10^8~\msun$ (second and fourth columns), the median best fit is close to the true value.

\FloatBarrier

\section{Orbit integration method}
\label{sec:oi}

The analytic model is extremely fast for generating simulated data. However it has some major limitations: 1) it only supports circular orbits and a spherical host potential; 2) it uses the impulse approximation which requires the impact geometry to satisfy Eq.~\ref{eq:analytic_assumption},  and breaks down when we have a large subhalo scale radius $r_s$ (e.g. a CDM subhalo with $M_\mathrm{sh}=10^9~\msun$ and $r_s=5$ kpc), small stream distance $r_0$ (e.g. streams at a distance of 5 kpc from us) or small subhalo velocity perpendicular to the stream $w_\perp$ (e.g. subhalo moving in the same or opposite directions as the stream); 3) it only works for subhalos with a Plummer potential; 4) it has an exact degeneracy between subhalo velocity and subhalo mass, which prevents us from independently inferring all the subhalo parameters. 

To probe the effects of subhalo impact beyond these constraints, we explore a much more flexible method to generate the impacted data -- orbit integration (OI). In this method, we assume points on the orbit of the progenitor to be the stream stars, and perform orbit integration on these stream stars under the host potential as well as the subhalo potential. Thus it works for any stream orbit, any host potential, any subhalo potential, and any encounter geometry. While it still makes an approximation that the stream star is very close to the progenitor orbit, it is much less expensive than simulating all stream stars. Dispersions can be still obtained by simulating the stream in the absence of an impact.

\begin{figure*}[t]
\centering
\includegraphics[width=0.49\textwidth]{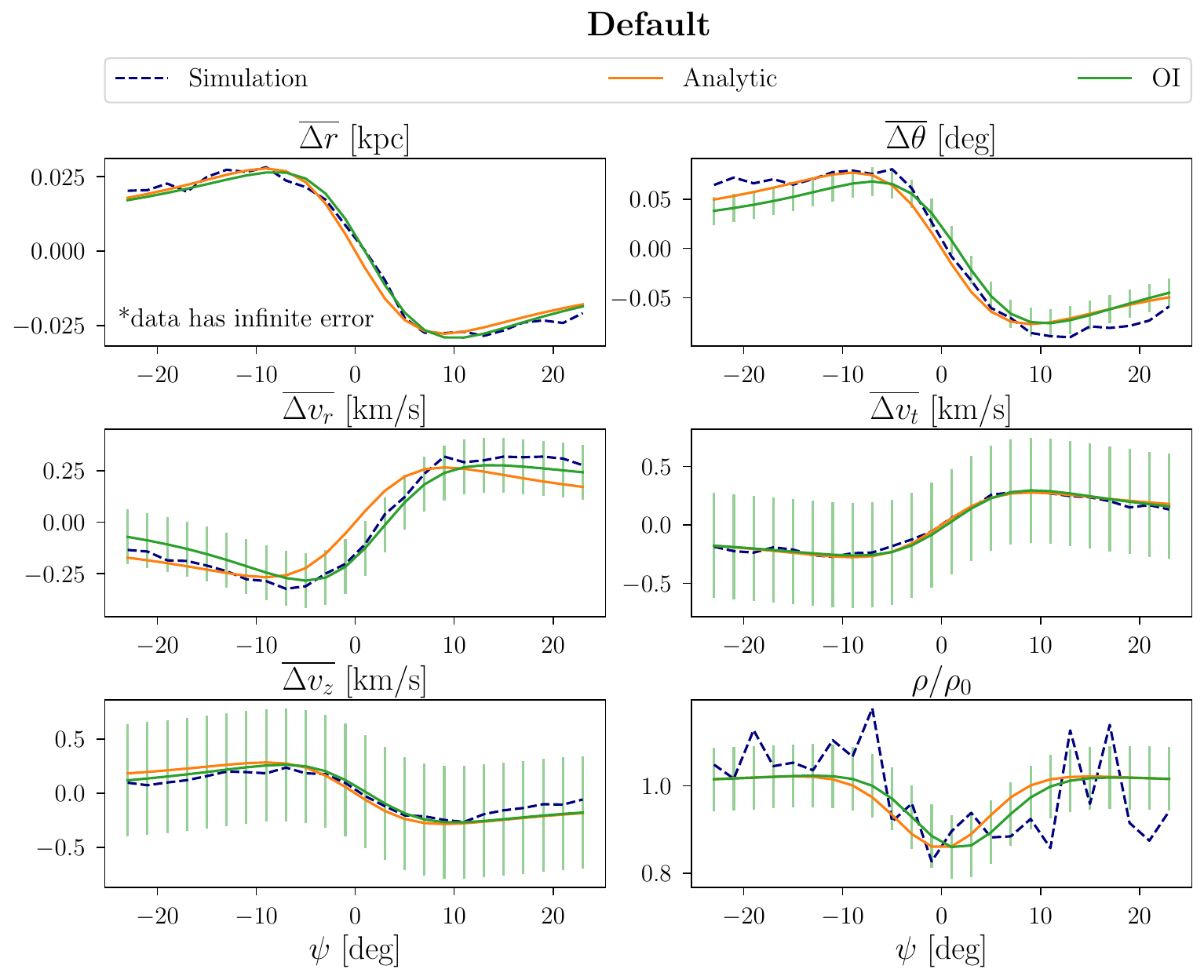}
\includegraphics[width=0.49\textwidth]{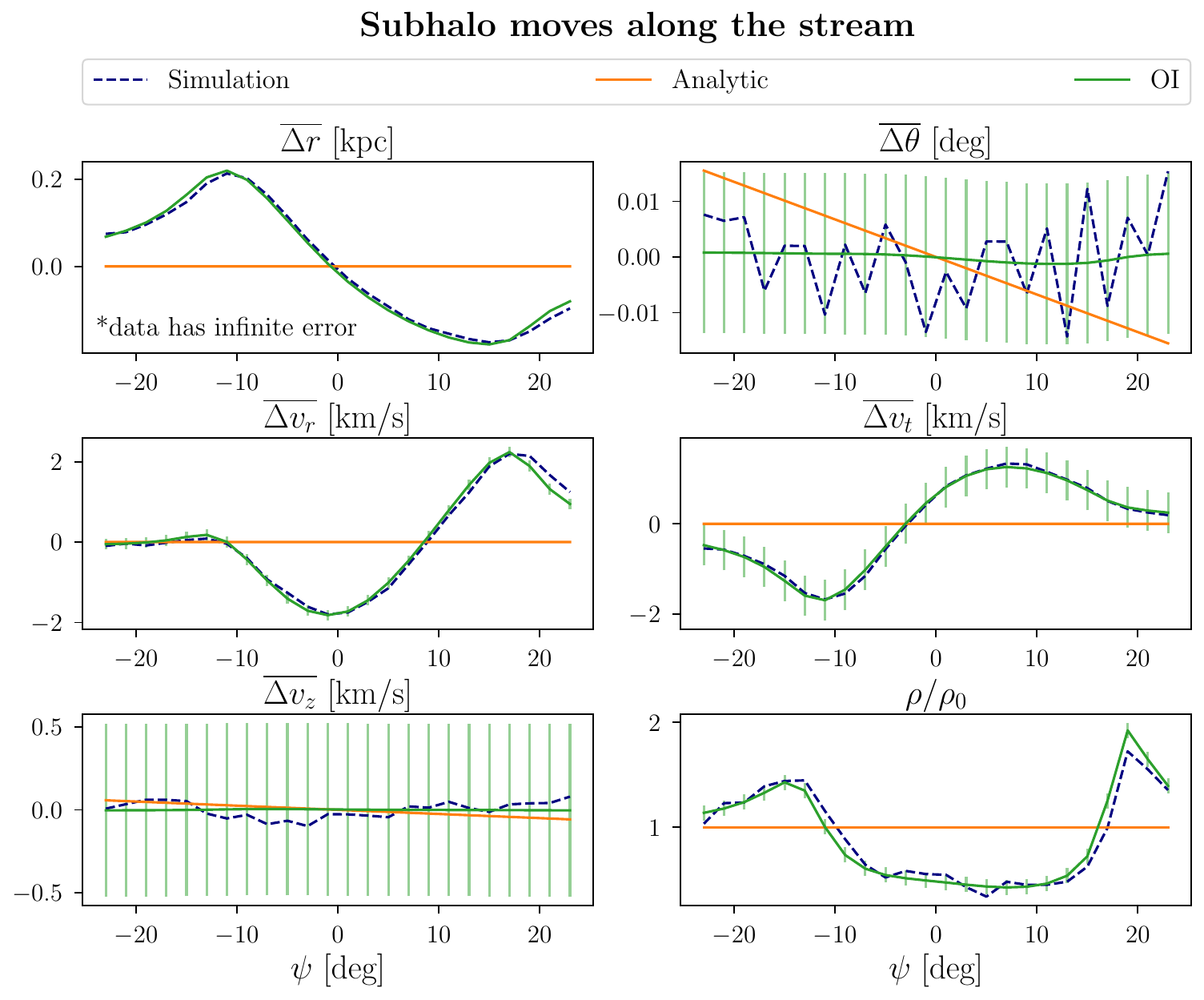}\\
\hspace{0.1in}\\
\includegraphics[width=0.49\textwidth]{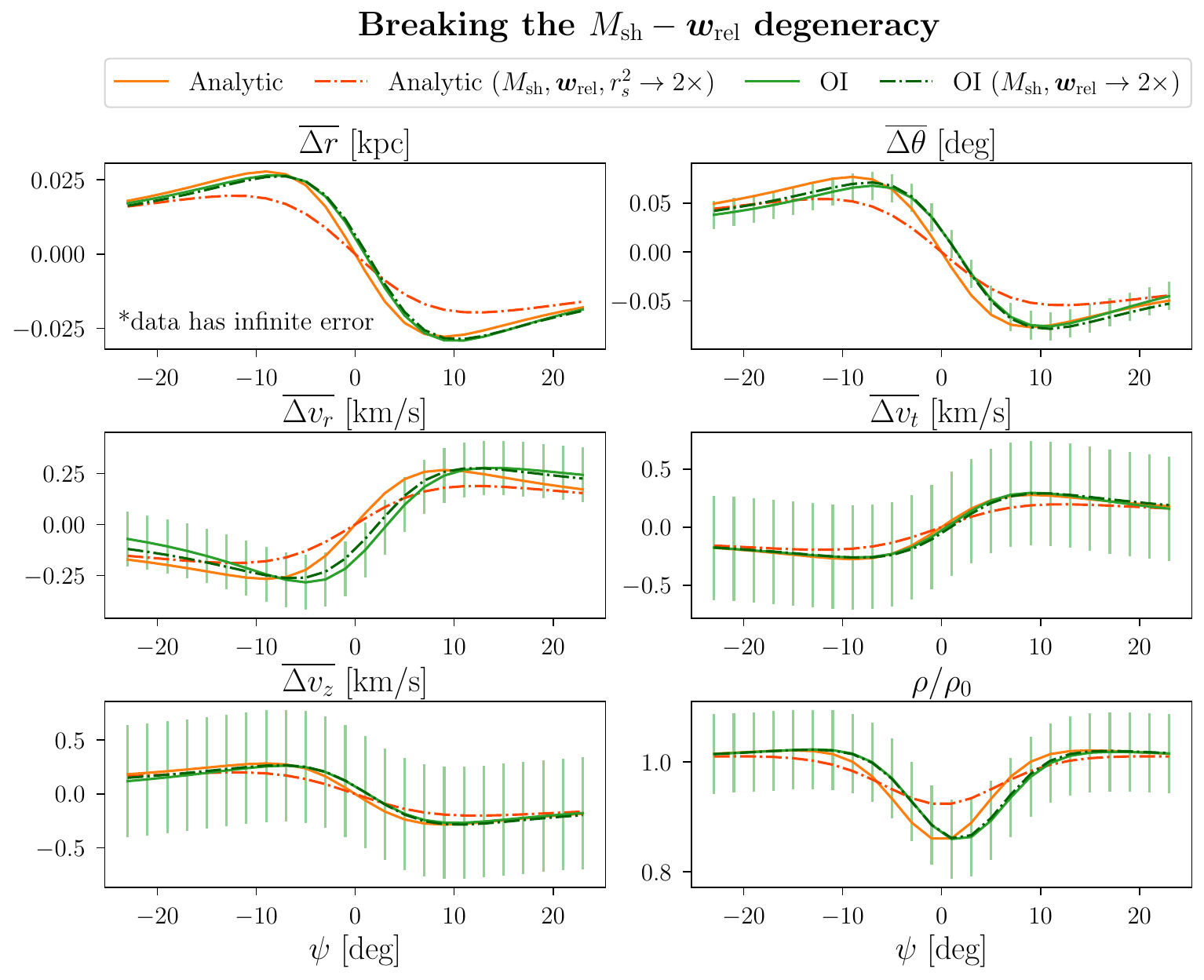}
\includegraphics[width=0.49\textwidth]{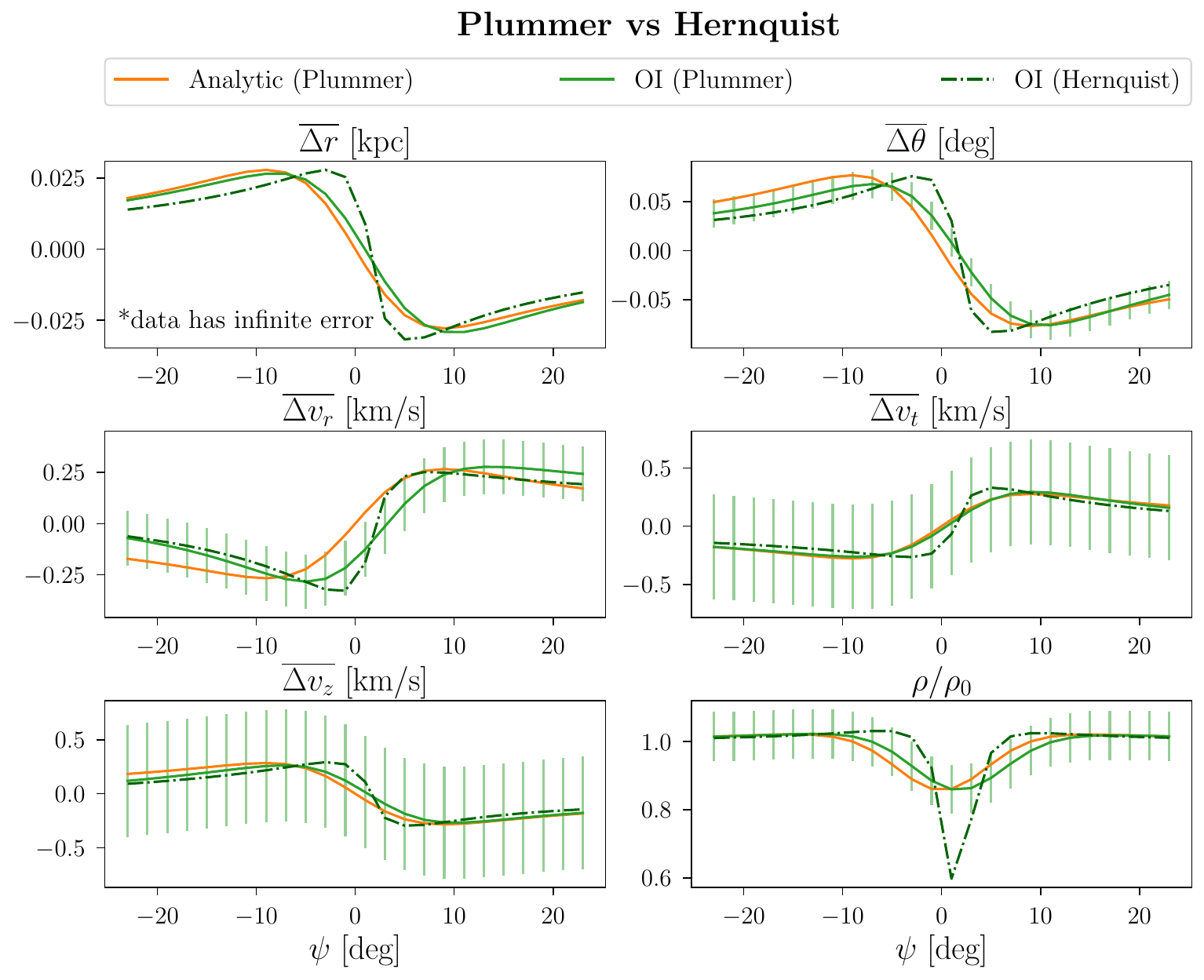}
\caption{
\textbf{Upper left: } A comparison of observables from simulation (dashed navy), analytic model (orange) and orbit integration (OI; green) under a subhalo impact of $M_\mathrm{sh}=10^{7.5}~\msun$. Other parameters are the default values given in Tab.~\ref{table:default_impact}, and in particular the subhalo moves in the default $\hat z$ direction. 
\textbf{Upper right: } Same as upper left, but with subhalo moving along the stream ($w_r=0$ km/s, $w_t=180$ km/s and $w_z=0.1$ km/s), in which case the analytic model fails completely.
\textbf{Lower left: } Same as upper left, but instead showing how the orbit integration result (solid green) changes when multiplying $M_\mathrm{sh}$ and $\bm{w}_\mathrm{rel}=(w_r, w_\parallel, w_z)$ by a factor of 2 (dash-dotted dark green). Changing $M_\mathrm{sh}$ and $\bm{w}_\mathrm{rel}$ simultaneously with the same factor does not lead to any difference in the analytic model, while orbit integration does show a small difference, breaking the degeneracy. To break the degeneracy in the analytic model, we need to assume a fixed $r_s(M_\mathrm{sh})$ relation according to Eq.~\ref{eq:mass_radius}, thus multiplying $r_s$ by a factor of $\sqrt{2}$ when multiplying $M_\mathrm{sh}$ and $\bm{w}_\mathrm{rel}$ by a factor of 2 (dash-dotted dark orange).
\textbf{Lower right: } Same as upper left, but instead showing how the orbit integration result (solid green) changes when using a Hernquist potential for the subhalo (dash-dotted dark green). The Hernquist potential is not supported by the analytic model for $b=0$.
In all panels, the error bars assume stream properties from a GD1-like stream and the best case observational scenario. For clarity, error bars are only shown on the default orbit integration case and the observables here do not include Gaussian noise. }
\label{fig:oi}
\end{figure*}

To illustrate the difference between the analytic model and the method of orbit integration, we show their side-by-side comparisons in four different cases in Fig.~\ref{fig:oi}. In the upper left panel, the subhalo moves in the default $z$ direction. We see that both analytic model (orange) and orbit integration (green) are close to the simulation results (dashed navy), meaning both work very well. While orbit integration is slightly closer to the simulation results, the differences are not large enough to significantly impact our results on subhalo detectability. However, there are some extreme cases where the assumption required by the analytic model Eq.~\ref{eq:analytic_assumption} breaks down. An example case is shown in the upper right panel of Fig.~\ref{fig:oi}, where the subhalo moves along the stream. We see that the orbit integration result (green) remains very close to the simulation results (dashed navy), while the analytic model (orange) is completely off.

Furthermore, we illustrate the ability of orbit integration to break the mass-velocity degeneracy of the analytic model, discussed below Eq.~\ref{eq:mass_radius} and in Sec.~\ref{sec:discussion}. In the lower left panel of Fig.~\ref{fig:oi}, we show two sets of results from orbit integration with the solid green one from $M_\mathrm{sh}=10^{7.5}~\msun$, $w_\mathrm{rel}=184$ km/s and the dash-dotted dark green one from $M_\mathrm{sh}=2\times10^{7.5}~\msun$, $w_\mathrm{rel}=2\times184$ km/s. With orbit integration, we see some small differences in the observables, which will not show up at all if we use the analytic model due to its exact degeneracy. This means with future surveys with small observational errors, we can in principle infer subhalo mass and subhalo radius at the same time using the method of orbit integration. (See also \cite{hilmi2024inferringdarkmattersubhalo} which demonstrated this using simulated streams.) To break the degeneracy in the analytic model, we use the $r_s(M_{\rm sh})$ relationship in Eq.~\ref{eq:mass_radius} (dash-dotted dark orange). While it appears that assuming $r_s(M_{\rm sh})$ breaks the degeneracy more strongly, this is a multi-dimensional parameter space and other parameters can partially compensate the change in observables. Fig.~\ref{fig:CI_subhalo_potential} shows that confidence intervals on $M_{\rm sh}$ are actually smaller when using orbit integration.

Lastly, in the lower right panel of Fig.~\ref{fig:oi}, we show a comparison using Plummer potential and Hernquist potentials for the subhalo with the method of orbit integration. The solid green line assumes a Plummer potential with CDM mass-radius relation according to Eq.~\ref{eq:mass_radius}. The dashed dotted dark green one assumes a Hernquist potential with CDM mass-relation according to Eq.~\ref{eq:mass_radius_hernquist}. As we can see, the difference is mostly within the $1\sigma$ range, with some exceptions around the center part of the impact, meaning the assumption of subhalo potential type won't change our main results much.

In the main text, we use this method to estimate how results change when the CDM mass-radius relation is not enforced or when the Hernquist potential is used (Fig.~\ref{fig:CI_subhalo_potential}). Because it is still much slower than the analytic model, we limit ourselves to results on the confidence intervals for subhalo mass. Applying this method for the minimum detectable mass is much slower since it requires many statistical realizations, and we leave this for future work focused more on realistic stream orbits.\\[3mm]

\end{document}